\documentclass[10pt]{article}

\usepackage{amsmath}
\usepackage{amssymb}

\usepackage{graphicx}

\usepackage{cite}

\usepackage{color} 

\usepackage[labelfont=bf,labelsep=period,justification=raggedright]{caption}

\makeatletter
\renewcommand{\@biblabel}[1]{\quad#1.}
\makeatother

\date{}

\usepackage{xspace}
\usepackage{hyperref}
\usepackage{placeins}
\usepackage{lscape}
\usepackage{appendix}

\newcommand{\nn}{\nonumber}

\newcommand{\Avg}[1]{\langle #1\rangle}
\newcommand{\TPAvg}[1]{\overline{\langle  {#1}\rangle_\phi}}
\newcommand{\ETPAvg}[1]{\overline{\langle \langle  {#1}\rangle_\phi\rangle_e}}
\newcommand{\HC}{n_{max}} 
\newcommand{\unit}[1]{\ensuremath{\, \mathrm{#1}}}
\newcommand{\umsps}{\unit{\mu m^2/s}}
\newcommand{\pcEL}{\unit{\% EL}}
\newcommand{\mum}{\unit{\mu m}}
\newcommand{\mysubsubsection}[1]{\textbf{#1.}}
\newcommand{\Fig}{Fig.\xspace}
\newcommand{\subfig}[1]{\textbf{(#1)}}
\newcommand{\SItext}{Text S1\xspace}
\catcode`_=\active
\def_#1{\ensuremath{\sb{\mathrm{#1}}}}
\newcommand{\sprm}[1]{\ensuremath{\sp{\mathrm{#1}}}}
\newcommand{\Bcd}{Bcd\xspace}
\newcommand{\Hb}{Hb\xspace}

\newcommand{\Cad}{Cad\xspace}
\newcommand{\Kni}{Kni\xspace}

\newcommand{\G}{G\xspace}
\newcommand{\bcd}{\textit{bcd}\xspace}
\newcommand{\hb}{\textit{hb}\xspace}

\newcommand{\kni}{\textit{kni}\xspace}
\newcommand{\kr}{\textit{kr}\xspace}
\newcommand{\gt}{\textit{gt}\xspace}

\begin{document}

\begin{flushleft} {\Large \textbf{Mutual Repression enhances the
      Steepness and Precision of Gene Expression Boundaries} }
\\
Thomas R. Sokolowski$^{1}$, 
Thorsten Erdmann$^{2}$, 
Pieter Rein ten Wolde$^{3,\ast}$
\\
\bf{1} FOM Institute AMOLF, Science Park 104, 1098 XG Amsterdam, The Netherlands
\\
\bf{2} University of Heidelberg, Institute for Theoretical Physics, Philosophenweg 19, 69120 Heidelberg, Germany
\\
\bf{3} FOM Institute AMOLF, Science Park 104, 1098 XG Amsterdam, The Netherlands
\\
$\ast$ E-mail: tenwolde@amolf.nl
\end{flushleft}

\section*{Abstract}
Embryonic development is driven by spatial patterns of gene expression
that determine the fate of each cell in the embryo. While gene
expression is often highly erratic, embryonic development is usually
exceedingly precise. In particular, gene expression boundaries are
robust not only against intra-embryonic fluctuations such as noise in
gene expression and protein diffusion, but also against
embryo-to-embryo variations in the morphogen gradients, which provide
positional information to the differentiating cells. How development
is robust against intra- and inter-embryonic variations is not
understood. A common motif in the gene regulation networks that
control embryonic development is mutual repression between pairs of
genes.
To assess the role of mutual repression in the robust formation
of gene expression patterns, we have performed large-scale stochastic
simulations of a minimal model of two mutually repressing gap genes 
in {\it Drosophila},  {\it hunchback} (\hb) and {\it knirps} (\kni).
Our model includes not only mutual repression between \hb and \kni,
but also the stochastic and cooperative activation of \hb by the anterior
morphogen  Bicoid (\Bcd) and of \kni by the posterior morphogen Caudal
(\Cad), as well as the diffusion of \Hb and \Kni between neighboring
nuclei.
Our analysis reveals that mutual repression can markedly increase the 
steepness and precision of the gap gene expression boundaries. 
In contrast to other mechanisms such as spatial averaging and cooperative 
gene activation, mutual repression thus allows for gene-expression 
boundaries that are both steep and precise. Moreover,
mutual repression dramatically enhances their robustness against
embryo-to-embryo variations in the morphogen levels. Finally, our
simulations reveal that diffusion of the gap proteins plays a critical
role not only in reducing the width of the gap gene expression
boundaries via the mechanism of spatial averaging, but also in
repairing patterning errors that could arise because of the
bistability induced by mutual repression.


\pagebreak

\section*{Introduction}
\pdfbookmark[1]{Introduction}{BookmarkIntro}
The development of multicellular organisms requires spatially
controlled cell differentiation. The positional information for the
differentiating cells is typically provided by spatial concentration gradients of
morphogen proteins.  In the
classical picture of morphogen-directed patterning, cells translate
the morphogen concentration into spatial gene-expression domains via
simple threshold-dependent readouts
\cite{Wolpert1969,Wolpert1994,Driever1988a,Driever1988b}.  Yet, while
embryonic development is exceedingly precise, this mechanism is not
very robust against intra- and inter-embryonic variations
\cite{Houchmandzadeh2002,Gregor2007,Gregor2007b}: the spatial patterns
of the target genes do not scale with the size of the embryo and the
boundaries of the expression domains are susceptible to fluctuations
in the morphogen levels and to the noise in gene
expression. Intriguingly, the target genes of morphogens often
mutually repress each other, as in the gap-gene system of the fruit
fly {\em
  Drosophila} \cite{Jackle1986,Clyde2003,
  Jaeger2004,Surkova2008,Manu2009PlosCompBiol,Manu2009PlosBiol,Vakulenko2009}.
To elucidate the role of mutual repression in the robust formation of
gene expression patterns, we have performed extensive
spatially-resolved stochastic simulations of the gap-gene system of
{\em Drosophila melanogaster}. Our results show that mutual repression between
target genes can markedly enhance both the steepness and the precision
of gene-expression boundaries. Furthermore, it makes them robust against
embryo-to-embryo variations in the morphogen gradients.

The fruit fly \textit{Drosophila melanogaster} is arguably the
paradigm of morphogenesis. During the first 90 minutes after
fertilization it is a syncytium, consisting of a cytoplasm that
contains rapidly diving nuclei, which are not yet encapsulated by
cellular membranes.  Around cell cycle 10 the nuclei migrate towards
the cortex of the embryo and settle there to read out the
concentration gradient of the morphogen protein Bicoid (\Bcd), which
forms from the anterior pole after fertilization \cite{Driever1988a}.
One of the target genes of \Bcd is the gap gene {\it hunchback} (\hb),
which is expressed in the anterior half of the embryo. In spite of
noise in gene expression, the midembryo boundary of the \hb expression
domain is astonishingly sharp. By cell cycle 11, the \hb mRNA
boundary varies by about one nuclear spacing only \cite{Porcher2010,
He2011, Perry2011}, while by cell cycle 13 a similarly sharp
oundary is observed for the protein level \cite{Houchmandzadeh2002,Gregor2007,He2008}.
This precision is higher than the best achievable precision for a 
time-averaging based readout mechanism of the Bcd gradient \cite{Gregor2007}.
Interestingly, the
study of Gregor {\it et al.} revealed that the Hb concentrations in
neighboring nuclei exhibit correlations and the authors suggested that
this implies a form of spatial averaging that enhances the precision
of the posterior Hb boundary \cite{Gregor2007}. Two recent simulation
studies suggest that the mechanism of spatial averaging is based on
the diffusion of Hb itself \cite{Erdmann2009,Okabe-Oho2009}; as shown
analytically in \cite{Erdmann2009}, Hb diffusion between neighboring
nuclei reduces the super-Poissonian part of the noise in its
concentration. In essence, diffusion reduces noise by washing out
bursts in gene expression. However, the mechanism of spatial averaging
comes at a cost: it tends to lessen the steepness of the expression
boundaries.

Bcd induces the expression of not only \hb, but a number of gap genes,
and pairs of gap genes tend to repress each other
mutually. Interestingly, repression between directly neighboring gap
genes is weak, whereas repression between non-adjacent genes is strong
\cite{Kraut1991}.  \hb forms a strongly repressive pair with {\it knirps}
(\kni) which is expressed further towards the posterior pole; 
both genes play a prominent role in the later positioning of
downstream pair-rule gene stripes \cite{Clyde2003}.
It has
been argued that mutual repression can enhance robustness to
embryo-to-embryo variations in morphogen levels
\cite{Manu2009PlosCompBiol,Manu2009PlosBiol,Vakulenko2009} and sharpen
a morphogen-induced transition between the two mutually 
repressing genes in a non-stochastic background \cite{Saka2007,Ishihara2008}.
However, mutual repression can also lead to bistability
\cite{Cherry2000,Kepler2001,Warren2004,Warren2005,Papatsenko2011}. While bistablity may buffer
against inter-embryo variations and rapid intra-embryo fluctuations in
morphogen levels, it may also cause stochastic switching between
distinct gene expression patterns, which would be highly detrimental.
Therefore, the precise role of mutual repression in the robust
formation of gene-expression patterns remains to be elucidated.

While the role of antagonistic interactions in the formation of
gene-expression patterns has been studied using mean-field models
\cite{Manu2009PlosCompBiol,Papatsenko2011,Ishihara2005,Zinzen2006,Zinzen2007},
to address the question whether mutual repression enhances the
robustness of these patterns against noise arising from the inherent
stochasticity of biochemical reactions a stochastic model is
essential. We have therefore performed large-scale stochastic
simulations of a minimal model of mutual repression between \hb and
\kni. Our model includes the stochastic and cooperative activation
of \hb by \Bcd and of \kni by the posterior morphogen Caudal (\Cad)
\cite{Rivera-Pomar1995,Schulz1995}. Moreover, \Hb and \Kni can diffuse
between neighboring nuclei and repress each other's expression,
generating two separate spatial domains interacting at midembryo (see
\Fig \ref{Fig1}). We analyze the stability of these domains by
systematically varying the diffusion constants of the \Hb and \Kni
proteins, the strength of mutual repression and the \Bcd and \Cad
activator levels. To quantify the importance of mutual
repression, we compare the results to those of a system containing
only a single gap gene, which is regulated by its morphogen only;
this is the ``system without mutual repression''. While our model is
simplified---it neglects, {\it e.g.}, the interactions of \hb and \kni
with {\it kr\"{u}ppel} (\kr) and {\it giant} (\gt)
\cite{Jaeger2011}---it does allow us to elucidate the mechanism by
which mutual repression can enhance the robust formation of gene
expression patterns.

One of the key findings of our analysis is that mutual repression
enhances the robustness of the gene expression domains against
intra-embryonic fluctuations arising from the intrinsic
stochasticity of biochemical reactions. Specifically, mutual
repression increases the precision of gene-expression
boundaries: it reduces the variation $\Delta x$ in their
positions due to these fluctuations.  At the same time, mutual
repression also enhances the steepness of the expression boundaries.
To understand the interplay between steepness, precision and
intra-embryonic fluctuations (biochemical noise), it is instructive
to recall that the width $\Delta x$ of a boundary of the expression
domain of a gene $g$ is, to first order, given by
\begin{equation}
\label{EqDeltaX}
\Delta x =\frac{\sigma_G (x_t)}{|\Avg{G(x_t)}^\prime|},
\end{equation}
where $\sigma_G(x_t)$ is the standard deviation of the copy number $G$
of protein \G and $|\Avg{G(x_t)}^\prime|$ is the magnitude of the
gradient of $G$ at the boundary position $x_t$
\cite{Gregor2007,Tostevin2007,Erdmann2009}. Steepness thus refers to
the slope of the average concentration profile,
$|\Avg{G(x_t)}^\prime|$, while precision refers to $\Delta x$,
which is the standard deviation in the position at which $G$
crosses a specified threshold value, here taken to be the
half-maximal average expression level of $G$. 

The simulations reveal, perhaps surprisingly, that mutual repression
hardly affects the noise $\sigma_G(x_t)$ at the expression boundaries
of \hb and \kni. Moreover, mutual repression can strongly enhance the
steepness $|\Avg{G(x_t)}^\prime|$ of these boundaries: the steepness
of the boundaries in a system with mutual repression can, depending on
the diffusion constant, be twice as large as that in the system
without mutual repression. Together with Eq. \ref{EqDeltaX}, these
observations predict that mutual repression can significantly enhance
the precision of the boundaries, i.e. decrease $\Delta x$, which
is indeed precisely what the simulations reveal. Interestingly, there
exists an optimal diffusion constant that minimizes the boundary width
$\Delta x$, as has been observed for a system without mutual
repression \cite{Erdmann2009}. While the minimal $\Delta x$ of the
system with mutual repression is only marginally lower than that of
the system without it, this optimum is reached at a lower value of the
diffusion constant, where the steepness of the boundaries is much
higher. We find that these observations are robust,
i.e. independent of the precise parameters of the model, such as
maximum expression level, size of the bursts of gene expression, and
the cooperativity of gene activation.

Our results also show that mutual repression can strongly buffer
against embryo-to-embryo variations in the morphogen levels by
suppressing boundary shifts via a mechanism that is akin to that of
\cite{Howard2005, Morishita2009}.  A more detailed analysis reveals
that when the regions where \Bcd and \Cad activate \hb and \kni
respectively overlap, bistability can arise in the overlap zone. Yet,
the mean waiting time for switching is longer than the lifetime of the
morphogen gradients, which means that the \hb and \kni expression
patterns are stable on the relevant developmental time scales. This
also means, however, that when errors are formed during development,
these cannot be repaired. Here, our simulations reveal another
important role for diffusion: without diffusion a spotty phenotype
emerges in which the nuclei in the overlap zone randomly express either
\Hb or \Kni; diffusion can anneal these patterning defects, leading
to well-defined expression domains of \Hb and \Kni. 
Finally, we also study a scenario where \hb and \kni are
activated by \Bcd only. While this scheme is not robust against
embryo-to-embryo variations in the morphogen levels, mutual
repression does enhance boundary precision and steepness also
in this scenario.

\section*{Results}
\pdfbookmark[1]{Results}{BookmarkResults}
\subsection*{Model}
\pdfbookmark[2]{Model}{BookmarkModel}
We consider the embryo in the syncytial blastoderm stage at late cell cycle
14, ca. $2~\unit{h}$ after fertilization. In this stage the majority of the nuclei
forms a cortical layer and \hb and \kni expression
can be detected \cite{Surkova2008}. Our model is an extension of the
one presented in \cite{Erdmann2009}. It is based on a cylindrical
array of diffusively coupled reaction volumes which represent the
nuclei, with periodic boundary conditions in the angular ($\phi$)
and reflecting boundaries in the axial ($x$) direction.  The
dimensions of the cortical array are $N_x=N_\phi=64$, with equal
spacing of the nuclei $\ell=8.5 \mum$ in both directions. For a given embryo length
$L$, this implies a cylinder radius
$R=\frac{L}{2\pi}\simeq\frac{L}{6}$, which is close to the
experimentally observed ratio.
The resulting number of $N=4096$ nuclei roughly corresponds to the
expected number of cortical nuclei at cell cycle 14 if non-dividing
polyploid yolk nuclei are taken into account \cite{Foe1983} (see
\SItext for details); we also emphasize, however, that none of
the results presented below depend on the precise number of nuclei.

\begin{figure}[!ht]
\begin{center}
\includegraphics[width=3.27in]{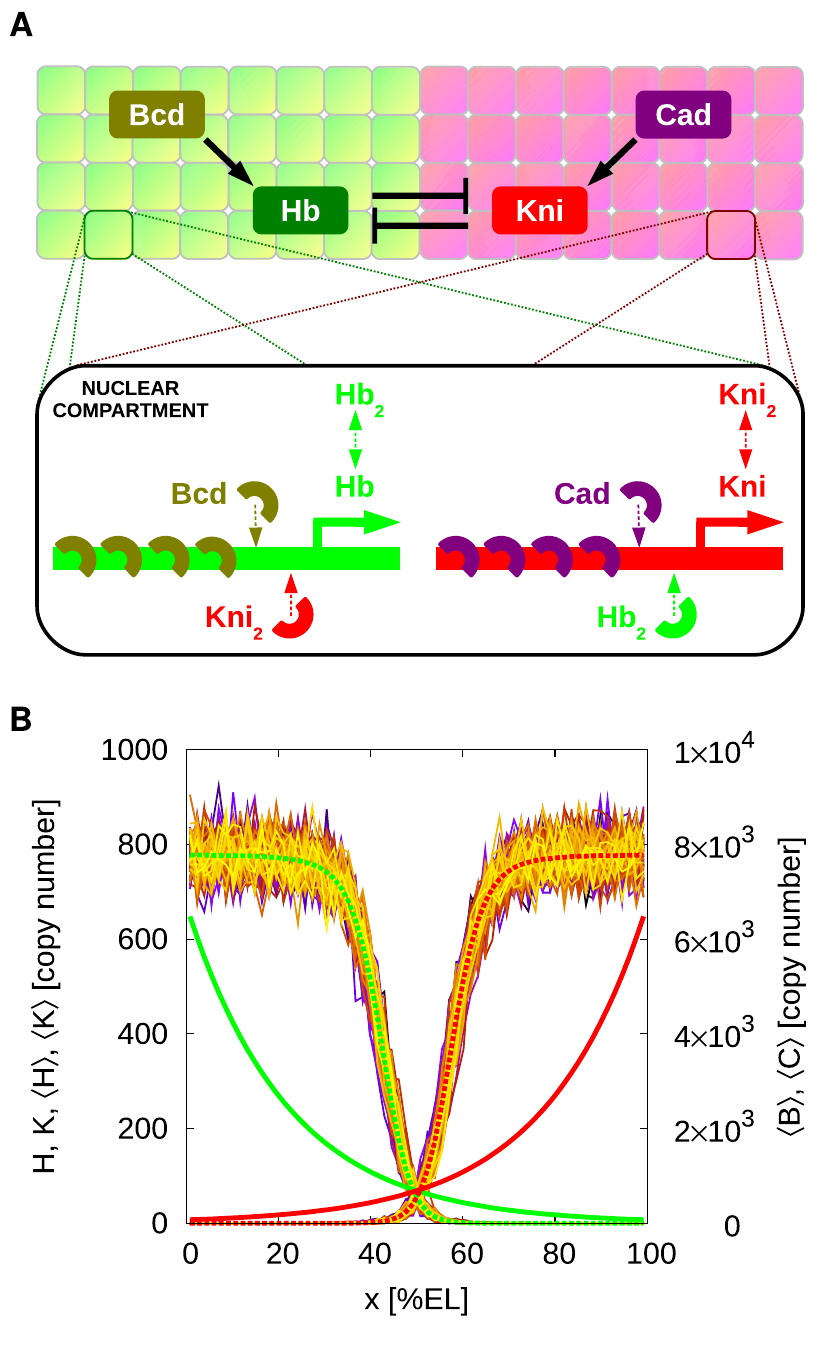}
\end{center}
\caption{ {\bf The model.}  \subfig{A} Cartoon of our model.  \Bcd
  activates \hb, while its antagonist \kni is activated by \Cad. The
  gap genes \hb and \kni repress each other mutually.  In each nuclear
  compartment we simulate the genetic promoters of both \hb and
  \kni. Activation is cooperative: In the default setting, 5 morphogen proteins have to bind to
  the promoter to initiate gene expression. \Hb and \Kni both form
  homodimers, which can bind to the other gene's promoter to totally
  block expression, irrespective of the number of bound morphogen
  proteins.  Both dimers and monomers travel between neighboring
  nuclear compartments via diffusion.  \subfig{B} Protein copy number
  profiles along the AP axis in a typical simulation in steady state,
  with parameter values as in Table S2 in \SItext.  Plotted are
  the morphogen gradients \Bcd ($\Avg{B}$, solid green line) and \Cad 
  ($\Avg{C}$, solid red line) and the resulting \Hb ($H$) and \Kni ($K$) total copy 
  number profiles for different times. The dashed green and red lines show the
  \Hb ($\Avg{H}$) and \Kni ($\Avg{K}$) profiles averaged over time and the 
  circumference of the (cylindrical) system.
  }
\label{Fig1}
\end{figure}

In each nuclear volume we simulate the activation of the gap genes \hb
and \kni by the morphogens \Bcd and \Cad, respectively, and mutual
repression between \hb and \kni (see \Fig \ref{Fig1}). In what
follows, we will refer to \Hb and \Kni as repressors and to \Bcd and
\Cad as activators.  Our model of gene regulation bears
similarities to those of
\cite{Bolouri2003,Janssens2006,Zinzen2006,Zinzen2007,Papatsenko2011},
in the sense that it is based on a statistical mechanical model of
gene regulation by transcription factors, allowing the computation
of promoter-site occupancies. However, the models of
\cite{Bolouri2003,Janssens2006,Zinzen2006,Zinzen2007,Papatsenko2011}
are mean-field models, which cannot capture the effect of
intra-embryonic fluctuations due to biochemical noise arising from
the inherent stochasticity of biochemical reactions. This requires a
stochastic model; moreover, it necessitates a model in which the
transitions between the promoter states are taken into account
explicitly, since these transitions form a major source of noise in
gene expression, as we will show. To limit the number of
combinatorial promoter states, we have therefore studied a minimal model
that only includes \Bcd, \Cad, \Hb and \Kni.
Following \cite{Erdmann2009}, we assume that \Bcd and \Cad bind
stochastically and cooperatively to $n_{max}$ sites on their target
promoters.  To obtain a lower bound on the precision of the \hb and
\kni expression domains, we assume that the activating morphogens \Bcd
and \Cad bind to their promoters with a diffusion-limited rate $k_{\rm on}^{\rm A}=4\pi\alpha D_A/V$,
where $\alpha$ is the dimension of a binding site, $D_A$ is the 
diffusion constant of the morphogen, and $V$ is the nuclear volume 
(see ``Materials \& Methods'' for parameter values). Since
the morphogen-promoter association rate is assumed to be diffusion
limited, cooperativity of \hb and \kni activation is tuned via the
dissociation rate $k_{\rm off,n}^{\rm A}=a/b^n$, which decreases with
increasing number $n$ of promoter-bound morphogen molecules. The
baseline parameters are set such that the half-maximal activation
level of \hb and \kni is at midembryo, and the effective Hill
coefficient for gene activation is around 5 \cite{Erdmann2009}; while
we will vary the Hill coefficient, this is our baseline
parameter. Again to obtain a lower bound on the precision of the
gap-gene expression boundaries, transcription and translation is
concatenated in a single step. Mutual repression between \hb and \kni
occurs via binding of \Hb to the \kni promoter, which blocks the
expression of \kni irrespective of the number of bound \Cad molecules,
and vice versa.  To assess the importance of bistability, \Hb and \Kni
can homodimerize and bind to their target promoters only in their
dimeric form, which is a prerequisite for bistability in the
mean-field limit \cite{Cherry2000}.  Both the monomers and dimers
diffuse between neighboring nuclei and are also degraded; the
effective degradation rate $\mu_{\rm eff}$ is such that the gap-gene
expression domains can form sufficiently rapidly on the time scale of
embryonic development ($\approx 10-20~\unit{min}$ \cite{Foe1983}).
In the absence of mutual repression, our model
behaves very similarly to that of \cite{Erdmann2009}, even though our
model contains both monomers and dimers instead of only monomers.

Motivated by experiment \cite{Driever1988a,Houchmandzadeh2002,Gregor2007b},
and in accordance with the diffusion-degradation model,
we adopt an exponential shape for the stationary \Bcd profile; we thus 
do not model the establishment of the gradient \cite{Sample2010}.
To elucidate the role of mutual repression,
it will prove useful to take our model to be
symmetric: the \Cad profile is the mirror image of the \Bcd
profile, and \hb and \kni repress each other equally
strongly. Diffusion of \Bcd and \Cad between nuclei induce
fluctuations in their copy numbers on the time scale
$\tau_d=\ell^2/(4D_A)\simeq 6~\unit{s}$. Because $\tau_d$ is much
smaller than the time scale for promoter binding, $1/k_{\rm
  on}^{\rm A}\simeq 360~\unit{s}$, fluctuations in the copy number
of \Bcd and \Cad are effectively averaged out by slow binding of \Bcd
and \Cad to their respective promoters, \hb and \kni
\cite{Erdmann2009}.  To elucidate the importance of the threshold
positions for \hb and \kni activation, we will scale the morphogen
gradients by a global dosage factor $A$; this procedure will also
allow us to study the robustness of the system against
embryo-to-embryo variations in the morphogen levels.

We simulate the model using the Stochastic Simulation Algorithm (SSA)
of Gillespie \cite{Gillespie1976, Gillespie1977}.  Diffusion is
implemented into the scheme via the next-subvolume method used in
MesoRD \cite{Elf2004, Hattne2005}.
A recent version of our code is available at GitHub and can be
accessed via \url{http://ggg.amolf.nl} .

\subsection*{Characteristics of gap-gene expression boundaries}
\pdfbookmark[2]{Characteristics of gap-gene expression boundaries}{BookmarkBoundaries}
Three key characteristics of gene expression boundaries are 1) the
noise in the protein concentration at the boundary; 2) the steepness
of the boundary; 3) the width of the boundary. While these quantities
may make intuitive sense, their definitions are not unambiguous.  Equally important, different definitions will
reveal different properties of the system.

\begin{figure}[!ht]
\begin{center}
\vspace{0.33in}
\includegraphics[width=3.27in]{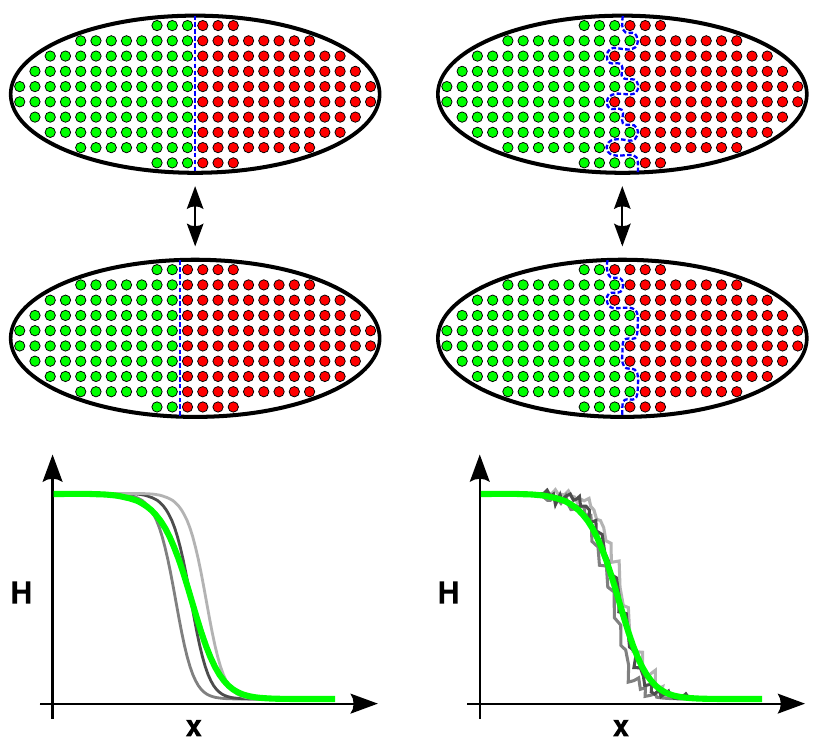}
\end{center}
\caption{ {\bf Two different contributions to the intra-embryonic
      variance in the boundary position.}  The total variance of the
    gap gene expression boundary position $x_t$ due to intra-embryonic
    fluctuations, $\sigma^2_{x_t, intra}$, can be decomposed into two
    contributions: $\sigma^2_{\Avg{x_t}_\phi}$, the variance in time
    of the circumferential mean of $x_t$, and
    $\overline{\sigma^2_{x_t}}$, the time-average of the variance of
    $x_t$ along the circumference of the embryo.  The sketch illustrates
    two extremal cases: If the boundary is very smooth along the
    circumference at any moment in time, concerted movements of the boundary
    will dominate the total variance, i.e. $\sigma^2_{x_t, intra}
    \simeq \sigma^2_{\Avg{x_t}_\phi}$ (left side).  If, in contrast,
    the boundary is rough but its mean position does not fluctuate
    much in time, then $\sigma^2_{x_t, intra} \simeq
    \overline{\sigma^2_{x_t}}$ (right side).  Naturally, a combination
    of the two types of fluctuations is possible.  }
\label{Fig2}
\end{figure}

\mysubsubsection{Decomposing the noise} Let's consider the variance in the copy number $G$ of protein \G
  at position $x$ along the anterior-posterior (AP) axis.  We define
  its mean copy number, averaged over all embryos, circumferential positions
  $\phi$ and all times, at the anterior-posterior position $x$ as
\begin{eqnarray}
\ETPAvg{G}(x)&\equiv&\frac{1}{N_e}\frac{1}{T}\frac{1}{N_\phi} \sum_{e=0}^{N_e-1}\sum_{t=0}^{T-1}\sum_{\phi=0}^{N_\phi-1}G_e(\phi,x,t),
\end{eqnarray}
where $G_e(x,\phi,t)$ is the copy number of protein \G in embryo $e$ at position $x$
 and angle $\phi$ in the
circumferential direction (perpendicular to the AP-axis) at time
$t$. Here, we introduce the convention that the overline denotes
an average in time, while the ensemble brackets with
a subscript $\phi$ 
denote an average along the $\phi$ direction and that with a
subscript $e$ an average over all embryos. The variance in the copy number
$G\equiv G_e(x,\phi,t)$ is then given by
\begin{eqnarray}
\sigma^2_G(x)&=&\ETPAvg{(G-\ETPAvg{G})^2}\\
&=&\Avg{\TPAvg{G^2}}_e-\Avg{\overline{\Avg{G}_\phi^2}}_e+\Avg{\overline{\Avg{G}_\phi^2}}_e-\Avg{\TPAvg{G}^2}_e+\Avg{\TPAvg{G}^2}_e-\ETPAvg{G}^2\\
&=&\overbrace{\Avg{\overline{\sigma^2_G}}_e(x)+\Avg{\sigma^2_{\Avg{G}_\phi}}_e(x)}^{\text{mean
    intra-embryonic
    noise}}\hspace{2.76cm}+\overbrace{\sigma^2_{\Avg{\overline{G}}_\phi}(x)}^{\text{inter-embryonic variations}}
\label{EqNoiseDecom}
\end{eqnarray}
The total variance in the copy number can thus be decomposed into
intra-embryonic fluctuations averaged over all embryos and
inter-embryonic variations. The former can, furthermore, be decomposed into
$\Avg{\overline{\sigma^2_G}}_e(x)$, which is the time-averaged mean of
the variance in $G$ along the circumferential direction,
$\overline{\sigma^2_G}(x)$, averaged over all embryos, and
$\Avg{\sigma^2_{\Avg{G}_\phi}}_e(x)$, which is the variance in time
over the mean of $G$ along the circumferential direction,
$\sigma^2_{\Avg{G}_\phi}(x)$, again averaged over all embryos. These
intra-embryonic terms capture different types of dynamics. If the
expression boundary is rough
but its average position does not fluctuate in time,
then $\overline{\sigma^2_G}(x)$ will be large yet
$\sigma^2_{\Avg{G}_\phi}(x)$ will be small. Conversely, when the
boundary is smooth but its average position does fluctuate in time, then
$\overline{\sigma^2_G}(x)$ will be small yet
$\sigma^2_{\Avg{G}_\phi}(x)$ will be large. Naturally, a combination
of the two is also possible. The third term,
$\sigma^2_{\Avg{\overline{G}}_\phi}(x)$, captures the embryo-to-embryo
variations in the average over time and $\phi$ of the protein-copy
number.  Similarly, we can decompose the fluctuations in the boundary
position $x_t$ as
\begin{eqnarray}
\Delta x &=&\sigma_{x_t}\\
&=&\sqrt{\Avg{\overline{\sigma^2_{x_t}}}_e+\Avg{\sigma^2_{\Avg{x_t}_\phi}}_e+\sigma^2_{\Avg{\overline{x_t}}_\phi}}
\label{EqXtDecom}
\end{eqnarray}
The two different contributions to the intra-embryonic variance, $\Avg{\overline{\sigma^2_{x_t}}}_e+\Avg{\sigma^2_{\Avg{x_t}_\phi}}_e$, are illustrated
 in \Fig \ref{Fig2}.
Here and in the next section, we will study the robustness of the
system against intra-embryonic fluctuations, while in the section
``Robustness to inter-embryonic variations: Mutual repression can
buffer against correlated morphogen level variations'' we will study
the robustness against inter-embryonic variations in the morphogen levels.

{\mysubsubsection{Intra-embryonic fluctuations} \Fig S2 in \SItext
  shows the decomposition of the noise in the \Hb copy number
  $H$ and the threshold position $x_t$ of the \Hb boundary, as a
  function of the diffusion constant. We show the intra-embryonic
  fluctuations for one given embryo (with the baseline parameter set);
  how $\Delta x$ (the boundary variance originating from
  intra-embryonic fluctuations) changes with embryo-to-embryo
  variations in the morphogen levels is addressed in section ``Overlap
  of morphogen activation domains does not corrupt robustness to
  intrinsic fluctuations''. \Fig S2 shows that by far the dominant
  contribution to the intra-embryonic noise in the copy number and
  threshold position is the time average of the variance in these
  observables along the circumferential direction; the variance in
  time of the $\phi$-average of these quantities is indeed very
  small. The picture that emerges is that the expression boundary is
  rough, even when the diffusion constant $D$ is large,
  i.e. $D=1~\umsps$. An analysis of the spatial correlation function
  at midembryo $\TPAvg{\delta H(0)\delta H(\phi)}(x_t)$, where $\delta
  H(\phi)=H(x_t,\phi,t)-\TPAvg{H}$, revealed that the correlation
  length $\xi_\phi$ is on the order of a few nuclei, which corresponds
  to the diffusion length $\lambda=\sqrt{D/\mu_{eff}}$ a protein can
  diffuse with diffusion constant $D$ before it is degraded with a
  rate $\mu_{eff}$; the correlation length is thus small compared to
  the circumference. One possible source of coherent
  fluctuations in the mean copy number $\Avg{X}_\phi$ and boundary
  position $\Avg{x_t}_\phi$ are temporal variations of the morphogen
  profiles. However, in our model, these profiles are static---we
  argued that the morphogen fluctuations are fast on the timescale
  of gene expression, and are thus effectively integrated out. The
  small correlation length $\xi_\phi$ then indeed means that the
  varations in the mean over $\phi$, $\Avg{\dots}_\phi$, will be
  small.   This leads to an interesting implication for
  experiments, which we discuss in the Discussion section.

  \mysubsubsection{The boundary steepness} Now that we have
    characterized the fluctuations in the copy number and the boundary
    position, the next question is how fluctuations in the copy number
    affect the steepness of the boundary.  In particular, a
    gene-expression boundary can be shallow either because at each
    moment in time the interface is shallow, or because at each moment
    in time the interface is sharp yet the interface fluctuates in
    time, leading to a smooth profile.  The question is thus how much
    the gradient of the mean concentration profile,
    $\TPAvg{G}^\prime$, and the mean of the gradient,
    $\TPAvg{G^\prime}$, differ (here the prime denotes the spatial
    derivative). \Fig S3 in \SItext shows both quantities as a function of the
    diffusion constant. It is seen that while the average of the
    gradient is larger than the gradient of the average (as it
    should), the difference is around a factor of 2. We thus conclude that
    the steepness of the expression boundary at each moment in time
    does not differ very much from the steepness of the average
    concentration profile.

In the rest of the manuscript, we will predominantly focus on the
properties of individual embryos, and average quantities are
typically averages over time and the circumference. For brevity, therefore, 
$\Avg{\dots}=\TPAvg{\dots}$, unless stated otherwise.
\subsection*{Robustness to intra-embryonic fluctuations: Mutual repression allows for steeper profiles without raising the noise level at the boundary}
\pdfbookmark[2]{Robustness to intra-embryonic fluctuations: Mutual repression allows for steeper profiles without raising the noise level at the boundary}{BookmarkRobustnessIntra}

\mysubsubsection{Mutual repression shifts boundaries apart}
\Fig \ref{Fig3}A shows the average \Hb and \Kni steady-state
profiles along the anterior-posterior (AP) axis as a function of their
diffusion constant $D$ for a system with mutual repression.  The inset
shows the morphogen-activation profiles, which are the spatial profiles of
the probability that the \hb and \kni promoters have 5 copies of their
respective morphogens bound. Without mutual repression, thus when \Hb
and \Kni cannot bind to their respective target promoters, these profiles
describe the probability that \hb and \kni are activated by their
respective morphogens.  Indeed, without mutual repression and without
\Hb and \Kni diffusion, the \Hb and \Kni concentration profiles would
be proportional to their respective morphogen-activation profiles
\cite{Erdmann2009}, which means that they would precisely intersect at
midembryo.  In contrast, \Fig \ref{Fig3}A shows that the \Hb and \Kni
concentration profiles are shifted apart in the system with mutual
repression.  There is already a finite separation for $D=0$, which
increases further as $D$ is increased.

\begin{figure}[!ht]
\begin{center}
\includegraphics[width=6in]{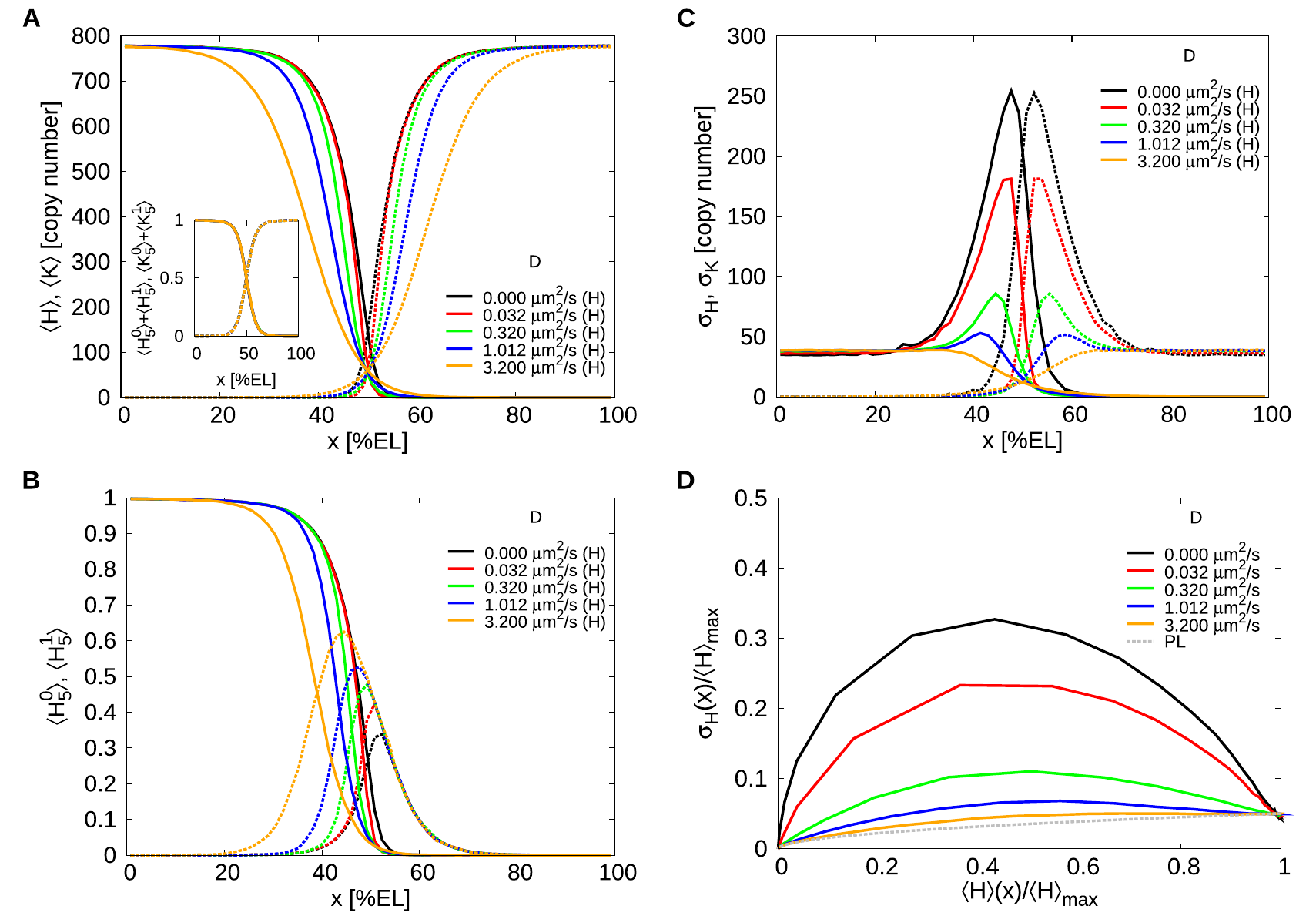}
\end{center}
\caption{ {\bf The effect of mutual repression on the average protein
    concentrations and their standard deviations.}  \subfig{A} Time-
  and circumference-averaged \Hb ($\Avg{H}$, solid lines) and \Kni
  ($\Avg{K}$, dashed lines) total protein copy number profiles along
  the AP axis for various diffusion constants $D$ in a system with
  mutual repression.  The inset shows for both the \hb and the \kni
  promoter the probability that the promoter binds 5 morphogen
  proteins irrespective of whether the antagonistic gap protein is
  bound to it (meaning that the promoter is activated by the
  morphogen, even though it may be repressed by the antogonistic gap
  protein); these ``morphogen-activation'' profiles are identical for
  all $D$ values.  \subfig{B} Profiles of the probability
  $\Avg{H^0_5}$ that the \hb promoter is induced, meaning that it has
  5 copies of \Bcd bound to it and no \Kni dimer (solid lines), and the
  probability $\Avg{H^1_5}$ that \hb is activated by \Bcd yet
  repressed by \Kni, in which case \hb is indeed not expressed (dashed
  lines).  \subfig{C} AP profiles of the time- and
  circumference-averaged standard deviation of the total gap protein
  copy number for \Hb ($\sigma_H$, solid lines) and \Kni ($\sigma_K$,
  dashed lines).  \subfig{D} Normalized standard deviation
  $\sigma_{H}(x)/\Avg{H}_{max}$ versus the normalized mean
  $\Avg{H}(x)/\Avg{H}_{max}$; $\Avg{H}(x)$ is the averaged total
  \Hb copy number at $x$ and $\Avg{H}_{max}$ is the maximum of this
  average over all $x$.  The grey dashed line represents the
  Poissonian limit (PL) given by
  $\sqrt{(1+f_D)\Avg{H}(x)}/\Avg{H}_{max}$, where $f_D$ is the
  fraction of proteins in dimers.  }
\label{Fig3}
\end{figure}

In \Fig \ref{Fig3}B we show the profile of the probability
$\Avg{H^0_5}$ that the \hb promoter is induced, meaning that it has 5
copies of \Bcd bound to it and no \Kni, and the profile of the
likelihood $\Avg{H^1_5}$ that \hb is activated by \Bcd, yet repressed
by \Kni, in which case \hb is not expressed. It is seen that
repression by \kni almost fully inhibits \hb expression beyond the
half-activation point, where \hb would be expressed without \kni
repression (see inset Panel A).  Indeed, mutual repression effectively
cuts off protein production beyond midembryo. The production
probability therefore changes more abruptly along the AP axis, leading
to a higher steepness of the protein profiles near midembryo.  For
$D>0$, repressor influx over the midplane increases, and as a result
the regions of expression inhibiton are enlarged and the concentration
profiles shift apart further.

\mysubsubsection{Noise reduction via spatial averaging}
\Fig \ref{Fig3}C shows the standard deviation of the protein copy
number along the AP axis for both \Hb ($\sigma_H$) and \Kni
($\sigma_K$).  It is seen that the noise increases close to the
half-activation point where promoter-state fluctuations are strongest
\cite{Tkacik2008,VanZon2006,So2011}.  This is also observed in \Fig
\ref{Fig3}D, which shows the normalized standard deviation $\sigma_H /
\Avg{H}_{max}$ versus the normalized mean $\Avg{H}/\Avg{H}_{max}$
of the average \Hb copy number; here, $\Avg{H}_{max}$ is the
maximum average concentration of \Hb. The noise maximum close to mid
embryo diminishes with increasing $D$, approaching the
Poissonian limit.  Note that the Poissonian limit here is given by
$\sigma_P=\sqrt{(1+f_D)\Avg{H}}$, where $f_D=2\Avg{H_D}/\Avg{H}$ is
the fraction of dimerized \Hb proteins with respect to the total \Hb
copy number (see \SItext for details).  Clearly, the
spatial averaging mechanism described in
\cite{Erdmann2009,Okabe-Oho2009} reduces the noise also in our system,
which differs from those in \cite{Erdmann2009,Okabe-Oho2009}
by the presence of both gap gene monomers and dimers instead
of monomers only.

\mysubsubsection{Mutual repression reduces the boundary width by
  increasing the steepness}
\Fig \ref{Fig4} quantifies the impact of spatial averaging and
mutual repression on the \Hb boundary width $\Delta x$, comparing it
to that of the system without mutual repression. To first order, the
boundary precision $\Delta x$ is related to the standard
deviation in the protein copy number at the boundary, $\sigma_H(x_t)$,
and the steepness of the boundary, $|\Avg{H(x_t)}^\prime|$, via
Eq. \ref{EqDeltaX} \cite{Tostevin2007,Gregor2007,Erdmann2009}. The
noise $\sigma_H(x_t)$ decreases with increasing $D$ due to spatial
averaging in an almost identical manner for the systems with and
without mutual repression (\Fig \ref{Fig4}, top panel); indeed,
perhaps surprisingly, mutual repression has little effect on the noise
at the boundary. Increasing $D$ also lessens the steepness of the
protein profiles, thus reducing the slope $|\Avg{H(x_t)}^\prime|$
(\Fig \ref{Fig4}, middle panel).  While without mutual repression this
reduction is monotonic, in the case with mutual repression the
steepness first rises because increasing $D$ increases the influx of
the antagonistic repressor into the regions where the gap genes are
activated by their respective morphogens, which, for low values of
$D$, {\em steepens} the effective gene-activation profile
$\Avg{H^1_5}(x)$ by most strongly reducing gene expression near
midembryo; after the steepness has reached its maximum at
$D=0.032~\umsps$, it drops for higher diffusion constants, because
the diffusion of the gap-gene proteins now flattens their
concentration profiles.  Most importantly, with mutual repression
$|\Avg{H(x_t)}^\prime|$ reaches significantly higher values for all
$D\leq 1.0~\umsps$.  At $D=0.032~\umsps$ the profile is roughly twice
as steep as in the case without repression. Interestingly, for
$D\lesssim 0.1~\umsps$, our simulation results for the steepness of
the profiles as normalized by their maximal values agree with those
measured experimentally by Surkova {\em et al.} in cell cycle 14
\cite{Surkova2008}: In both simulation and experiment, the
concentration drops from 90\% to 10\% of the maximal values over
5-10\% of the embryo length.

\begin{figure}[!ht]
\begin{center}
\includegraphics[width=3.27in]{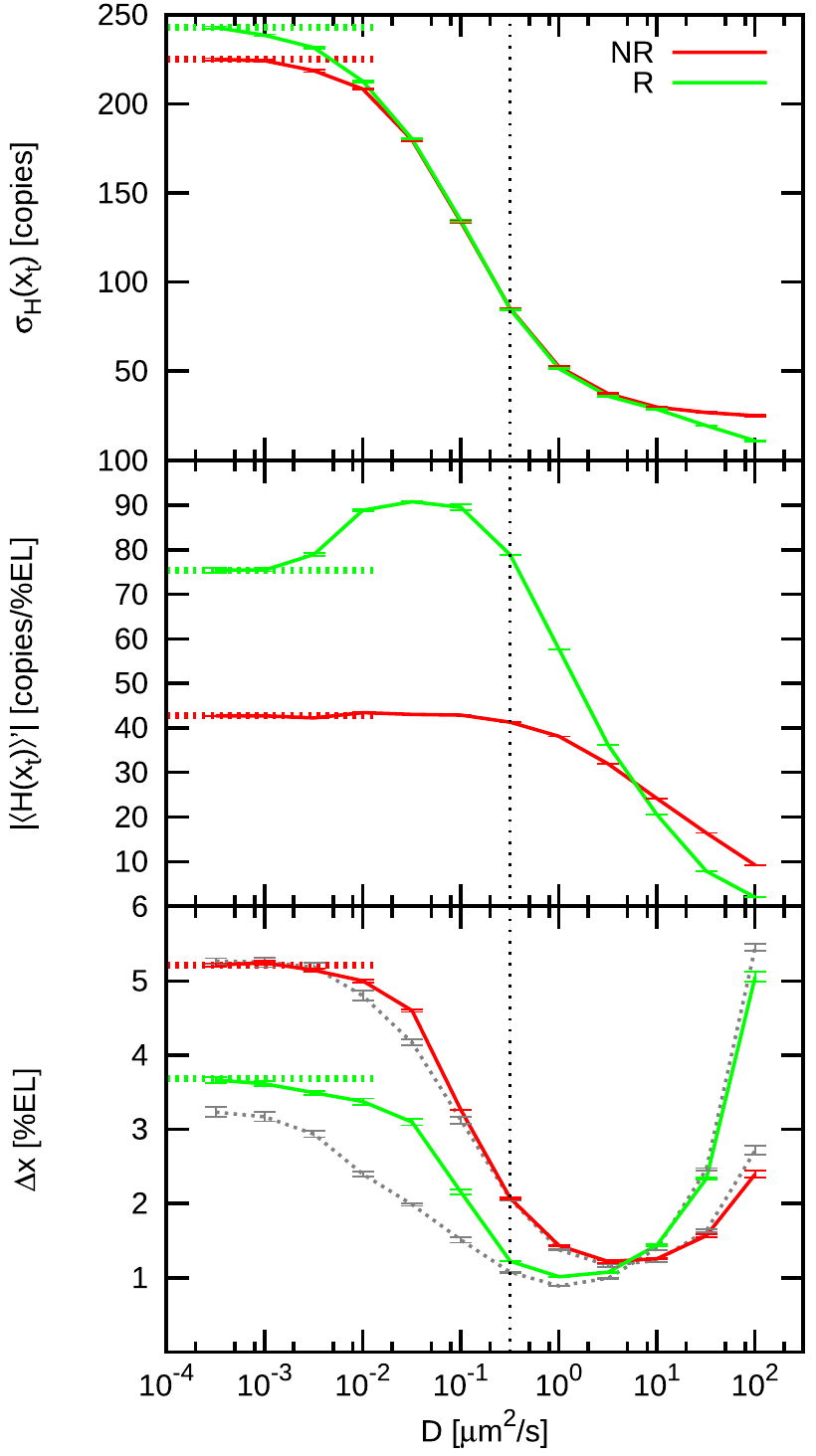}
\end{center}
\caption{ {\bf The effect of mutual repression on the precision and
    steepness of the \Hb boundary.}  The figure shows the
  time- and circumference-average of the standard deviation of the
  total \Hb copy number at the boundary $\sigma_H(x_t)$ (upper panel),
  the slope of the total \Hb copy number profile at the boundary
  $|\Avg{H}'(x_t)|$ (middle panel) and the \Hb boundary width $\Delta
  x$ (lower panel) as a function of the diffusion constant $D$ of the
  gap proteins.  Red solid lines show the case without (NR) and green
  solid lines the case with mutual repression (R); the red and green
  dashed lines show the limiting values without diffusion of the gap
  proteins. The grey dashed lines in the boundary width plot are the
  values based on the approximation $\Delta
  x=\sigma_H(x_t)/|\Avg{H(x_t)}'|$.  Note that for $D<3.2~\umsps$,
  mutual repression enhances the steepness of the boundary, which in
  turn enhances the precision of the boundary.
  The black dotted line marks the $D$-value
  where the boundary is both steep and precise due to mutual repression.}
\label{Fig4}
\end{figure}

Both with and without \Hb-\Kni mutual repression the trade-off between
noise and steepness reduction leads to an optimal diffusion constant
$D_{min}$ that maximizes boundary precision, i.e. minimizes $\Delta x$
(\Fig \ref{Fig4}, lower panel).  Mutual repression enhances the
precision for $D\leq 1.0~\umsps$ because in this regime decreasing $D$
increases the steepness markedly while it has only little effect on
the noise as compared to the system without mutual repression.
Conversely, $\Delta x$ is increased by mutual repression for $D\geq
10~\umsps$ because it reduces the steepness.  The minimum in the case
with repression is marginally lower than that without
($D_{min,R}/D_{min,NR}\simeq0.86$), but located at a lower $D$-value
($1.0~\umsps$ vs. $3.2~\umsps$).  Most importantly, at
$D=0.32~\umsps$, the system with mutual repression produces a profile
that is twice as steep as that of the system without it at
$D_{min,NR}=3.2~\umsps$, whereas the precision $\Delta x$ is
essentially the same in both cases.  Clearly, mutual repression can
strongly enhance the steepness of gene-expression boundaries without
compromising their precision.

\mysubsubsection{Influence of Hill coefficient}
A key parameter controlling the precision of the gap-gene
expression boundaries, is the degree of cooperativity by which the
gap genes are activated by their respective morphogens---this
determines the profile steepness of the average gap-gene
promoter activity. To investigate this, we have lowered the
effective Hill coefficient from its baseline value of 5 by reducing
the number $n_{\rm max}$ of morphogen molecules that are required to
bind the promoter to activate gene expression. To isolate the effect
of varying the {\em mean} gene-activation profiles $\Avg{H_{n_{\rm
max}}^0}(x)$ and $\Avg{K_{n_{\rm max}}^0}(x)$, we varied, upon
varying $n_{\rm max}$, the association and dissociation rates such
that 1) the average gene activation probabilities near midembryo,
$\Avg{H_{n_{\rm max}}^0}(L/2)$ and $\Avg{K_{n_{\rm max}}^0}(L/2)$, are
unchanged and 2) the waiting-time distribution for the gene
on-to-off transition is unchanged (since the average activation
probability is fixed, the mean off-to-on rate is also unchanged, although
the waiting-time distribution is not; see also \Fig S5 in \SItext).
We observe that  mutual repression
markedly enhances the steepness of the gap-gene expression
boundaries, also with a
lower Hill coefficient for gene activation (\Fig S6 in \SItext). However, lowering the Hill coefficient
reduces the steepness of the gene-activation profiles, causing the
two antagonistic gene-activation profiles to overlap more. As a
result, in each of the two gap-gene expression domains, more of the
antagonist is present, which tends to increase the noise in gene
expression by occassionally shutting off gene production. This, as
explained in more detail later, is
particularly detrimental when the diffusion constant is low. Indeed,
when the effective Hill
coefficient of gene activation is 3 or lower, mutual repression {\em
increases} $\Delta x$ when the diffusion constant is low,
i.e. below approximately $0.1~\umsps$. Nonetheless, the {\em minimal} $\Delta
x$ is still lower with mutual repression, and, consequently, also
with a lower Hill coefficient for gene activation, mutual repression
can enhance both the steepness and the precision of gene-expression
boundaries.

\mysubsubsection{Influence of the repression strength}
As a standard we assume very tight binding of the \Hb and \Kni dimers,
``the repressors'', to their respective promoters. To test how this
assumption affects our results we performed simulations in which we
systematically varied the repressor-promoter dissociation rate $k^{\rm
  R}_{off}$ in the range $[5.27\cdot 10^{-4}\unit{/s},5.27\cdot
10^{2}\unit{/s}]$, keeping the diffusion constant at $D=1.0~\umsps$
(the value that minimizes the boundary width at $k^{\rm
  R}_{off}=5.27\cdot 10^{-3}\unit{/s}$) and all other parameters the
same as before.  \Fig \ref{Fig5} shows the noise, steepness and
boundary precision as a function of the repressor-promoter
dissociation rate. For high dissociation rates, these quantities equal
those in the system without mutual repression (dashed lines). Yet, as
the dissociation rate is decreased, the steepness rises markedly at
$k^{\rm R}_{off}=1/{\rm s}$. In contrast, the noise $\sigma_H(x_t)$
first decreases with decreasing $k^{\rm R}_{off}$, passing through
a minimum at $k^{\rm R}_{off}=0.1/{\rm s}$ before rising to a level that
is higher than that in a system without mutual repression. This
minimum arises because on the one hand increasing the affinity of
the repressor (the antagonist) makes the operator-state fluctuations
of the activator (the morphogen) less important---increasing
repressor binding drives the concentration profiles of \Hb and \Kni
away from midembryo, where the promoter-state fluctuations of the
activators are strongest; on the other hand, when the repressor
binds too strongly, then slow repressor unbinding leads to
long-lived promoter states where gene expression is shut off,
increasing noise in gene expression; this phenomenon is similar to
what has been observed in Refs. \cite{VanZon2006} and
\cite{Morelli2008}, where slower binding of the gene regulatory
proteins to the promoter increases noise in gene expression and
decreases the stability of a toggle switch, respectively. The interplay between the
noise and the steepness yields a marked reduction of the boundary
width $\Delta x$; indeed, even in the limit of very tight repressor binding,
mutual repression significantly enhances the precision of the
boundary.

\begin{figure}[!ht]
\begin{center}
\includegraphics[width=3.27in]{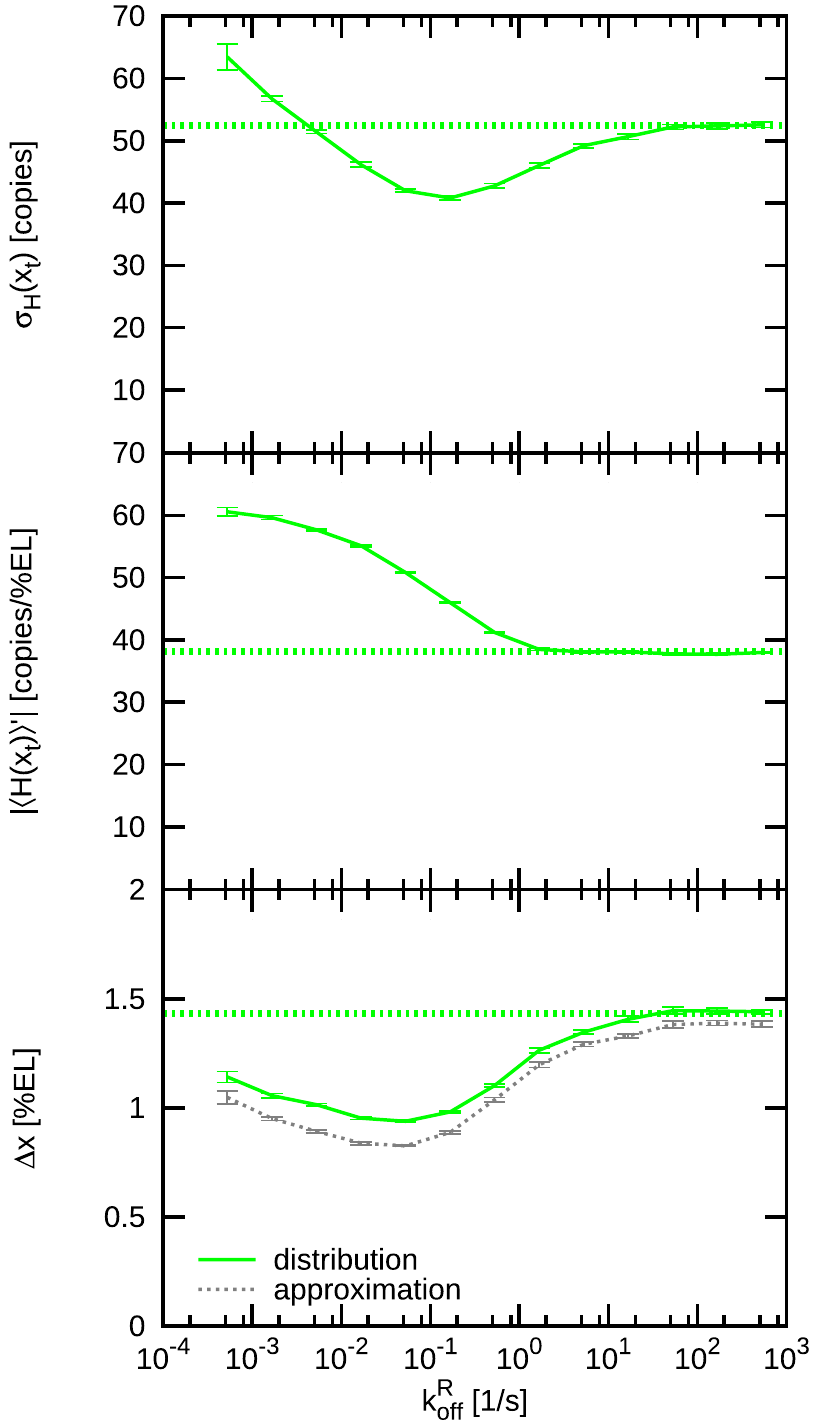}
\end{center}
\caption{ {\bf The effect of varying repression strength on the precision
    and steepness of the \Hb boundary.}  Shown are the time- and
  circumference average of the standard deviation of the total \Hb copy
  number at the boundary $\sigma_H(x_t)$ (upper panel), the steepness
  of the boundary $|\Avg{H}'(x_t)|$ (middle panel) and the \Hb
  boundary width $\Delta x$ (lower panel) as a function of
  $k_{\rm off}^{\rm R}$, the promoter-dissociation rate of \Hb and \Kni.  The solid green line are values obtained from the
  boundary position distribution, the dashed grey line the ones
  calculated from the approximation $\Delta
  x=\sigma_H(x_t)/|\Avg{H(x_t)}'|$.  Straight dashed lines mark
  the limits for the case without mutual repression ($k_{on}^{\rm R}=k_{off}^{\rm R}=0$).}
\label{Fig5}
\end{figure}

\mysubsubsection{Influence of expression level}
Since the precise gap protein expression level is not known, we also varied the
maximal protein copy number $N$ by varying the maximal expression
rate $\beta$ (see \SItext). \Fig S9 in \SItext shows the output
noise and slope at the boundary position, and the boundary precision $\Delta x$, 
as a function of the diffusion constant for three different
expression levels. It is seen that for low diffusion constant, the
precision is independent of $N$, while for higher diffusion constant it scales roughly with
$1/\sqrt{N}$. This can be understood by noting that the steepness of
the gene-expression boundary scales to a good approximation
with $N$ independently of $D$, while the noise $\sigma$ scales with
$N$ when the diffusion constant is small, but with $\sqrt{N}$ when
the diffusion constant is large (see also Eq. \ref{EqDeltaX}). The
scaling of the noise with $N$ is due to the fact that for low $D$
the noise in the copy number is dominated by the noise coming from
the promoter-state fluctuations, which scales linearly with $N$,
while for high $D$, diffusion washes out the expression bursts resulting from
the promoter-state flucutations, leaving only the noise coming from
the Poissonian fluctuations arising from transcription and
translation, which scales with the square root of $N$
\cite{Erdmann2009}. In \SItext we also study the importance of
bursts arising in the transcription-translation step (see \Fig S8 in \SItext); however, we
find that for a typical burst size, these bursts do not dramatically
affect boundary precision.

\FloatBarrier

\subsection*{Robustness to inter-embryonic variations: Mutual repression can buffer against correlated morphogen level variations}
\pdfbookmark[2]{Robustness to inter-embryonic variations: Mutual repression can buffer against correlated morphogen level variations}{BookmarkRobustnessInter}
Although the \Bcd copy number at midembryo has been determined
experimentally \cite{Gregor2007}, the measured value is not
necessarily the half-activation threshold of \hb.  Indeed, in vivo the
\Hb profile is shaped by other forces, like mutual repression. In the
\kni-\kr double mutant, the Hb boundary at midembryo shifts
posteriorly \cite{Manu2009PlosBiol}. Moreover, gap gene domain
formation has been observed at strongly reduced \Bcd levels,
suggesting that \Bcd might be present in excess
\cite{Ochoa-Espinosa2009}.  Also from a theoretical point of view it
is not obvious that a precisely centered morphogen-activation
threshold is optimal, in terms of robustness against both
intra-embryonic fluctuations and inter-embryonic variations. Here, we
study the effect of changing the threshold position where \hb and \kni
are half-maximally activated by their respective morphogens, \Bcd and
\Cad.  While the threshold positions could be varied by changing the
threshold morphogen concentrations for half-maximal gap-gene
activation (for example by changing the morphogen-promoter
dissociation rates), we will vary these positions by changing the
amplitude of the morphogen profiles by a factor $A$.  This procedure
not only preserves the promoter-activation dynamics at the
boundaries---a key determinant for the noise at the boundaries---but
also allows us to study the importance of mutual repression in
ensuring robustness against embryo-to-embryo variations. Indeed, we
will examine not only how changing the threshold position affects the
precision of the gap-gene expression boundaries, $\Delta x(A)$,
but also how the average boundary positions vary with morphogen
dosage, $x_t(A)$, and how the latter gives rise to embryo-to-embryo
variations in the boundary position $\Delta x_t(\Delta A)$ due to
embryo-to-embryo variations in the morphogen dosage $\Delta A$.

\mysubsubsection{Double-activation induces bistability}
We first consider the scenario in which the amplitudes of both
morphogens are scaled by the same factor $A$. When $A=1$, the
position at which \hb and \kni are half-maximally activated by their
respective morphogens coincide at midembryo, meaning that the
domains in which \hb and \kni are activated beyond half-maximum are
adjoining, but do not overlap---this is the scenario discussed in
the previous sections. When $A>1$, the position at which \hb is
half-maximally activated by its morphogen is shifted posteriorly,
while that of \kni is shifted anteriorly, creating an overlap between
the two regions where \hb and \kni are activated.
In this ``double-activated region'' both \hb and \kni are
activated by their respective morphogens, yet they also mutually
repress each other. This may lead to bistability.  To probe whether
this is the case, we performed a bifurcation analysis of the
mean-field chemical-rate equations of isolated nuclei, implying that
$D=0$ (see \Fig S1 in \SItext). In addition, we performed stochastic
simulations of isolated nuclei with different morphogen levels
corresponding to different positions along the AP axis. All other
parameter values were the same as in the full-scale simulation.  We
recorded long trajectories of the order parameter $\Delta N \equiv
H-K$, the difference between the total \Hb and total \Kni copy numbers,
in the stationary state.  From each trajectory we computed the
distribution $P(\Delta N)$ of the probability that the system is in a
state with copy number difference $\Delta N$.  This defines a ``free
energy'' $G(\Delta N)\equiv -\ln P(\Delta N)$, with minima of
$G(\Delta N)$ corresponding to maximally probable values of $\Delta
N$ \cite{Warren2004,Warren2005}. For a bistable system, $G(\Delta
N)$ resembles a double-well potential with minima located at a
positive value of $\Delta N=\Delta N_{\rm H}$ and a negative value of
$\Delta N=\Delta N_{\rm K}$, respectively. At midembryo the morphogen
levels of \Bcd and \Cad are the same and hence the biochemical network
in the nuclei in the midplane is symmetric, which means that, if this
network is bistable, $G(\Delta N)$ resembles a symmetric double-well
potential with $\Delta N_H=-\Delta N_K$ and $\Delta G \equiv G(\Delta
N_H) - G(\Delta N_K)=0$. Away from the middle, the morphogen levels
differ, and one state will become more stable than the other; if the
other state is, however, still metastable, then $G(\Delta N)$ will
resemble an asymmetric double-well potential, with $\Delta G$ being
negative if the \hb-dominant state is more stable than the
\kni-dominant state, and vice versa. The emergence of such a
``spatial switch'' along the AP axis is also captured by our
mean-field, bifurcation analysis (see \SItext) and was recently
also shown in the mean-field analysis of Papatsenko and Levine for the same pair of
mutually repressing genes\cite{Papatsenko2011}.

\begin{figure}[!ht]
\begin{center}
\includegraphics[width=3.27in]{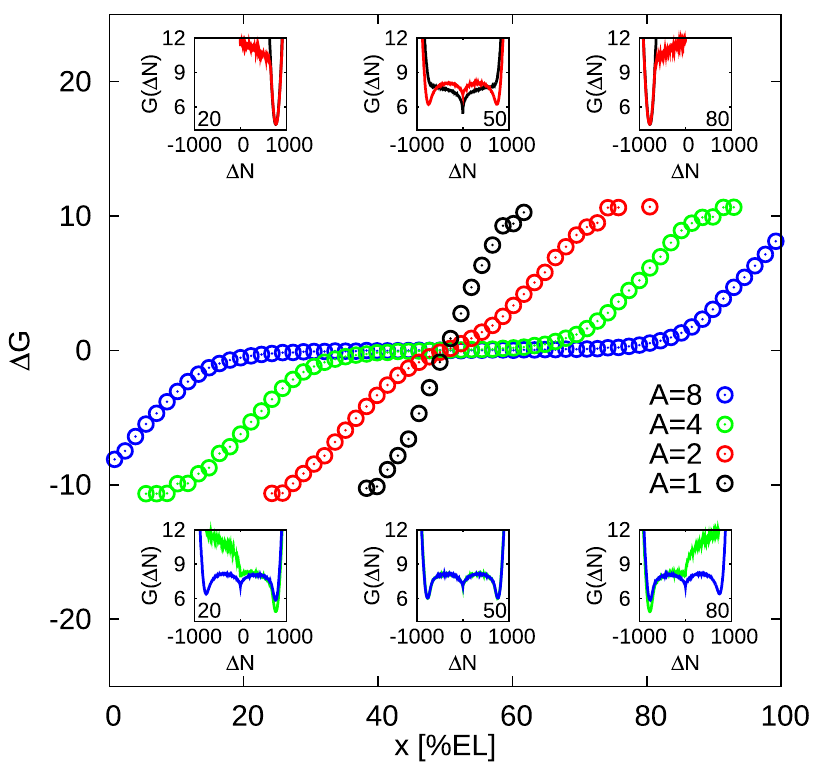}
\end{center}
\caption{ {\bf Emergence of bistability in double-activated regions.}
  The ``free energy'' difference $\Delta G\equiv G(\Delta N_H) -
  G(\Delta N_K)$ as a function of $x$, the distance of the nucleus
  from the anterior pole, for different amplitudes of the morphogen
  gradients $A$; here, $G(\Delta N)\equiv -\ln(P(\Delta N))$, where
  $P(\Delta N)$ is the stationary distribution of the order parameter
  $\Delta N=H-K$; $\Delta N_H\simeq -\Delta N_K \approx 800$ correspond to
  the minima of $G(\Delta N)$.  Negative values of $\Delta G$
  represent a strong bias towards the high-\Hb state, while positive
  values correspond to high-\Kni states.  The insets shows $G(\Delta
  N)$ as a function of $\Delta N$ at the positions indicated by the
  numbers in their corners (values in [\pcEL]; colors correspond to
  main plot). The data is obtained from simulations of single nuclei
  with morphogen levels corresponding to the ones at position $x$ in
  the full system; this is equivalent to the full system without
  diffusion between neighboring nuclei. Note the bistable behavior in
  a wide region of the embryo for higher $A$ values.  }
\label{Fig6}
\end{figure}

\Fig \ref{Fig6} shows $\Delta G$ as a function of the position along
the AP axis, for different amplitudes $A$ of the morphogen gradients. The
inset shows the energy profiles $G(\Delta N)$ for different positions
along the AP axis.  For $A=1$, $G(\Delta N)$ always exhibits one
minimum only, irrespective of the position along the AP axis; at midembryo,
this minimum is located at $\Delta N=0$, while moving towards
the anterior (posterior) the energy minimum rapidly shifts to $\Delta
N \approx +800 (-800)$, reflecting that in the anterior (posterior) half of
the embryo \hb (\kni) is essentially fully expressed. For $A=2$,
$G(\Delta N)$ develops into a double-well potential at midembryo,
with two pronounced minima at $\Delta N\approx 800$ and $\Delta
N\approx-800$, respectively. These two minima correspond to a state in
which \hb is highly expressed ($\Avg{H}\approx 800$) and \kni is
strongly repressed ($\Avg{K}\approx 0$) and another state in which
\kni is highly expressed and \hb strongly repressed, respectively. The
fact that the two energy mimima are equal indicates that both of these
states are equally likely. Moving away from midembryo, however, one
gap-gene expression state rapidly becomes more stable than the other,
and bistability is lost, yielding a potential with one minimum located
at $\Delta N\approx 800$ in the anterior half and a potential with one minimum located
at $\Delta N\approx -800$ in the posterior half of the embryo.
Interestingly, for $A=4$
and $A=8$ a wide region of bistability  develops
around midembryo. In this region, $\Delta G \approx 0$, meaning that 
the high-\hb---low-\kni state and the low-\hb---high-\kni state are
equally stable. These two states are equally likely because in this
region both the \hb and \kni promoters are fully activated by their
respective morphogens. It can also be seen that the width of this
bistable region increases with the amplitude of the morphogen
gradients, as expected.
\pagebreak

\mysubsubsection{Slow switching ensures a low noise level while diffusion avoids error locking}
The bistability observed for $A>1$ and $D=0$ raises an important
question, namely whether the nuclei can switch between the two
gap-gene expression states on the time scale of embryonic
development. This question is particularly pertinent for the higher
morphogen amplitudes, where these two states are equally likely ($\Delta
G\approx 0$) over a wide region of the embryo (\Fig \ref{Fig6}): random
switching between the two distinct gap-gene expression states in this
wide region would then lead to dramatic fluctuations in the positions
of the \hb and \kni expression boundaries, which clearly would be
detrimental for development. We therefore computed \cite{Warren2005} from the recorded
switching trajectories the average waiting time for switching,
$\tau_{s}$, at midembryo  ($\Delta
G\simeq 0$) for different values of $A$; for $A\geq 2$, we find
$\tau_{s} \simeq
6~\unit{h}$ (see Table S1 in \SItext). During cell cycle 14, approximately 2-3 hours
after fertilization, the \Bcd gradient disappears \cite{Drocco2011}, suggesting that the
spontaneous switching rate is indeed low on the relevant time scale of
development.

With diffusion of \Hb and \Kni between neighboring nuclei ($D>0$), the
time scale for switching will be even longer. Diffusion couples
neighboring nuclei, creating larger spatial domains with the same
gap-gene expression state. This reduces the probability that a nucleus
in the overlap region flips to the other gap-gene expression
state. The latter can be understood from the extensive studies on the
switching behavior of the ``general toggle switch''
\cite{Warren2004,Warren2005,Allen2005,Lipshtat2006,Loinger2007,Morelli2008}, which is highly
similar to the system studied here---indeed, the toggle switch
consists of two genes that mutually repress each other. These studies
have revealed that the ensemble of transition states, which separate
the two stable states, is dominated by configurations where both
antagonistic proteins are present in low copy numbers. Clearly, the
probability that in a given nucleus not only the minority gap protein,
but also the majority gap protein reaches a low copy number, is
reduced by the diffusive influx of that majority species from the
neighboring nuclei, which are in the same gap-gene expression
state. In essence, diffusion increases the effective system size, with
its spatial dimension given by $\lambda=\sqrt{D/\mu_{\rm eff}}$; in
fact, since the stability of the toggle switch depends exponentially
on the system size \cite{Warren2004,Warren2005}, we expect the stability $\tau_{s}$
to scale with the diffusion constant as $\tau_{s}\sim e^D$. We thus
conclude that random switching between the two gap-gene expression
states, the high-\hb---low-\kni and low-\hb---high-\kni states, is not
likely to occur on the time scale of early development.

The observation that the switching rate is low raises another
important question: if errors are formed during development, can they
be corrected? We observe in the simulations with $D=0$ that when we
allow the gap-gene expression patterns to develop starting from
initial conditions in which the \Hb and \Kni copy numbers are both
zero, in the overlap (bistable) region a spotty gap-gene expression
pattern emerges, consisting of nuclei that are either in the
high-\hb---low-\kni state or in the low-\hb---high-\kni state. When
the diffusion constant of \Hb and \Kni is zero, then these defects are
essentially frozen in, precisely because of the low switching
rate. Interestingly, however, we find in the simulations that a finite
diffusion constant {\em can} anneal these defects. This may seem to
contradict the statement made above that diffusion lowers the
switching rate. The resolution of this paradox is that while diffusion
lowers the switching rate for nuclei that are surrounded by nuclei
that are in the same gap-gene expression state, it enhances the
switching rate for nuclei that are surrounded by nuclei with a
different gap-gene expression state; this is indeed akin to spins in
an Ising system below the critical point. The mechanism for the
formation of the gap-gene expression patterns, then, depends on the
diffusion constant. When $D$ is small yet finite, $0<D<0.1 \umsps$, in
the overlap region first small domains are formed consisting of nuclei
that are in the same gap-gene expression state; these domains then
coarsen analogously to Ostwald ripening of small crystallites in a
liquid below the freezing temperature; ultimately, they combine with
the \hb or \kni expression domains that have formed in the meantime
outside the overlap region, where \hb and \kni are activated by their
respective morphogens yet do not repress each other (see Videos S1 and S2).
For $D\gtrsim 0.1 \umsps$, no ``crystallites'' are formed in
the overlap region (both the \Hb and \Kni copy numbers are low yet
finite and \hb and \kni simultatenously repress each other); instead,
the \hb and \kni domains formed near the poles slowly invade the
overlap region (see Videos S3 and S4). Interestingly, even while in the
absence of \Hb and \Kni diffusion $\Delta G\approx 0$ in the overlap region, the
interface between the \hb and \kni expression domains does slowly
diffuse towards midembryo when $D>0$ and $A\leq 4$, 
due to the diffusive influx of \Hb and \Kni from the regions outside
the overlap region. When $A=8$, the \hb and \kni expression boundaries
are not pinned to the middle of the embryo, and their positions
exhibit slow and large fluctuations, presumably because the energetic driving
force is small, and the diffusive influx of \Hb and \Kni from the
regions near the poles is negligible.  We will investigate this effect in
more detail in a forthcoming publication.

\mysubsubsection{Mutual repression inhibits boundary shifts}
\Fig \ref{Fig7}A shows the average gap-gene expression profiles for
$A\in\lbrace 1,2,4\rbrace$ and $D=1.0~\umsps$, which minimizes the
boundary width $\Delta x$ when $A=1$ (see \Fig \ref{Fig4}). While
the morphogen-activation thresholds shift beyond midembryo as $A$ is
increased beyond unity, leading to an overlap of the domains where the
gap genes are activated by their respective morphogens (see inset),
the gap-gene expression boundaries overlap only marginally. This is
quantified in panel B, which shows the \Hb boundary position $x_t$ as
a function of $A$ and as a function of $\Delta x_A\equiv
x_{A,Kni}-x_{A,Hb}$, which is defined as the separation between the
positions $x_{A,Kni}$ and $x_{A,Hb}$ where \Kni and \Hb are
half-maximally activated by their respective morphogens; for $A=1$,
with adjoining morphogen activation regions, $\Delta x_A=0$ and for
$A>1$, with overlapping activation regions, $\Delta x_A$ is
negative. Without mutual repression (red data), the \Hb boundary
position $x_t$ tracks the shift of the \hb activation threshold, as
expected.  In contrast, with mutual repression (green data) the
boundary does not move beyond the position for $A=1$ as $A$ is
increased. The same robustness was also observed for other values
of the Hill coefficient of gap-gene activation (see \Fig S7 in \SItext).

\begin{figure}[!ht]
\begin{center}
\includegraphics[width=6in]{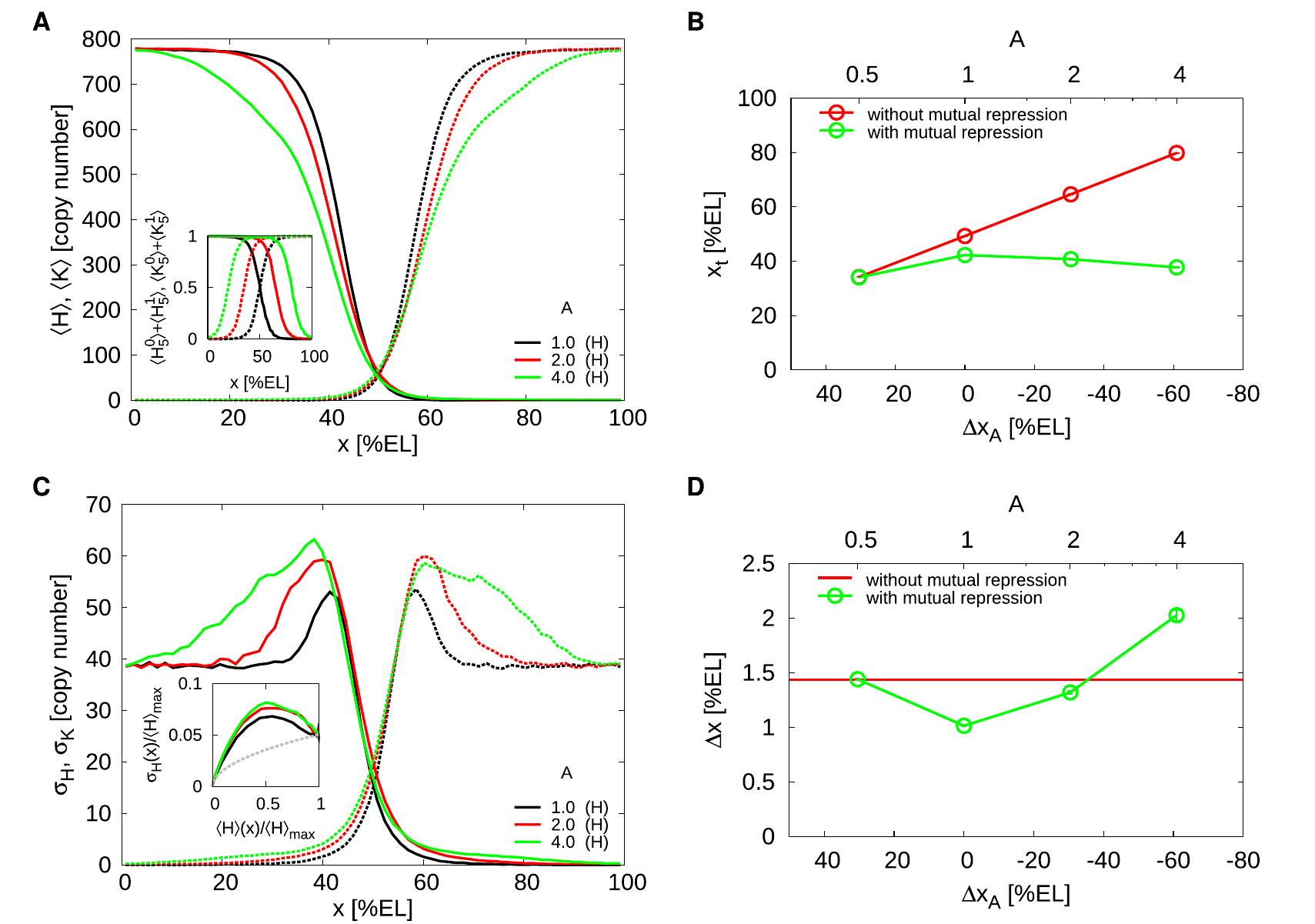}
\end{center}
\caption{ {\bf Mutual repression buffers against correlated variations
    in the activator levels.}  \subfig{A} Time- and
  circumference-averaged \Hb ($\Avg{H}$, solid lines) and \Kni
  ($\Avg{K}$, dashed lines) total copy-number profiles along the AP
  axis for various morphogen dosage factors $A$.  Inset: the
  corresponding average occupancy of the promoter states with five
  bound morphogen molecules as a function of $x$.  \subfig{B} The average 
  \Hb boundary position $x_t$ as a function of $\Delta x_A$, the distance
  between the \Hb and \Kni boundaries without mutual repression, for
  the system with mutual repression (green,) and without it (red);
  $\Delta x_A$ is varied by changing the morphogen dosage factor
  $A$. Note that mutual repression makes the gap-gene expression
  boundaries essentially insensitive to correlated changes in
  morphogen levels when $A>1$.  \subfig{C} AP profiles of the average
  standard deviation of the total \Hb ($\sigma_H$, solid lines) and
  \Kni ($\sigma_K$, dashed lines) copy numbers.  Inset:
  $\sigma_{H}(x)/\Avg{H}_{max}$ as a function of
  $\Avg{H}(x)/\Avg{H}_{max}$, where $\Avg{H}(x)$ is the average \Hb
  copy number at $x$ and $\Avg{H}_{max}$ its maximum over $x$.  The
  grey dashed line represents the Poissonian limit.\subfig{D} The \Hb
  boundary width $\Delta x$ as a function of $\Delta x_A$ with (green)
  and without (red) mutual repression. For $A=4$, it was impossible to
  obtain a reliable error bar on $\Delta x$, because of the weak
  pinning force on the \hb and \kni expression boundaries. }
\label{Fig7}
\end{figure}

\mysubsubsection{Mutual repression enhances robustness to
  embryo-to-embryo variations}
The fact that mutual repression can pin expression boundaries,
dramatically enhances the robustness against embryo-to-embryo
variations in the morphogen levels. We did not sample inter-embryo
variations in $A$ explicitly, but made an estimate using $\Delta x_t =
(dx_t/dA) \Delta A$, where $d x_t/dA$ was taken from \Fig \ref{Fig7}B.
A correlated symmetric variation $\delta_A\equiv\Delta A/A=0.1$ of
both morphogen levels then would lead to $\Delta
x_t(\delta_A)\simeq0.82~\pcEL$ at $A=1$ and $\Delta
x_t(\delta_A)\simeq0.25~\pcEL$ at $A=2$.  Without mutual repression
$\Delta x_{t,NR}(\delta_A)\simeq2.2~\pcEL$. This analysis thus
suggests that mutual repression reduces boundary variations due to
fluctuations in the morphogen levels by almost a factor of 10 if the
half-activation threshold is slightly posterior to midembryo
(e.g. $A=2$). If, on average, $A=1$, then mutual repression
still reduces $\Delta x_t$ by inhibiting posterior shifts in those
embryos in which $A>1$.  These results are consistent with those of
\cite{Howard2005,Vakulenko2009}.

\mysubsubsection{Overlap of morphogen activation domains does not
  corrupt robustness to intrinsic fluctuations}
While mutual repression proves beneficial in buffering against
embryo-to-embryo variations in morphogen levels, the question arises
whether overlapping morphogen-activation domains does not impair
robustness to intrinsic fluctuations arising from noisy gene
expression and diffusion of gap gene proteins. We found that this
depends on the Hill coefficient of gap-gene
activation, which depends on the number $n_{\rm max}$ of
morphogen binding sites on the promoter. \Fig \ref{Fig7}C shows, for $n_{\rm max}=5$, that even though
mutual repression increases the noise in gap-gene expression away
from the boundaries, it has little effect on the noise at the boundaries
when $A\leq 2$. For $A> 2$, the noise does increase significantly;
in fact, it was impossible to obtain reliable error bars, because of
the weak pinning force of the \hb-\kni interface. Moreover,
overlapping morphogen activation domains decrease the steepness of
the expression boundaries (panel A), and this increases the boundary
width $\Delta x$ (panel D). Indeed, when $n_{\rm max}=5$,
mutual repression can enhance the precision of gene-expression
boundaries, but only if the activation domains are adjoining
($A=1$), or have a marginal overlap ($1<A<2$). For lower values of
$n_{\rm max}$, however, this enhancement of precision extends over a much
broader range of $A$ values; in fact, when $n_{\rm max}<3$, mutual repression enhances
precision even up to $A=4$ (see \Fig S7 in \SItext).

\subsection*{Boundaries shift upon uncorrelated variations in morphogen levels, yet intrinsic noise remains unaltered}
\pdfbookmark[2]{Boundaries shift upon uncorrelated variations in morphogen levels, yet intrinsic noise remains unaltered}{BookmarkUncorrelated}
Since correlated upregulation of both morphogen levels
is a special case, we also studied the effect of uncorrelated
activator scaling.  To this end, only the \Bcd level was multiplied by
a global factor $A\in\lbrace0.5,1,2,3,4\rbrace$, while
other parameters were left unchanged.  Again we investigated the \Hb boundary
position $x_t$, its variance $\Delta x_t(\Delta A)$ due to extrinsic
(embryo-to-embryo) variations in $A$ and the variance due to
intrinsic (intra-embryo) fluctuations $\Delta x(A)$.  Results for
$D=1.0\umsps$ are summarized in \Fig \ref{Fig8}.

\begin{figure}[!ht]
\begin{center}
\includegraphics[width=6in]{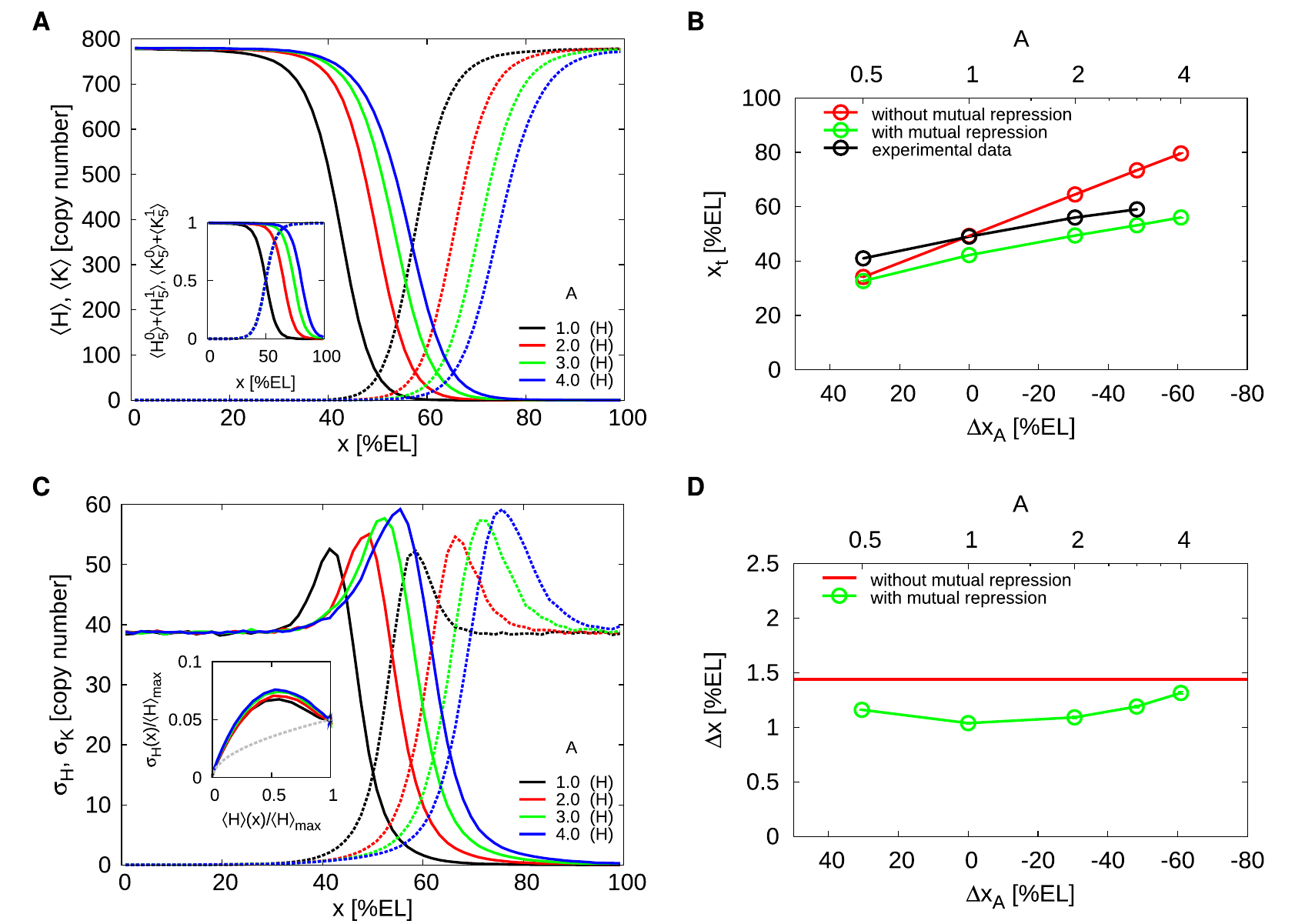}
\end{center}
\caption{ {\bf Robustness of the gap-gene expression boundaries to
    variations in the \bcd gene dosage.}  \subfig{A} Time- and circumference-averaged \Hb
  ($\Avg{H}$, solid lines) and \Kni ($\Avg{K}$, dashed lines)
  total copy-nymber profiles along the AP axis for various \bcd gene dosage
  factors $A_{Bcd}=A\in\lbrace 0.5, 1, 2, 3, 4\rbrace$ and
  $D=1.0~\umsps$.  Inset: the average occupancy of the promoter states
  with five bound morphogen molecules as a function of $x$.
  \subfig{B} Comparison of the boundary position
  $x_t$ as a function of $A$ for $D=1.0~\umsps$ to values measured by
  Houchmanzadeh et al. \cite{Houchmandzadeh2002} (black line).  The
  red line shows the simulation results for the system without mutual
  repression. Note the good agreement between the experimental data
  and the simulation data of the system with mutual repression.   \subfig{C} Profiles of the average standard deviation of the total \Hb
  ($\sigma_H$, solid lines) and \Kni ($\sigma_K$, dashed lines) copy
  number.  Inset: $\sigma_{H}(x)/\Avg{H}_{max}$ as a function of
  $\Avg{H}(x)/\Avg{H}_{max}$.  The grey dashed line represents the
  Poissonian limit.
  \subfig{D} The \Hb boundary width $\Delta x$ as a function of $A$ and
  $\Delta x_A$, the separation between the \Hb and \Kni boundaries in
  a system without mutual repression, for the system with (green)
  and without (red) mutual repression. $\Delta x_A$ is varied by
  multiplying the \Bcd level by $A_{Bcd}$.
  }
\label{Fig8}
\end{figure}

\mysubsubsection{The \Hb boundary shifts less with mutual repression}
\Fig \ref{Fig8}A shows that the \hb expression boundary shifts posteriorly
with increasing $A$, in contrast to the case of correlated
activator scaling. The \Kni profile retracts in concert with the
advance of the \Hb domain.  In \Fig \ref{Fig8}B we compare the \Hb
boundary $x_t(A)$ to the data of Houchmandzadeh et
al. \cite{Houchmandzadeh2002}, assuming a 100\% efficiency of the
additional \bcd gene copies. It is seen that the agreement between
simulation and experiment is very good: while $x_t(A)$ of the simulations
has a marginal offset as compared to the experimental data, the slope
of $x_t(A)$ is essentially the same. Moreover, the slope is much lower
than that obtained without mutual repression, showing that mutual
repression can indeed buffer against uncorrelated variations in
morphogen levels. These results parallel those of 
\cite{Howard2005}.

\mysubsubsection{Robustness to inter-embryo fluctuations} To estimate
the boundary variance due to inter-embryo variations in morphogen
levels, we fitted a generic logarithmic function $x_{t,fit}(A):=a \log
(A) + b$ to the simulation data, giving $a\lesssim 15~\pcEL$ for all
values of $D$ studied.  Hence $\Delta x_t(\Delta A)\lesssim
15~\pcEL~\Delta A/A$.  A $10\%$ variability in $A$ around $A=1$ thus
would result in $\Delta x_t(\Delta A)\lesssim 1.5~\pcEL$, which is
half as much as predicted by the model in \cite{Howard2005} for that
case.  Nevertheless, it is yet too large to correspond to the
experimental observations of Manu et al. that variations in the \Bcd
gradient of $\Delta A/A\approx 20\%$ correspond to variations in the
\Hb boundary position of $\Delta x_t(\Delta A)\lesssim 1.1~\pcEL$
\cite{Manu2009PlosBiol}. 
Our results therefore support their conjecture that
higher levels of \Bcd are correlated with upregulation of \Kni and \Cad.

\mysubsubsection{Robustness to intra-embryo fluctuations} The output
noise at the \Hb boundary remains largely unaffected (\Fig \ref{Fig8}C
and inset) by \Bcd upregulation, whereas the slope is reduced by
approximately $10\%$ per doubling of $A$ (data not shown).  As a
result, the boundary width $\Delta x$ stays close to $1~\pcEL$ for all
considered $A$ (green data; \Fig \ref{Fig8}D), remaining lower than that obtained
without mutual repression (red data; \Fig \ref{Fig8}D).

\subsection*{Mutual repression with one morphogen gradient}
\pdfbookmark[2]{Mutual repression with one morphogen gradient}{BookmarkOneGradient}
In the mutual repression motif discussed above, the two antagonistic
genes were activated by independent morphogens, one emanating from the
anterior and the other from the posterior pole. An alternative mutual
repression motif is one in which the two genes are activated by the
same morphogen, {\it e.g.} \hb and \kni both being activated by \Bcd
\cite{Saka2007,Cotterell2010}. 

We simulated a system in which \hb and \kni mutually repress each
other, yet both are activated by \Bcd, with \kni having a lower \Bcd
activation threshold than \hb. This generates a \Hb and \Kni domain,
with the latter being located towards the posterior of the former (see
\Fig S4 in \SItext). We systematically varied the mutual repression strength and
the diffusion constant, to elucidate how mutual repression and spatial
averaging sculpt stable expression patterns in this motif. Our
analysis reveals that since \hb and \kni are both activated by the
same morphogen gradient, \hb should repress \kni more strongly than
vice versa: with equal mutual repression strengths either a spotty
gap-gene expression pattern emerges in the anterior half, namely when
the \Hb and \Kni diffusion constant are low ($D<0.1 \umsps$), or \Kni
dominates or even squeezes out \Hb, namely when their diffusion
constant is large.  Nonetheless, for unequal mutual repression
strengths and sufficiently high $D$, the repression of \hb by \kni does enhance the precision
and the steepness of the \Hb boundary, although the effect is smaller
than in the two-gradient motif (\Fig S4 in \SItext). Clearly, while the
one-morphogen-gradient motif cannot provide the robustness against
embryo-to-embryo variations in morphogen levels that the
two-morphogen-gradient motif can provide, mutual repression can
enhance boundary precision also in this motif.

\pagebreak
\section*{Discussion}
\pdfbookmark[1]{Discussion}{BookmarkDiscussion}
Using large-scale stochastic simulations, we have examined the role of
mutual repression in shaping spatial patterns of gene expression, with
a specific focus on the \hb-\kni system. 
Our principal findings are that mutual
repression enhances the robustness both against intra-embryonic
fluctuations due to noise in gap-gene expression and
embryo-to-embryo variations in morphogen levels.

To investigate the importance of mutual repression in shaping
gene-expression patterns, we have systematically varied a large
number of parameters: the strength of mutual repression, the
diffusion constant of the gap proteins, the maximum expression
level, the Hill coefficient of gap-gene activation, and the
amplitude of the morphogen gradients. To elucidate how varying
these parameters changes the precision of the gap-gene boundaries,
we examined how they affect both the steepness of the
gene-expression boundaries and the expression noise at these boundaries (see
Eq. \ref{EqDeltaX}). The
effect on the steepness is, to a good approximation, independent
of the noise, and would therefore be more accessible
experimentally. We find that the steepness increases with
decreasing diffusion constant, but increases with increasing strength of mutual repression,
maximum expression level, and Hill coefficient of gap-gene
activation. Moreover, mutual repression shifts the expression
boundaries apart and makes the system more robust to
embryo-to-embryo variations in the morphogen levels. In contrast,
the noise at the expression boundaries decreases with increasing
diffusion constant,  decreasing expression level,    and decreasing Hill
coefficient, while the dependence on the strength of mutual
repression is non-monotonic, albeit not very large.
The interplay between noise and steepness means that the precision of the
gap-gene expression boundaries  increases (i.e., $\Delta
x$ decreases) with increasing expression level.
The dependence of $\Delta x$ on the diffusion constant and the strength of mutual
repression, on the other hand, is non-monotonic: there is an optimal diffusion
constant and repression strength that maximizes precision.
The effect of the Hill coefficient is conditional on the strength of mutual
repression: without mutual repression, the precision
slightly decreases with increasing Hill coefficient, 
while with mutual repression the precision increases with
increasing Hill coefficient.

While mutual repression has only a weak effect on the noise in the
expression levels at the gene-expression boundaries, it does markedly
steepen the boundaries, especially when the diffusion constant is low.
Indeed, mutual repression can enhance the precision of gene expression
boundaries by steepening them.  Nonetheless, even with mutual
repression spatial averaging \cite{Okabe-Oho2009,Erdmann2009} appears
to be a prerequisite for achieving precise expression boundaries:
without diffusion of the gap proteins, the width of the \hb expression
boundary is larger than that observed experimentally
\cite{Gregor2007}. Hence, while previous mean-field analysis found
diffusion not be important for setting up gene-expression patterns
\cite{Papatsenko2011,Manu2009PlosCompBiol}, our analysis underscores
the importance of diffusion in reducing copy-number
fluctuations. In addition, diffusion can anneal patterning defects
that might arise from the bistability induced by mutual
repression. Diffusion is, indeed, a potent mechanism for
reducing the effect of fluctuations, such that mean-field analyses
can accurately describe mean expression profiles.

Interestingly, the minimum boundary width at the optimal diffusion
constant in a system with mutual repression is not much lower than
that in one without mutual repression. Yet, in the latter case the
boundary width is already approximately one nuclear spacing, and there
does not seem to be any need for reducing it further. However, with mutual
repression, the same boundary width can be obtained at a lower
diffusion constant, where the steepness of the boundaries is much
higher, approximately twice as high as that without mutual
repression. Our results thus predict that mutual repression allows for
gap-gene expression boundaries that are both precise and steep. In
fact, the width and steepness of the boundaries as prediced by our
model are in accordance with those measured experimentally
\cite{Surkova2008}.

Our observation that mutual repression increases the steepness
of gene-expression boundaries without significantly raising the
noise, makes the mechanism distinct from other mechanisms for
steepening gene expression boundaries, such as lowering diffusion
constants \cite{Erdmann2009} or increasing the cooperativity of
gene activation (see \Fig S6 in \SItext). These mechanisms typically involve a
trade off between steepness and noise: lowering the diffusion
constant or increasing the Hill coefficient of gene activation
steepens the profiles but also raises the noise in protein
levels at the expression boundary. In fact, increasing the Hill
coefficient (without mutual repression) {\em decreases} the
precision of gene-expression boundaries. This is because increasing the
Hill coefficient increases the width of the distribution of times
during which the promoter is off, leading to larger promoter-state
fluctuations and thereby to larger noise in gene expression (see \Fig S5 in \SItext).

Another important role of mutual repression as suggested by our
simulations is to buffer against inter-embryonic variations in the
morphogen levels. Houchmandzadeh {\it et al.} observed that in
\bcd overdosage experiments the \Hb boundary does not shift as far
posteriorly as predicted by the French flag model
\cite{Houchmandzadeh2002}. One possible explanation that has been put
forward is that \Bcd is inactivated in the posterior half of the
embryo via a co-repressor diffusing from the posterior pole
\cite{Howard2005}. More recently, it has been proposed that gap gene
cross regulation underlies the resilience of the gap-gene expression
domains towards variations in the \bcd gene dosage
\cite{Manu2009PlosBiol,Manu2009PlosCompBiol}. Our analysis supports the latter
hypothesis. In particular, our results show that when the regions in
which \hb and \kni are acitvated by their respective morphogens
overlap, the boundary positions are essentially insensitive to
correlated variations in both morphogen levels, and very robust
against variations of the \Bcd level only, with the latter being in
quantitative agreement with what has been observed experimentally
\cite{Houchmandzadeh2002}. Moreover, when this overlap is about 0-20\%
of the embryo length, mutual repression confers robustness not only
against inter-embryonic variations in morphogen levels, but also
intra-embryonic fluctuations such as those due to noise in gene
expression.

Manu {\it et al.} found that in the \kr;\kni double mutant,
which lacks the mutual repression between \hb and \kni/\kr, the \Hb
midembryo boundary is about twice as wide as that in the wild-type
embryo \cite{Manu2009PlosBiol}. This could be due to a reduced
robustness against embryo-to-embryo variations in morphogen levels,
but it could also be a consequence of a diminished robustness against
intra-embryonic fluctuations. The analysis of Manu {\it et al.}
suggests the former \cite{Manu2009PlosBiol,Manu2009PlosCompBiol}, and
also our results are consistent with this hypothesis. However, our
results also support the latter scenario: for $D\approx 0.3\umsps$,
the \Hb boundary width in the system without mutual repression is
about twice as large as that in the system with mutual repression (see
\Fig \ref{Fig4}C). Clearly, new experiments are needed to establish
the importance of intra-embryonic fluctuations versus inter-embryonic
variations in gene expression boundaries.

To probe the relative magnitudes of intra- vs inter-embryonic
variations, one ideally would like to measure an ensemble of embryos
as a function of time; one could then measure the different
contributions to the noise in the quantity of interest following
Eq. \ref{EqNoiseDecom}.  This, however, is not always possible;
staining, e.g., typically impedes performing measurements as a
function of time.  The question then becomes: if one measures
different embryos at a given moment in time, are embryo-to-embryo
variations in the mean boundary position or protein copy number
(thus averaged over the circumference) due to intra-embryonic
fluctuations in time or due to systematic embryo-to-embryo
variations in e.g. the morphogen levels? Experiments performed on different embryos but at one
time point cannot answer this question. Our analysis, however,
suggests that the intra-embryonic fluctuations in the mean copy
number or boundary position (i.e. averaged over $\phi$) over time are very
small, and that hence embryo-to-embryo variations in the mean
quantity of interest are really due to systematic embryo-to-embryo
variations; these variations then correspond to 
$\sigma^2_{\Avg{\overline{G}}_\phi}$ or
$\sigma^2_{\Avg{\overline{x_t}}_\phi}$ in Eq. \ref{EqNoiseDecom}
or Eq. \ref{EqXtDecom}, respectively.  The
intra-embryonic fluctuations, $\Avg{\overline{\sigma^2_G}}_e(x)$
or $\Avg{\overline{\sigma^2_{x_t}}}_e(x)$, can then be measured
by measuring the quantity of interest, $G$ or $x_t$, as a
function of $\phi$, and averaging the resulting variance over
all embryos. We expect that these observations, in particular
the critical one that intra-embryonic fluctuations in the mean
quantity of interest are small, also hold for
non-stationary systems, although this warrants further investigation.

Our model does not include self-activation of the gap
genes. Auto-activation has been reported for \hb, \kr
and \gt, but there seems to be no evidence in case of \kni
\cite{Treisman1989,Jaeger2011}.  The self-enhancement of gap genes has
the potential to steepen and sharpen expression domains even more by
amplifying local patterns \cite{Lopes2008,Holloway2011}. Our
results suggest, however, that auto-activation is not necessary to reach the
boundary steepness and precision as observed experimentally. 

 Our results provide a new
perspective on the Waddington picture of development
\cite{Waddington1942, Waddington1959}. Waddington argued that
development is ``canalized'', by which he meant that cells
differentiate into a well-defined state, despite variations and
fluctuations in the underlying biochemical processes. It has been
argued that canalization is a consequence of multistability
\cite{Manu2009PlosBiol,Manu2009PlosCompBiol,Papatsenko2011},
which is the idea that cells are driven towards attractors, or basins of attraction in state
space. To determine whether a given system is multistable, it is
common practice to perform a stability analysis at the level of
single cells or nuclei. Our results show that this approach should be
used with care: diffusion of proteins between cells or nuclei within
the organism can qualitatively change the energy landscape;
specifically, a cell that is truly bistable without diffusion might be
monostable with diffusion. Indeed, our results highlight that a
stability analysis may have to be performed not at the single cell
level, but rather at the tissue level, taking the diffusion of
proteins between cells into account.

Finally, while our results have shown that mutual repression can
stabilize expression patterns of genes that are activated by morphogen
gradients, one may wonder whether it is meaningful to ask the converse
question: do morphogen gradients enhance the stability of expression
domains of genes that mutually repress each other? This question
presupposes that stable gene expression patterns can be generated
without morphogen gradients.
Although it was shown that confined (though aberrant) gap gene patterns
form in the absence of \Bcd \cite{Huelskamp1989,Huelskamp1990,Struhl1992} and that
\Hb can partly substitute missing \Bcd in anterior embryo patterning\cite{Simpson-Brose1994},
it is not at all obvious how precise domain positioning could succeed in such a scenario.
In particular, one might expect that with mutual repression
only, thus without morphogen gradients, there is no force that pins
the expression boundaries. Our results for the large overlapping
morphogen-activation domains, with $A=8$, illustrate this problem: in
the overlap region, both \hb and \kni are essentially fully activated
by their respective morphogens, as a result of which the morphogen
gradients cannot determine the positions of the gap-gene boundaries
within this region; indeed, mutual repression has to pin the
expression boundaries of \hb and \kni. Yet, our results show that in
this case the positions of the \hb and \kni expression boundaries
exhibit large and slow fluctuations, suggesting that mutual repression
alone cannot pin expression boundaries. Interestingly, however, with
$A=4$, the region in which both genes are activated is still quite
large, about 50\% of the embryo, and yet even though the underlying
energy landscape is flat in this region, the interfaces do
consistently move towards the middle of the embryo, due to diffusive
influx of \Hb and \Kni from the polar regions. It is tempting to
speculate that mutual repression and diffusion can maintain stable
expression patterns, while morphogen gradients are needed to set up
the patterns, {\it e.g.} by breaking the symmetry between the possible
patterns that can be formed with mutual repression only.

\pagebreak
\section*{Materials and Methods}
\pdfbookmark[1]{Materials and Methods}{BookmarkMatMeth}
In the following we describe details of our parameter choice
and sampling technique.
To unravel the mechanisms by which mutual repression shapes
gene-expression patterns, it is useful to take the \Cad-\Kni-system
to be a symmetric copy of the \Bcd-\Hb-system. \Cad thus
inherits its parameters from \Bcd and \Kni from \Hb, if not otherwise
stated.  Table S2 in \SItext
gives an overview of our standard parameter values.  Data from
experiments was used whenever possible.  When it was unavailable we
made reasonable estimates.

\mysubsubsection{Binding rates are diffusion limited}
We assume all promoter binding rates to be diffusion limited and
calculate them via $k_{on}^{\rm X} =4\pi\alpha D_X/V$.  Here
$\alpha=10~\unit{nm}$ is the typical size of a binding site, $D_X$ is
the intranuclear diffusion constant of species $X$ and $V=143.8~\mu{\rm m}^3$
is the nuclear volume. The precise values of $D_X$ for the different species
in our system are not known.  Gregor et al. have shown experimentally that the
nuclear concentration of \Bcd is in permanent and rapid dynamic
equilibrium with the cytoplasm \cite{Gregor2007b}, suggesting that
nuclear and cytoplasmic diffusion constants can be taken for equal.
They have found $D_{Bcd}\simeq 0.32~\umsps$ by FRAP measurements.
This value has been subject to controversy because it is too low to
establish the gradient before nuclear cycle 10 ($\simeq
90~\unit{min}$) by diffusion and degradation only,  prompting
alternative gradient formation models \cite{Bergmann2007, Bialek2008,
  Barkai2009, Spirov2009, Hecht2009, Porcher2010b}.  A more recent
study revisited the problem experimentally via FCS, yielding
significantly higher values for $D_{Bcd}$ up to $10~\umsps$ with a
lower limit of $1~\umsps$ \cite{Abu-Arish2010}.  We therefore have
chosen a 10x higher value of $D_{Bcd}=D_{Cad}\equiv D_A=3.2~\umsps$ as
compared to the earlier choice in \cite{Erdmann2009}.  For simplicity,
this value is taken for all binding reactions occuring in our model,
except for the dimerization reaction rate $k_{on}^{\rm D}$, which is taken
to be higher by a factor of 2 to account for the fact that both reaction
partners diffuse freely. 

To model cooperative activation of \hb and \kni by their respective
morphogens, the morphogen-promoter dissociation rate is given by
$k_{off,n}^{\rm A}=a/b^n / {\rm s}$, where $n$ is the number of
morphogen molecules that are bound to the promoter; for our standard
cooperativity $n_{max}=5$ the values of
$a=410$ and $b=6$ have been chosen such that the threshold
concentration for promoter activation (in the absence of repression)
equals the observed average number of morphogen molecules at midembryo (when
$A=1$, see below). $n_{max}$ is varied in some simulations; we describe in
\SItext how $a$ and $b$ are chosen in these cases.
The promoter unbinding rate of \hb and \kni (the
repressor-promoter unbinding rate) $k_{off}^{\rm R}$ is a parameter
that we vary systematically. To study the potential role of
bistability we decided to set $k_{off}^{\rm R}$ to a value which
ensures bistable behavior when both \hb and \kni are fully activated
by their respective morphogens (meaning that all five binding
morphogen-binding sites on the promoter are occupied).  This requires
tight repression, yielding dissociation constants $\sim
10^{-2}~\unit{nM}$ (but see also below).  The dimer dissociation rate is set to be
$k_{off}^{\rm D}=k_{on}^{\rm D}/V$, which is motivated by the choice
for the toggle switch models studied in \cite{Warren2004, Warren2005} and
\cite{Morelli2008}, and asserts that at any moment in time the
majority of the gap proteins is dimerized.  This is a precondition for
bistability in the mean-field limit \cite{Cherry2000, Warren2004, Warren2005}.

The parameters of the exponential morphogen gradients are chosen such
that the number of morphogen molecules at midembryo and the decay
length of the gradient are close to the experimentally observed values
for \Bcd, 690 and $\lambda=119.5~\unit{\mu m}$, respectively
\cite{Gregor2007}.

\mysubsubsection{Production and degradation dynamics}
The copy numbers of both monomers and dimers and the effective gap
gene degradation rate $\mu_{eff}$ depend in a nontrivial manner on
production, degradation and dimerization rates.  However, for constant
production rate $\beta$, without diffusion and neglecting promoter
dynamics, an analytical estimate for the monomer and dimer copy
numbers can be obtained from steady state solutions of the rate
equations (see \SItext).  Based on this we have made a
choice for $\beta$ and the monomeric ($\mu_M$) and dimeric ($\mu_D$)
decay rates that leads to reasonable copy numbers and $\mu_{eff}$ (see
Table S2 in \SItext).  The latter is defined as the mean of
$\mu_M$ and $\mu_D$ weighted by the species fractions.  $\mu_M$ and
$\mu_D$ are set such that $\mu_{eff}\simeq
4.34\cdot10^{-3}~\unit{1/s}$, which corresponds to an effective
protein lifetime of $\sim 4~\unit{min}$.  This is close to values used
earlier \cite{Howard2005,Erdmann2009} and allows for the rapid
establishment of the protein profiles observed in experiments.  The
dimers have a substantially lower degradation rate than monomers,
which enhances bistability \cite{Buchler2005}.  The lower decay rate
of the dimers may be attributed to a stabilizing effect of
oligomerization (cooperative stability) \cite{Buchler2005}.

\mysubsubsection{Free parameters}
One of the key parameters that we vary systematically is the
internuclear gap gene diffusion constant $D$, which defines a nuclear
exchange rate $k_{ex}=4D/\ell^2$ ($\ell$ = internuclear distance).  To
study the effect of embryo-to-embryo variations in the morphogen
levels, the latter are scaled globally by a dosage factor $A$.  We
considered two scenarios: scaling both gradients by the same $A$
(``correlated variations'') or scaling the \Bcd gradient only
(``uncorrelated variations'').  To test how strongly the assumption of
strong repressor-promoter binding affects our results, we also varied
the repressor-promoter dissociation rate $k_{off}^{\rm R}$. Moreover,
to study the dependence of our results on the gap-gene copy numbers,
we also increased the protein production rate $\beta$. These
simulations are much more computationally demanding; therefore we limited
ourselves to simulations with $\beta=2\beta_0$ and $\beta=4\beta_0$ where
$\beta_0$ is our baseline value. Finally we
also studied a system where both gap genes are activated by the same
gradient (\Bcd), varying both the diffusion constant $D$ and the \Kni
repressor off-rate $k_{off}^{\rm R,\Kni}$, while keeping $k_{off}^{\rm
  R,\Hb}$ at the standard value.

\mysubsubsection{Algorithmic details}
All simulations are split into a relaxation and a measurement run.
During the relaxation run we propagate the system towards the steady
state without data collection.  To reach steady state, as a standard
we run $1\cdot10^9-3\cdot10^9$ Gillespie steps (ca. $2\cdot10^5-7\cdot 10^5$ updates per
nucleus).  The measurement run is performed with twice the number of
steps ($2\cdot10^9-6\cdot10^9$). The simulations are started from exponential
morphogen gradients and step profiles of the gap proteins; however,
we verified that the final result was independent of the precise
initial condition, and that the system reached
steady state after the equilibration run. The results for
$A=4$ (\Fig \ref{Fig7}) form, however, an exception: here it was impossible to obtain a
reliable error bar, because of the weak pinning force on the \hb and
\kni expression boundaries.

In steady state, we record for each row of nuclei and with a
measurement interval of $\tau_{m}=100~\unit{s}$ the \Hb boundary
position $x_t$, i.e. the position where $H$ drops to half of the average
steady-state value measured at its plateau close to the anterior pole,
which in our simulations is equal to the maximum average total \Hb level $\Avg{H}_{max}$.
From the corresponding histogram we obtain the boundary width $\Delta x$ by
computing the standard deviation.  Additionally, after runtime we calculate 
an approximation for $\Delta x$ from the standard deviation
of $H$ divided by the slope of the averaged $H$ profile, both
quantities taken at $x_t$, see Eq. \ref{EqDeltaX} \cite{Gregor2007,
  Tostevin2007, Erdmann2009}.  Further details of boundary measurement
are described in \SItext.

Error bars for a given quantity are estimated from the standard
deviation among $N_B=10$ block averages (block length $6\cdot10^8$)
divided by $\sqrt{N_B-1}$, following the procedure described in
\cite{Frenkel+Smit}.  We verified that estimates with smaller and
larger block sizes yield similar estimates for a representative set of
simulations.

\section*{Acknowledgments}
\pdfbookmark[1]{Acknowledgements}{BookmarkAcknowlegdements}
We thank N. Becker for fruitful discussions and a critical reading 
of the manuscript.

\pagebreak
\pdfbookmark[1]{References}{BookmarkReferences}
\bibliography{Drosophila}

\renewcommand{\thetable}{S\arabic{table}}

\newcounter{myc} 
\renewcommand{\thefigure}{S\arabic{myc}} 

\newpage

\section*{SUPPORTING TEXT S1}
\pdfbookmark[0]{SUPPORTING TEXT S1}{BookmarkSupportingText}

\section{Methodic details}
\subsection{Measurement of the boundary width}
By default we determine the boundary width in the following two ways:

Let $c\sprm{m}_{n,s}$ be the copy number of species $s$ in a nucleus
with angular index $m<N_\phi$ and axial index $n<N_x$, where
$N_\phi$ is the number of rows around the circumference of the
cylinder, and $N_x$ is the number of colums in the axial direction
along the AP axis. To compute the boundary width of the expression
domain of a gap protein $s$, we compute for each row $m$
$T\sprm{m}_{n,s}=(c\sprm{m}_{n,s}-\theta_s)\cdot(c\sprm{m}_{n+1,s}-\theta_s)$ as
a function of $n$, where $\theta_s$ is half the copy number expected
at full activation. A boundary position $x\sprm{m}_t=x\sprm{m}(n_t+\frac{1}{2})$ is defined as the
position (nucleus) where $T\sprm{m}_{n_t,s}<0$. 
The values of $x\sprm{m}_t$ are recorded in a
histogram; here, the positions for the different rows $m$ are put in
the same histogram. The histogram is normalized at the end of the
simulation, and the boundary width $\Delta x$ is calculated as the
standard deviation of this histogram.

Secondly, at the end of the simulation, the slope of the average, $\Avg{H(x_t)}^\prime$,
and the standard deviation of the total \Hb copy number $\sigma_H(x_t)$ at the
\Hb boundary position $x_t$ are calculated from the time- and
$\phi$-averaged profiles.  From this, an approximation for the
boundary width given by $\Delta x \approx
\frac{\sigma(x_t)}{|\Avg{H(x_t)}^\prime|}$ is obtained, following
\cite{Tostevin2007,Gregor2007,Erdmann2009}.  To this end, first
$x_t$ is determined in the same way as in the runtime measurements,
only now working on the (both time- and circumference-) averaged
profile.
We describe in the following section how the steepness $\Avg{H(x_t)}^\prime$ is measured.

\subsection{Measurement of the profile steepness}
In our discrete system the measurement of a local derivative at the boundary position $x_t$ is
a process prone to even small stochastic variations if a naive measurement technique
is chosen. If the average boundary position $x_t$ for a set of different samples
with identical initial conditions always is in between two
particular nuclear positions $x(n_0)$ and $x(n_0+1)$, then using
linear differences to determine the steepness $\Avg{H(x_t)}^\prime$ at the boundary position
may give a reasonable estimate.
If, however, $x_t$ fluctuates around a particular nuclear position $x(n_0)$
among different samples and $\Avg{H(x(n_0-1))}-\Avg{H(x(n_0))}$ significantly differs from
$\Avg{H(x(n_0))}-\Avg{H(x(n_0+1))}$, the linear differences method will produce a large error
bar and also markedly affect the mean of $x_t$ among these samples.
As a result both the measured steepness and the quality of that measurement
for a given set of parameters depends on whether $x_t$ accidently happens
to predominantly vary in the interval between the same nuclear positions or not.
To overcome this illness we measure the boundary steepness from the average protein
profile by a two-step polynomial fitting procedure:
First we fit a polynomial of 3rd degree to a region of the data around $x_t$ that
contains at least four points (nuclei).
The derivative of the polynomial at $x_t$ gives an initial estimate of the
boundary slope, which we use this to calculate the
approximative x-interval over which the profile falls from maximal to minimial
expression level.
If the latter is larger than the original fitting range (which usually is the case)
we repeat the fitting on the enlarged interval.
Since the profiles to a good approximation are sigmoidal functions this
improves the quality of the fit.
The measured boundary slope then is defined as the derivative of the
polynomial function at $x_t$ after the second fitting.

\subsection{Number of cortical nuclei at cell cycle 14}
The development of the Drosophila embryonic syncytium starts with a single
nucleus. The first 9 nuclear divisions happen in the yolk. During cell
cycles 7 to 10 a migration of the nuclei towards the cortex can be observed.
However, approximately 200 polyploid nuclei stay behind in the yolk
and stop dividing after their 10th cycle \cite{Foe1983}.
This quiescence persists during subsequent cell cycles, including cycle 14.
As an effect of this, the number of nuclei at the cortex in cycle 14 is
considerably lower than $2^{13}=8192$.
An estimate of the reduced number of cortical nuclei is given by:
\begin{align}
 N_{cortex} \simeq (2^9 - 200) \cdot 2^4 = 4992
\end{align}
This number indeed is closer to $2^{12}$ than to $2^{13}$.
Note that in our model the precise number of nuclei does not
matter, rather it is the distance between the nuclear compartments
and the diffusion correlation length that impact on the results.
Our values for both the internuclear distance and the
nuclear diameter correspond to the experimental values reported by
Gregor et al. \cite{Gregor2007, Gregor2007b}.

\subsection{Predicted copy numbers and effective protein lifetime}
\label{SecPredictedCopyNumbers}
Our main observables are the total copy numbers of \Hb and \Kni,
defined as follows:
\begin{align}
 H\equiv c\sprm{m}_{n,H} = c\sprm{m}_{n,H_M} + 2c\sprm{m}_{n,H_D} + 2\sum_{j=0}^5 c\sprm{m}_{n,K^1_j}	\nn\\
 K\equiv c\sprm{m}_{n,K} = c\sprm{m}_{n,K_M} + 2c\sprm{m}_{n,K_D} + 2\sum_{j=0}^5 c\sprm{m}_{n,H^1_j}	
\label{EqObservables}
\end{align}
Here, for $G\in\lbrace H,K\rbrace$, $c\sprm{m}_{n,G^1_j}=1$ if the
promoter of species G is binding $j$ morphogen molecules and one (repressing) gap
dimer; evidently, at any given moment in time $c\sprm{m}_{n,G_j^1}$
can be equal to one for only one $j \in \lbrace 0 .. 5 \rbrace$.

The ratio between the number of monomeric and the number of dimeric
proteins is a nontrivial function of the monomer production rate, the
monomer and dimer degradation rates and the parameters that determine
the dimerization and dedimerization reactions.  To obtain an estimate
for the expected copy numbers of monomers and dimers of gene $g$ we
solved the mean-field rate equations for a simplified model which
comprises monomer production, (de)dimerization and monomer and dimer
degradation only, i.e. in which promoter state fluctuations and
diffusion are neglected, in the steady state.  We assume here that
stochastic monomer production events can be accounted for by an
effective mean-field production rate $\Avg{\beta}=\beta \Avg{H_5^0}$
for \Hb and similarly for \Kni, which depends on promoter (un)binding
parameters and the particular morphogen and repressor levels.  This
yields the following prediction for the copy number of monomers
($G_{M,\Avg{\beta}}$) and dimers ($G_{D,\Avg{\beta}}$):
\begin{align}
G_{M,\Avg{\beta}} &= \frac{1}{4 k\sprm{D}_{on} \mu_D} \Bigg\lbrace 2 k\sprm{D}_{on} \mu_D -k\sprm{D}_{off} \mu_M  - \mu_M \mu_D 				\nn\\
	   & \hspace{1.9cm} + \sqrt{8\Avg{\beta} k\sprm{D}_{on} \mu_D \left(k\sprm{D}_{off} + \mu_D\right) 
			  + \left[ \mu_M \left(k\sprm{D}_{off} + \mu_D\right) -2 k\sprm{D}_{on} \mu_D \right ]^2 } \Bigg\rbrace		\nn\\
G_{D,\Avg{\beta}} &= \frac{1}{8 k\sprm{D}_{on} \mu_D^2} \Bigg\lbrace k\sprm{D}_{off} \mu_M^2 + \mu_D \left[4 \Avg{\beta} k\sprm{D}_{on} + \mu_M \left(\mu_M - 2 k\sprm{D}_{on} \right)\right] \nn\\
	    & \hspace{1.9cm} - \mu_M \sqrt{8\Avg{\beta} k\sprm{D}_{on} \mu_D \left(k\sprm{D}_{off} + \mu_D\right)
			  + \left[ \mu_M \left(k\sprm{D}_{off} + \mu_D\right) -2 k\sprm{D}_{on} \mu_D \right ]^2 } \Bigg\rbrace
\end{align}

Here $\mu_M$ ($\mu_D$) is the monomeric (dimeric) degradation rate and
$k_{on}^{\rm D}$ ($k_{off}^{D}$) are the dimerization forward (backward)
rates, respectively.
From this we calculate the total expected copy number $G_{\Avg{\beta}}:=2G_{D,\Avg{\beta}}+G_{M,\Avg{\beta}}$
at effective production rate $\Avg{\beta}$.
In particular in the full-activation case, i.e. when the probability to be fully activated and unrepressed $\Avg{G\sprm{0}_5}\approx 1$
and therefore $\Avg{\beta}\approx\beta$, the above estimates correspond to average values from our simulations very well.

We define the effective degradation as $\mu_{eff}=\mu_{eff}(G_M, G_D)=\frac{1}{G_M+2G_D}\left( \mu_MG_M + 2\mu_D G_D \right)$.
Our standard values result in $\mu_{eff}\approx 4.34\cdot10^{-3}\unit{/s}$ with $G_M=G_{M,\beta}$ and $G_D=G_{D,\beta}$.

\section{Additional analysis}
\subsection{Poissonian limit with dimerization}
In \cite{Erdmann2009}, it was shown that, when $D\to \infty$, the
variance in the protein concentration becomes equal to the mean
concentration: diffusion washes out bursts in gene expression, thus
reducing the non-Poissonian part of the noise. However, in that model
the proteins do not dimerize, 
in contrast to our model. With
dimerization, a different limit for the variance in the total protein
concentration is approached as $D\to \infty$.  To derive this limit,
first note that the total protein copy number $G$ of a protein G is
$G\approx2G_D+G_M$. Assuming that $\Avg{G_DG_M}\approx
\Avg{G_D}\Avg{G_M}$ (our simulations indicate that this
approximation is very accurate), we find that the variance
$\sigma^2_G$ in $G$ is:
\begin{align}
 \sigma^2_G \approx 4\sigma^2_{G_D} + \sigma^2_{G_M},
\end{align}
where $\sigma^2_{G_D}$ is the variance in the dimer level $G_D$ and
$\sigma^2_{G_M}$ is the variance in the monomer level $G_M$. 
Both monomers and dimers are subject to spatial averaging, and
therefore their variances can be written in the form
\cite{Erdmann2009}:
\begin{align}
 \sigma^2_{G_M} = G_M + \frac{1}{N}\left( \sigma^2_{0,G_M} - G_M \right)	\nn\\ 
 \sigma^2_{G_D} = G_D + \frac{1}{N}\left( \sigma^2_{0,G_D} - G_D \right)
\end{align}
Here $N$ is the number of nuclei contributing to the averaging, which
is proportional to $D$, and $\sigma^2_{0,G_{M/D}}$ is the variance in
the monomer and dimer levels in the absence of diffusion,
respectively.  The part preceded by $1/N$ represents the variance that
can be reduced by spatial averaging.  Plugging these expressions
into the previous and using $G= 2G_D + G_M$ we arrive at:
\begin{align}
 \sigma^2_{G} &= 4G_D + G_M + \frac{1}{N}\left[ 4\sigma^2_{0,G_D} + \sigma^2_{0,G_M} -4G_D -G_M \right]	\nn\\ 
	      &= \left( 1 + \frac{2G_D}{G}\right)G + \frac{1}{N}\left[ \sigma^2_{0,G} - \left( 1 + \frac{2G_D}{G}\right)G \right]	\nn\\
	      &=: \left( 1 + f_D\right)G + \frac{1}{N}\left[ \sigma^2_{0,G} - \left( 1 + f_D\right)G \right]
\end{align}
Note that $N$ is the same for both monomers and dimers because their
diffusion constant does not differ in our model.  Evidently, the lower
bound for $\sigma_G$ in the limit $N\rightarrow\infty$ is not
$\sqrt{G}$ any more, but given by $\sqrt{\left(1+f_D\right)G}$, where
$f_D$ is the fraction of proteins in the dimer state with respect to
the total protein number (implying $f_D\leq 1$).  This is indeed what
we observe in our data for $\sigma_G$.  In our simulations the
equilibrium is strongly shifted towards the dimerized state, so that
$f_D\approx 0.97$. We can understand the limit
$N,D\rightarrow \infty$ intuitively by noting that in this limit there is no noise
in the nuclear protein concentration due to the stochastic production
and decay of molecules in each of the nuclei---this is because the
synthesized molecules are immediately donated to a reservoir that is
infinitely large; instead, there is only noise in the nuclear protein
concentration due to the sampling of molecules from this reservoir,
which obeys Poissonian statistics: $\sigma^2_{G_M} = G_M$ and
$\sigma^2_{G_D}=G_D$. This yields, for $N,D\rightarrow \infty$,
$\sigma^2_G=4\sigma^2_{G_D}+\sigma^2_{G_M}=4G_D+G_M=(1+f_D)G$.

\subsection{Bifurcation analysis}
In order to predict the regions in which bistability can be expected
for different amplitudes $A$ of the morphogen gradients we performed a deterministic
mean-field bifurcation analysis for a simplified 1-dimensional version of our model
of mutual repression between \hb and \kni.
The analysis is based on the following two equations describing the
change of the mean-field total copy number of \Hb ($H(x)$) and \Kni ($K(x)$) at position $x$:
\begin{align}
 \partial_t H(x) &= \beta_H(x) \frac{K_R^2}{K_R^2 + \left[f_D K(x)\right]^2} - \mu_H H(x)	\\
 \partial_t K(x) &= \beta_K(x) \frac{K_R^2}{K_R^2 + \left[f_D H(x)\right]^2} - \mu_K K(x)
\end{align}

Here $\beta_H$ and $\beta_K$ represent the protein synthesis rates, $\mu_H$ and
$\mu_K$ the corresponding (effective) degradation rates, $K_R$ is the dissociation constant of
cooperative repressor binding to the promoter and $f_D$ is the fraction of proteins in the
dimerized state.
Note that, since the intermediate step of dimerization is neglected here, we
have to take $K_R=\sqrt{k^R_{off}/k^R_{on}}$ if $k^R_{on}$ and $k^R_{off}$ are
the binding rates of the dimers.
To facilitate calculations we make two further simplifying assumptions here:
\begin{enumerate}
 \item We neglect activation dynamics and resulting promoter state fluctuations, i.e.
we assume that certain constant levels of the activators at position $x$ lead to 
average constant production rates $\beta_H=\beta([Bcd](x))$ and $\beta_K=\beta([\Cad](x))$,
respectively.
In our standard case $\beta([Act](x))=[Act]^5(x)/([Act]^5(x)+690^5)$ for both $[Act]=[Bcd]$
and $[Act]=[Cad]$.

\item In our simulations we have different degradation rates for monomers and dimers
so that the effective total degradation rate depends on the monomer-to-dimer ratio, which
in turn varies with the total copy number (see section \ref{SecPredictedCopyNumbers}).
Thus, in principle, also $f_D$ and
$\mu_H$ and $\mu_K$ are functions of $x$, or the corresponding activator levels.
Since this introduces further nonlinearities
into the above equations and complicates their solution, we substitute the degradation
rates $\mu_H$ and $\mu_K$ by a constant value $\mu_{eff}$, which is the effective degradation
rate for the maximal expression level (full activation).
Also for $f_D$ we take the constant value for full activation, $f_D\simeq0.97$, which
reflects that the dimerization equilibrium in our simulations is strongly shifted
towards the dimerized state.
The predictions concerning the bifurcation behavior only change marginally if
$\mu_{eff}$ and $f_D$ values for lower expression levels are used.
\end{enumerate}

For each position $x$ with local activator levels corresponding to the ones
in the simulations we calculated fixed point solutions for the copy number pair $(H(x), K(x))$
starting from the steady-state assumption $\partial_t (H(x), K(x)) = (0,0)$.
The stability of the fixed points was determined starting from the Jacobian for the
above ODE system:
\begin{equation}
J(H,K) = \left(
\begin{array}{cc}
\partial_H \left[\partial_t H\right] & \partial_K \left[\partial_t H\right] \\
\partial_H \left[\partial_t K\right] & \partial_K \left[\partial_t K\right] \\
\end{array}
\right) \\
\end{equation}

Within the relevant parameter regime we obtained fixed points with either two
negative eigenvalues (i.e. stable fixed points) or one positive and one
negative eigenvalue (i.e. saddle points).
The determinant therefore completely characterizes the stability of the
fixed points. If $\det J(H_0,K_0) < 0$, then $(H_0,K_0)$ is a saddle point.
Otherwise it is stable.

\Fig \ref{FigBifurcationAnalysis} shows the fixed point solutions for \Hb and \Kni as
a function of $x$ for different activator amplitudes $A$. Stable solutions
are drawn with solid, unstable solutions with dashed lines.
Depending on the $A$ value, the system displays a saddle node bifurcation at a
point towards the anterior (\Hb) or posterior (\Kni) from midembryo.
Within the region confined by the bifurcation points two stable and one unstable
fixed points exist for each gene, implying bistability.
The region clearly widens for increasing $A$ and spans almost the whole embryo
length for $A=8$.
Our deterministic analysis therefore predicts the enlargement of the region of
bistability as observed in our single nucleus simulations.

\subsection{Estimation of switching times}
To quantify the swiching times in the presence of bistability
we performed simulations of isolated single nuclei featuring
the same set of reactions and parameters as in the full scale simulation.
To obtain estimates of switching times at different positions $x$ along
the AP axis we set the levels of \Bcd and \Cad in the given nucleus equal
to the ones at $x$ in the space-resolved simulations.
The switching time was estimated by calculating from long time trajectories of
the total \Hb and \Kni copy numbers the relaxation time $t_s$ of 
the average correlation function
\begin{align}
 \Avg{C(t)}_{t_0} \equiv \frac{ \Avg{I_H(t_0) I_K(t)}_{t_0} }{ \Avg{I_H(t_0)}_{t_0} }
\end{align}
where $I_H$ ($I_K$) are indicator functions which are one if the difference in
the total gap gene copy numbers $\Delta N = H-K$ is above (below) a certain threshold
$\Theta_N$ ($-\Theta_N$). $\Theta_N$ thus defines the regions within which the switch
is considered to have switched to the \Hb--high or \Kni--high states, respectively, and
serves to separate the stable attractor states from the transition region.
We found that $\Theta_N=200$ is a reasonable choice for our set of parameters.

We determined the switching times from one long sample for different positions $x$
and different activator amplitudes $A$ and find that $t_s$ is very similar within
the double-activated bistable regions for high $A$.
To obtain an error estimate we additionally calculated block averages of estimated
switching times among 10 long samples for various $A$ at midembryo ($x=L/2$).
Table \ref{TabSwitchingTimes} shows our results from the latter procedure.

\begin{center}
\begin{table}[ht!]
  \caption{
    \bf{Switching times at midembryo for different activator levels.}
  }
  \label{TabSwitchingTimes}

\begin{center}
\begin{tabular}{|c|c|}
\hline
Activator amplitude $A$ & Switching time $t_s [s]$ \\
\hline
(1) & (6343.7  $\pm$ 17.2) \\
2 & 20302.5 $\pm$ 74.5 \\
4 & 20957.3 $\pm$ 54.9 \\
8 & 20994.7 $\pm$ 67.2 \\
\hline
\end{tabular}
\end{center}

\end{table}
\end{center}

Note that for $A=1$ the system is not truly bistable yet because
for $A=1$ we have half-activation at midembryo and due to the lack of
diffusion large promoter-state fluctuations dominate over long-time
switching potentially induced by mutual repression.
Consequently, the given number does not reflect a switching time.
We cite it here for completeness, however.

\subsection{Analysis of statistical properties of the boundary}
Our measures for both the boundary steepness and the variance of the boundary
position are based on averages over both the time and the circumference of the
embryo which were calculated during runtime.
While the double-averaging procedure limits the amount of data that must be stored
and facilitates rapid acquisition of good statistics, it also discards information
about the microscopic properties of the boundary at a given time instance.
Based on the average data it is impossible to determine whether the blurring
of the boundary quantified by $\Delta x$ is due to concerted stochastic movements
of a steep and rather homogeneous instantaneous boundary or simply due to 
stochastic fluctuations of the boundary position in each nuclear row around
a well-defined constant mean boundary position (or due to both).
In the latter case the boundary will be rough at each given time instance, 
i.e. the time average of the boundary position variance in the cicumferential direction
will be large, but the time variance of its circumferential mean will be negligible.
The opposite will be the case in the other extreme.
These quantities therefore can be used to distinguish the two hypothetical situations.
The overall boundary width in both cases is given by the sum:
\begin{align}
 \Delta x^2 &= \overline{\sigma^2_{x_t(\phi)}} + \sigma^2_{\Avg{x_t}_\phi(t)}
\end{align}
Here $\Avg{\dots}_\phi$ denotes the average over the circumference, while the bar
denotes the time average.
An identical variance decomposition can be made for the fluctuations of the
\Hb copy number at any position $x$ along the AP axis.
Similarly, comparing the average of the profile steepness for a particular nuclear row
and time instance to the steepness of the time- and circumference average of
the copy number reveals whether the steepness of the average profile is due
to concerted movements of similarly shallow instantaneous profiles or due
to unconcerted fluctuations of steep instantaneous profiles.

In order to determine which of the portrayed blurring mechanisms is dominant in our
system we performed the described variance decomposition for a set of 100 instantaneous
outputs of the fully resolved 2D system in steady state, i.e. for 6400 different total \Hb
copy number profiles along the AP axis, for both the variances at the boundary and
for the steepness at the boundary and for both the system with and without mutual repression.
We focused on our standard parameter set (see Table \ref{TabParameters}) and a range of
gap protein diffusion constants $D$.

\subsubsection{At each time instance the boundary is rather rough}
\Fig \ref{FigProfileAnalysisNoise} shows 
for the systems with (\Fig \ref{FigProfileAnalysisNoise}A and C) and without 
(\Fig \ref{FigProfileAnalysisNoise}B and D) mutual repression
the variance decomposition for the variance of the \Hb copy number at the boundary
(\Fig \ref{FigProfileAnalysisNoise}A and B) and for the variance
of the boundary position $x_t$ (\Fig \ref{FigProfileAnalysisNoise}C and D) as
a function of the \Hb protein diffusion constant $D$.
As a control we compare the total variances calculated from the instantaneous profiles to
the variances accumulated during runtime and, in case of $\Delta x$, to the value obtained 
from the approximation $\Delta x=\sigma_H(x_t)/|\Avg{H(x_t)}'|$ (note that here
$\Avg{\dots}$ is the average over both time and $\phi$).
We see a good agreement between these quantities.
The plots reveal that both for $\sigma_H(x_t)$ and $\Delta x$ the variance over
the circumference at a fixed time is by far the dominant contribution to the overall
variance.
This implies that in our system the boundary is indeed very rough at each time point
and that concerted boundary movements do not occur.

\subsubsection{At each time instance the profiles are slightly steeper than their average}
The calculation of the variance decomposition is less straightforward for the slope.
In particular for low $D$, when spatial averaging is still inefficient, the instantaneous
profiles are very ragged and the boundary threshold value typically is crossed at
multiple positions along the AP axis.
This makes it impossible to uniquely define an instantaneous
boundary position as required to calculate the instantaneous boundary slope.
In order to perform the analysis at least on a subset of the data we introduced a 
protocol which only takes into account instantaneous profiles with a single boundary
crossing, rejecting all other profiles.
For low $D$, however, the rejection rates rise above $90\%$.
We therefore decided to smoothen the profiles by computing running averages between a
fixed number $\nu$ of nuclei along the AP axis before the analysis.
The averaging lowers the rejection rate dramatically, however it also decreases the profile steepness
and therefore manipulates the observable of interest.
Nevertheless we can make a qualitative statement on the base of the results obtained
for only slight smoothening of the profiles ($\nu=3$).
For simplicity and due to increased data abundance,
in this analysis we used simple finite differences to determine the slope.

In \Fig \ref{FigProfileAnalysisSlope} we plot the average of the instantaneous boundary steepness
for different degrees of smoothening (averaging over $\nu=3,5,7$ nuclei along the $x$-axis)
as a function of $D$ and compare to the steepness of the average \Hb profile for the system with (\ref{FigProfileAnalysisSlope}A)
and without (\ref{FigProfileAnalysisSlope}B) mutual repression.
While the data for $\nu>3$ clearly must be considered biased by the running averages,
the values for $\nu=3$ show that the instantaneous boundary slope on average is higher than
the slope of the average profile, in particular for low diffusion constants.

The variance decomposition for the boundary position $x_t$ shows that the variance
of the circumference mean of the boundary position in time is very small.
This implies that the steepness of the circumference-averaged profiles should be
approximately equal to the steepness of the time- and circumference-averaged profile.
As a control we therefore repeated the above analysis on the 100 $\phi$-averaged
instantaneous profiles of the same dataset.
The averaging along the circumference significantly reduces the number of profiles
with ambiguous boundary positions.
We therefore were able to obtain reasonable estimates of the observable without 
pre-smoothening of the profiles ($\nu=1$).
The results are shown in \Fig \ref{FigProfileAnalysisSlope}C for the system with mutual repression
and \Fig \ref{FigProfileAnalysisSlope}D for the system without mutual repression.
In the system with mutual repression the average slope of the $\phi$-averaged
profiles for $\nu=1$ agrees well with the slope of the both time- and $\phi$-averaged \Hb profile.
In the system without mutual repression the $\phi$-averaged profiles are slightly
steeper than the average.

\section{Additional simulations}
\subsection{Influence of the Hill coefficient}
\label{SecHillCoeff}
To address the influence of changing activator cooperativity on our results we
performed simulations with reduced number of activator binding sites $\HC$.
While in our model this is achieved by simply reducing the number of 
intermediate states between the empty promoter state and the producing promoter
state, the binding parameters have to be rescaled with care to preserve the
activation equilibrium at midembryo.
Since we assume the activator binding rates to be diffusion limited, the
necessary changes affect the unbinding rates $k\sprm{A}_{off,n}=a/b^n$.
However, even when preserving the equilibrium, the freedom in the choice of
these parameters allows for altering the time scale of transitions between
the different activation levels.
In order to rescale the rates in a unique fashion upon lowering $\HC$ we
imposed the following constraints:

\begin{enumerate}
 \item For all $\HC$ the effective activator dissociation constant at midembryo $K\sprm{A}_D=A_{1/2}=690$ is preserved,
       which implies that for all $\HC$ the average activation probability at midembryo is $1/2$.
 \item The waiting-time distribution for the unbinding from the producing state is the same for all $\HC$
       and, for comparison, equal to the one for the default cooperativity $\HC=5$,
       i.e. $\forall n: k\sprm{A}_{off,\HC} = const = k\sprm{A}_{off,5}$.
 \item The off-rate reduction per subsequent activator binding is always $1/b$, i.e. $\forall n: b(n)=b$.
\end{enumerate}

Note that for $\HC=1$ the first two conditions can be met together only if
$K\sprm{A}_D k\sprm{A}_{on}=k\sprm{A}_{off,5}$, which is not the case for our parameter set.
We therefore restricted ourselves to $\HC\in\lbrace2,3,4,5\rbrace$.
For each $\HC$, the above constraints were used to uniquely determine the parameters $a$ and $b$ from
the exact analytical solution for the average occupancy of the producing state,
which was obtained from steady-state mean-field solutions of the chemical mass-action ODEs.
Interestingly this results in only minor differences in $a$ among the different
$\HC$ values, while the reduction per binding step $1/b_\HC$ becomes significantly larger
for lower $\HC$.
This fact has an important implication for the noise charactistics of the
different promoters: If $a_\HC \simeq a=const$ for all $\HC$ then the unbinding
rate from the state binding $(\HC-1)$ activator proteins (the ``highest'' non-producing
activator state) is given by:
\begin{align*}
 k\sprm{A}_{off,(\HC-1)} &\simeq a/b_\HC^{(\HC-1)} = b_\HC k\sprm{A}_{off,5}
\end{align*}
Since $b_\HC$ markedly increases with decreasing Hill coefficient
the off-rate $k\sprm{A}_{off,(\HC-1)}$ for low $\HC$ will be higher than the
corresponding rate for high $\HC$.
This will favor rapid returns to the producing state with $\HC$ bound activator
molecules for high Hill coefficients, whereas for low Hill coefficients the promoter
is more likely to descent into the regime with less activator molecules bound.
The fact that this is less likely for higher Hill coefficients is compensated
by the fact that also the time to return to the producing state from the states
binding low numbers of activator molecules on average is longer for higher $\HC$.
Note that the mean off-time--just as the mean on-time--is the same for all $\HC$.
In short, for the promoters with higher Hill coefficients we expect an off-time
distribution with high probability weight on short off-times and a long low-probability
tail for long off-times, while the distribution for lower Hill coefficients should
resemble an exponential.

In order to illustrate this effect we recorded long time-trajectories of the occupancy
of the producing state in a single isolated nucleus close to midembryo for 
different $\HC$ and without mutual repression nor diffusion.
All other parameters were kept at the standard values.
From these trajectories we determined the on- and off-times of the promoter
and binned them into a histogram. The results are shown in \Fig \ref{FigHillCoeffDistrib}.
It can be seen that while for $\HC=2$ the two distributions are exponential with 
approximately equal mean, the off-times distribution increasingly deviates from
an exponential distribution as $\HC$ is increased; more probability is shifted
to very short off-times and very long off-times, causing the emergence of a long tail
in the distribution.

\subsubsection{Also for lower Hill coefficients mutual repression steepens profiles without corrupting boundary precision}
The broadening of the off-times distribution is expected to result in higher output
noise for high $\HC$ as compared to low $\HC$. This is confirmed by the simulations
of the full-scale spatially resolved system for different $\HC$.
\Fig \ref{FigHillCoeffBoundary} shows $\sigma_H(x_t)$, the average standard deviation of the total \Hb copy number
at the boundary position $x_t$ (upper panels), the steepness $|\Avg{H(x_t)}'|$ of the average 
\Hb profile at $x_t$ (middle panels) and the boundary width $\Delta x$ (lower panels)
as a function of the gap protein diffusion constant $D$ for $\HC\in\lbrace2,3,4,5\rbrace$.
$\sigma_H(x_t)$ is indeed decreasing upon lowering $\HC$, in particular in the regime of low diffusion
constants. For higher diffusion constants the decrease is less pronounced: spatial averaging is
efficient enough to lower the output noise down to the observed values irrespective of the width of
the off-time distribution.
The noise decreases less markedly for the systems with mutual repression. This is most likely
due to the fact that lowering $\HC$ also increases the probability of occasional repressor
production beyond midembryo, which in turn increases the noise.
The steepness plots reveal that, although the profiles naturally become less steep for lower $\HC$,
the steepness in the systems with mutual repression is markedly higher than the one
in the system without mutual repression.
In all systems the steepness as a function of $D$ shows a very similar behavior: Upon increasing
$D$ the steepness in the systems with mutual repression first increases towards a maximum before
it rapidly decreases.
Since both $\sigma_H(x_t)$ and $|\Avg{H(x_t)}'|$ change with $\HC$ in a similar fashion,
in particular in the region around $D=1~\umsps$, the width $\Delta x$ as a function of $D$
also looks very similar in this region for all $\HC$.
In all cases the profiles in the system with mutual repression are more precise and markedly
steeper as compared to the system without mutual repression at a $D$-value which is one 
order of magnitude less than the optimal value in the systems without mutual repression.
Therefore the basic effect observed in our simulations for $\HC=5$ persists in the
simulations for lower Hill coefficients.

\subsubsection{Lower Hill coefficients allow for stronger morphogen level variations}
Although lowering $\HC$ in our system reduces the protein production noise it
also markedly decreases the steepness of the gene activation profiles.
An important implication of this is that for lower $\HC$ the activation probability
beyond midembryo increases.
Lowering $\HC$ thus is similar to increasing the activator amplitude $A$ and,
in principle, might result in the creation of a bistable region around midembryo
already for lower $A$-values as compared to the system with $\HC=5$.
We analysed how the results for $\Delta x$ as a function of
$A$ for the case of correlated variations change as $\HC$ is decreased.
\Fig \ref{FigHillCoeffVarA} shows $\Delta x(A)$ for $\HC\in\lbrace2,3,4,5\rbrace$
and $D=1.0~\umsps$ for systems with and without mutual repression.
Overall, $\Delta x(A)$ is very similar for all considered $\HC$.
For $A\leq2$ the width $\Delta x$ in the systems with mutual repression
is always lower than in the systems without mutual repression.\
The minimal $\Delta x$ is attained at $A=1$ in all cases.
The main difference is in how $\Delta x$ changes with $A$ for $A>1$:
The lower $\HC$, the slower the width increases with $A$.
Thus, while lower Hill coefficients decrease the steepness of the profiles significantly,
they may prove beneficial by extending the range over which extrinsic
variations are successfully buffered without increasing intrinsic fluctuations of the boundary.

\subsection{Influence of the expression level}
In order to examine the influence of a changed signal-to-noise ratio on our
results we performed simulations with altered production dynamics.
We did this by
(1) introducing bursty production, i.e. producing 10 copies of the gap protein
monomer at a time with a 10 times lower production rate ($\beta=\beta_0/10$), and
(2) by keeping the burst size at one and changing the production rate $\beta$.
To preserve the binding equilibrium of the repression reaction at midembryo
upon changing $\beta$ we also changed the off-rate of the repressor dimers by a factor
$f_\beta\sprm{D}$, which is the ratio between the expected number of dimers at midembryo for the
altered production rate $\beta$ and the corresponding value for the standard production rate $\beta_0$.
Note that, since in our system the copy numbers of both monomers and dimers depend on $\beta$
in a nontrivial fashion (see section \ref{SecPredictedCopyNumbers}) the effective copy number
increase typically does not correspond to the ratio $\beta/\beta_0$.
Therefore $f_\beta\sprm{D}>\beta/\beta_0$ for $\beta>\beta_0$.

\subsubsection{Bursty production has only a marginal influence on the boundary properties}
In \Fig \ref{FigBurstSize} we plot the standard deviation of the total \Hb copy number
at the boundary, the steepness of the total \Hb copy number profile at the boundary and the
boundary width $\Delta x$ as a function of $D$ for the system with bursty production (burst size 10).
There is no significant difference as compared to the system with normal production (burst size 1,
compare to \Fig \ref{FigHillCoeffBoundary}(D) or \Fig 3 of main article).
For low $D$ the production noise is marginally higher with bursty production, resulting in
a slight increase of $\Delta x$ in this regime; the effect of varying $D$, however, is much
more important.
This is most likely a consequence of the fact that for the given Hill coefficient $\HC=5$ promoter
state fluctuations are already at a high level due to a very broad off-time distribution (see section \ref{SecHillCoeff}).

\subsubsection{Increased production rates reveal different noise scaling behavior for different regimes of the diffusion constant}
Upon increasing the production rate and consequently the total copy number of the gap proteins we
may expect a relative decrease in the output noise, but only if the latter is purely Poissonian.
In our system this corresponds to the limit of high gap protein diffusion constants.
In that limit, we expect $\sigma_G \propto \sqrt{G}$, where $\sigma_G$ is the noise in the
total gap gene copy number $G$.
However, in the abscence of spatial averaging, i.e. for the limit $D\rightarrow0$,
non-Poissonian noise prevails and the expected scaling is $\sigma_G \propto G$ \cite{Erdmann2009}.
If the copy number profile is scaled uniformly at each AP position $x$, which--to a good approximation--is
the case in our system, we expect for the scaling of the gradient at midembryo $G^\prime(x_t)\propto G(x_t)$.
The expected scaling for the boundary width $\Delta x$ then is $\Delta x\propto 1$ for low diffusion
constants and $\Delta x\propto 1/\sqrt{G}$ for high $D$.
While the overall characteristics of the boundary are very similar to the system with $\beta=\beta_0$,
a comparison roughly confirms the predicted scaling.
\Fig \ref{FigProdRate} compares for \Hb the standard deviation of the total copy number at the boundary (\ref{FigProdRate}A), 
the steepness at the boundary (\ref{FigProdRate}B) and the resulting boundary width
(\ref{FigProdRate}C) as a function of $D$ for increased production rates $\beta=2\beta_0$ and $\beta=4\beta_0$
to the corresponding values for the sytem with production rate $\beta/2$.
Thus, the values for $\beta=4\beta_0$ are compared to $\beta=2\beta_0$ and the values
for $\beta=2\beta_0$ are compared to $\beta_0$.
Blue lines mark the expected change of the quantities as predicted
by the scaling relations, where the corresponding copy number increase is
given by the factor $f_2 \equiv f_{2\beta_0}=2.22$ and $f_4\equiv f_{4\beta_0}/f_{2\beta_0}=2.16$.
Here $f_\beta\simeq f_\beta\sprm{D}$ is the predicted \textit{total} copy number at midembryo divided by
the corresponding value for $\beta=\beta_0$.

The plots show that while the slope ratio is roughly equal to $f_2$ ($f_4$) for all $D$
both in the system with (green) and without (red) mutual repression, the noise ratio
depends on the diffusion constant and also slightly differs for the systems with and
without mutual repression.
Nevertheless the predicted scaling behavior is confirmed in both cases:
in the low diffusion constant regime the noise ratio is roughly $f_2$ ($f_4$)
and approaches $\sqrt{f_2}$ ($\sqrt{f_4}$) as $D$ increases;
together this leads to a boundary width ratio of one for low $D$ which decreases
towards $1/\sqrt{f_2}$ ($1/\sqrt{f_4}$) for higher $D$.

\subsection{Activation of both gap genes by a single gradient}
In the one-morphogen gradient scenario, both \hb and \kni are
activated by the \Bcd gradient. Here, \kni is activated in the same
way as \hb, namely by 5-step cooperative binding, but with a lower
activation threshold. This results in the induction of both genes in
the anterior half of the embryo up to the posterior \Hb boundary and
of \kni in an additional region posterior to the \Hb boundary.  Given
that \hb represses \kni more strongly than vice versa in the
double-activated bistable region, this parameter choice will result in
the formation of two neighboring domains.  We chose the \kni
activation threshold to be lower by a factor of $1/2$, which causes an
offset of its half-activation point by approximately 10 nuclei
($83~\unit{\mu m}$) towards the posterior.  We varied the protein
diffusion coefficient $D$ and $k_{\rm off}^{\rm R,K}$, the off-rate of
the \Kni repressor dimers from the \hb promoter.  The rate for
dissociation of the \Hb dimers from the \kni promoter was kept at the
standard value $k_{off}^{\rm R,H}=5.27\cdot10^{-3}\unit{/s}$ in all
simulations.

The diffusion constant $D$ of the gap proteins and the dissociation
rate $k_{off}^{\rm R,K}$ of \Kni from the \hb promoter are indeed key
parameters. On the one hand, \hb must repress \kni more strongly than
the other way around, because otherwise there will be only one \kni
domain. On the other hand, when $k_{off}^{\rm R,K}$ is high, then \kni
is only significantly expressed when $D$ is low, because \kni
represses \hb more weakly than vice versa, which means that low amounts
of invading \Hb dimers are sufficient to shut off \Kni production
almost completely; indeed, in this regime, \kni has
hardly any effect on the precision of the \hb expression domain. We
found that when $k_{off}^{\rm R,H}/k_{off}^{\rm R,K}$ is roughly
between 0.1 and 1, both \hb and \kni domains are formed
robustly. In \Fig \ref{FigOneGradient} we display the case for $k_{off}^{\rm
  R,H}/k_{off}^{\rm R,K}=1/\sqrt{10}\approx1/3$.

\Fig \ref{FigOneGradient}A shows that the maximum of the average \Kni copy
number is lower than that of \Hb, even though for $x<60~{\rm \%EL}$ \kni
is essentially fully activated by \Bcd (\Fig \ref{FigOneGradient}B).
The lower maximum is due to
the fact that \hb represses \kni more strongly than vice
versa. Another point worthy of note is that  the fluctuations in the \Kni copy
number in the \Kni domain are higher than those of \Hb in the \Hb
domain (\Fig \ref{FigOneGradient}C). This is essentially due to the small width
of the \Kni domain: \kni is either fully activated by \Bcd yet still
repressed by \Hb or not repressed by \Hb yet stochastically activated
by \Bcd. 

Panels D-F show, respectively, the noise in the \Hb copy number at the
\hb expression boundary, the steepness of this boundary, and the width of
this boundary, as a function of the diffusion constant $D$ of the gap
proteins. It is seen that the results are highly similar to those of
the two-gradient motif. The noise in the \Hb copy number at the
boundary is not much affected by mutual repression (panel D), while
the steepness, and consequently boundary precision, is markedly enhanced by
mutual repression, especially when the diffusion constant is
small. Note that while for the two-morphogen gradient scenario the
approximation $\Delta x\approx\sigma_H(x_t)/|\Avg{H(x_t)}^\prime|$ is
in very reasonable agreement with $\Delta x$ as measured from the
distribution of threshold crossings $p(x)$, here the agreement is much
less.  This is due to sporadic repression events in the anterior
region where \hb and \kni are both fully activated, which leads to a
long tail of $p(x)$ extending towards the anterior pole; while $p(x)$
in the tail is small, the fact that the tail is long does markedly
increase the standard deviation $\Delta x$.
Given that the
approximation $\Delta x\approx\sigma_H(x_t)/|\Avg{H(x_t)}^\prime|$
works so well for all the other cases, we consider this approximation,
which does not suffer from sporadic but strong \hb repression events in the
anterior, to be more reliable. We therefore conclude that also in the
one-morphogen gradient scenario, mutual repression can enhance both
the steepness and the precision of gene-expression boundaries.


\renewcommand{\thetable}{S\arabic{table}}
\begin{landscape}
\begin{table}[!ht]
  \caption{
    \bf{The standard set of the most important parameters in our
      model.}
  }
  \label{TabParameters}
\begin{tabular}{|l|c|cc|c|}
\hline
Parameter / Quantity				& Symbol	& Value	& Unit		& Remarks\\
\hline

\textbf{Geometry}				&&&&\\
Number of nuclei in axial direction $x$		& $N_x$		& 64	&		&	\\
Number of nuclei in circumferential direction $\phi$	& $N_\phi$	& 64	&		&	\\
Internuclear distance (lattice spacing)		& $l$		& 8.5	& $\unit{\mu m}$& supported by \cite{Gregor2007}\\
- resulting embryo length			& $L$		& 544	& $\unit{\mu m}$&	\\
Nuclear radius					& $R$		& 6.5	& $\unit{\mu m}$& supported by \cite{Gregor2007b}\\
- resulting nuclear volume			& $V$		& 143.8	& $\unit{\mu m^3}$& assuming spherical shape \\
\hline

\textbf{Morphogen / activator gradients}	&&&&\\
Standard copy number at the poles		& $A_0$		& 6720	&		& varied in some simulations \\
- resulting copy number at half activation	& $A_{\frac{1}{2}}$ & 690 &		& supported by \cite{Gregor2007}	\\
Morphogen diffusion length			& $\lambda$	& 119.5	& $\unit{\mu m}$& supported by \cite{Gregor2007}	\\
\hline

\textbf{Activation dynamics}			&&&&\\
Intranuclear activator diffusion constant	& $D_{A}$	& 3.2	& $\umsps$	& supported by \cite{Abu-Arish2010}	\\
Activator binding site target size		& $\alpha$	& 10	& $\unit{nm}$		&	\\
- resulting activator binding rate		& $k\sprm{A}_{on}$	& 0.40 	& $\unit{\mu m^3/s}$	& diffusion limited, $k\sprm{A}_{on} =4\pi\alpha D_A$\\
Activator unbinding rate 			& $k\sprm{A}_{off,n}$ & $410/6^n$ & $\unit{1/s}$	& $n$ = no. of bound activator molecules	\\
Hill coefficient of activation 			& n_{max} 	& 5  	& 		& supported by \cite{Gregor2007}	\\
\hline

\textbf{Gap gene dynamics}			&&&&\\
Gap protein monomer production rate		& $\beta_0$	& 3.37			& $\unit{1/s}$	&	\\
Gap protein monomer degradation rate		& $\mu_M$	& 3.37$\cdot 10^{-2}$	& $\unit{1/s}$	&	\\
Gap protein dimer degradation rate		& $\mu_D$	& 3.37$\cdot 10^{-3}$	& $\unit{1/s}$	& 	\\
Dimerization forward rate			& $k\sprm{D}_{on}$	& 0.80			& $\unit{\mu m^3/s}$  & same as $2 k\sprm{A}_{on}$\\
Dimerization backward rate			& $k\sprm{D}_{off}$	& 5.59$\cdot 10^{-3}$	& $\unit{1/s}$	&	\\
- predicted total \Hb (\Kni) copy number 	&		&			&		&	\\
\hspace{0.14cm} at full activation and $D=0$ 	& $H_{\beta_0}$ ($K_{\beta_0}$)& 775	&	&	\\
Gap protein diffusion constant			& $D$		& varied		&	&	\\
\hline

\textbf{Repression dynamics}			&&&&\\
Repressing dimer binding rate			& $k\sprm{R}_{on}$	& 0.40 			& $\unit{\mu m^3/s}$	& same as $k\sprm{A}_{on}$ \\
Repressing dimer unbinding rate			& $k\sprm{R}_{off}$	& 5.27$\cdot 10^{-3}$ 	& $\unit{1/s}$	& equal to $0.1\cdot k\sprm{-}_{A,5}$; varied in some simulations \\

\hline
\end{tabular}
\end{table}
\end{landscape}

\pdfbookmark[1]{SUPPORTING FIGURES}{BookmarkSupportingFigures}
\begin{figure}[!ht]
\stepcounter{myc}
\begin{center}
\includegraphics[width=6in]{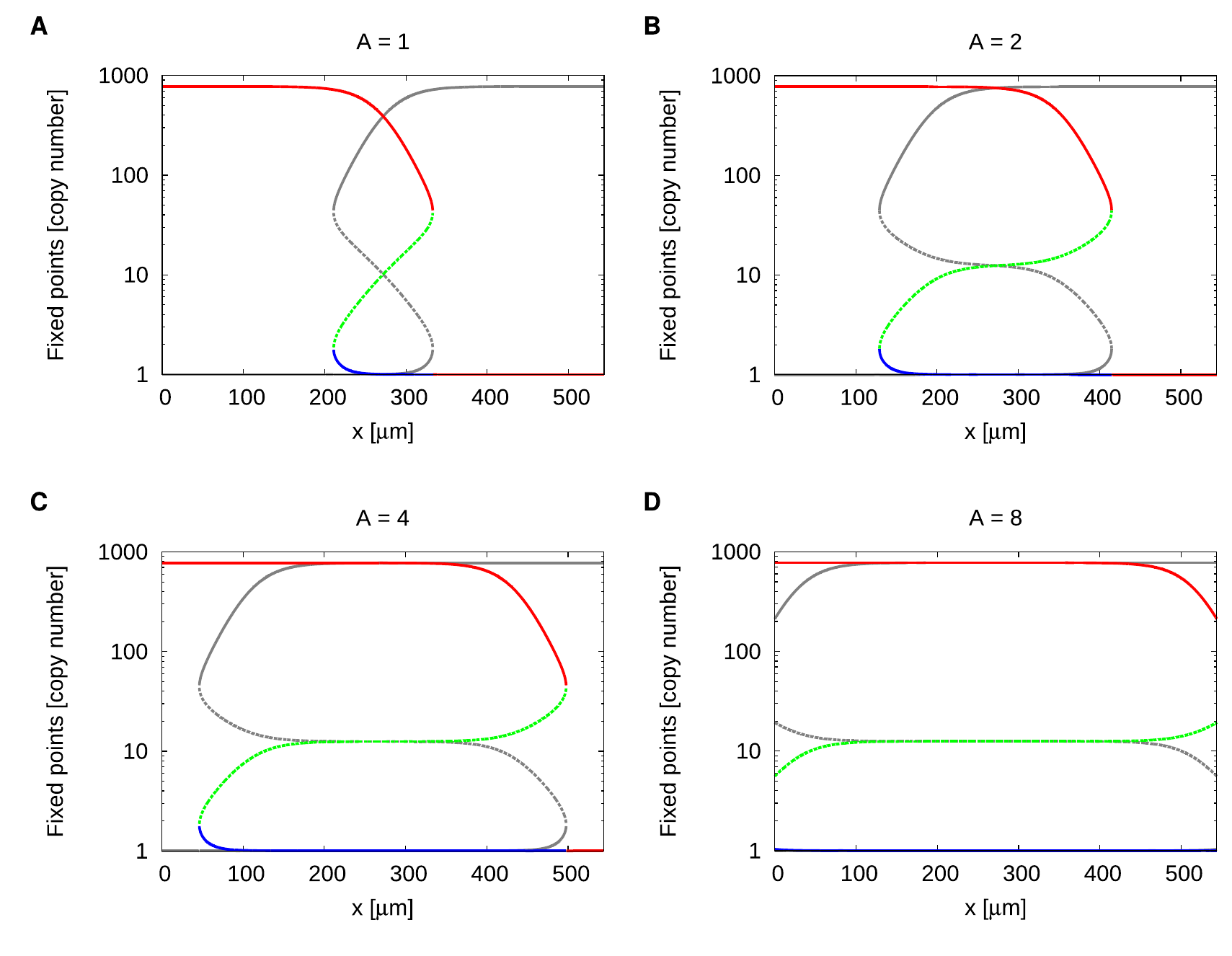}
\end{center}
\caption{ {\bf Bifurcation analysis of a one dimensional mean-field model of
  mutually repressing gap genes activated by morphogen gradients.}
  Plotted are the stable (solid lines) and unstable (dashed lines) fixed points
  of the copy number of \Hb (colored lines) and \Kni (grey lines) as a function
  of the AP position $x$ as predicted by a bifurcation analysis performed on a
  1D mean-field model in which \hb and \kni are activated cooperatively by
  their respective morphogens and mutually repressing each other.
  Different colors correspond to different fixed points.
  The different panels show the solutions for activator amplitudes
  \subfig{A} A=1, \subfig{B} A=2, \subfig{C} A=4 and \subfig{D} A=8.
  All other parameters values are the standard values from Table \ref{TabParameters}.
  Activator concentrations at position $x$ used in the mean-field analysis
  correspond to the ones in the 2D stochastic simulations.   
  Away from midembryo each gap protein level has only one stable fixed point
  and one of the two levels is always zero.
  For all $A$ there is a region around midembryo in which the protein
  levels have two stable and one unstable fixed point, implying bistability.
  In this region the analysis predicts bistable switching between the
  high-\Hb--low-\Kni and the low-\Hb--high-\Kni state.
  For clarity we color-code the \Hb fixed points only.
  The \Kni solutions are identical to the \Hb solutions mirrored with
  respect to midembryo.
  }
\label{FigBifurcationAnalysis}
\end{figure}

\begin{figure}[!ht]
\stepcounter{myc}
\begin{center}
\includegraphics[width=6in]{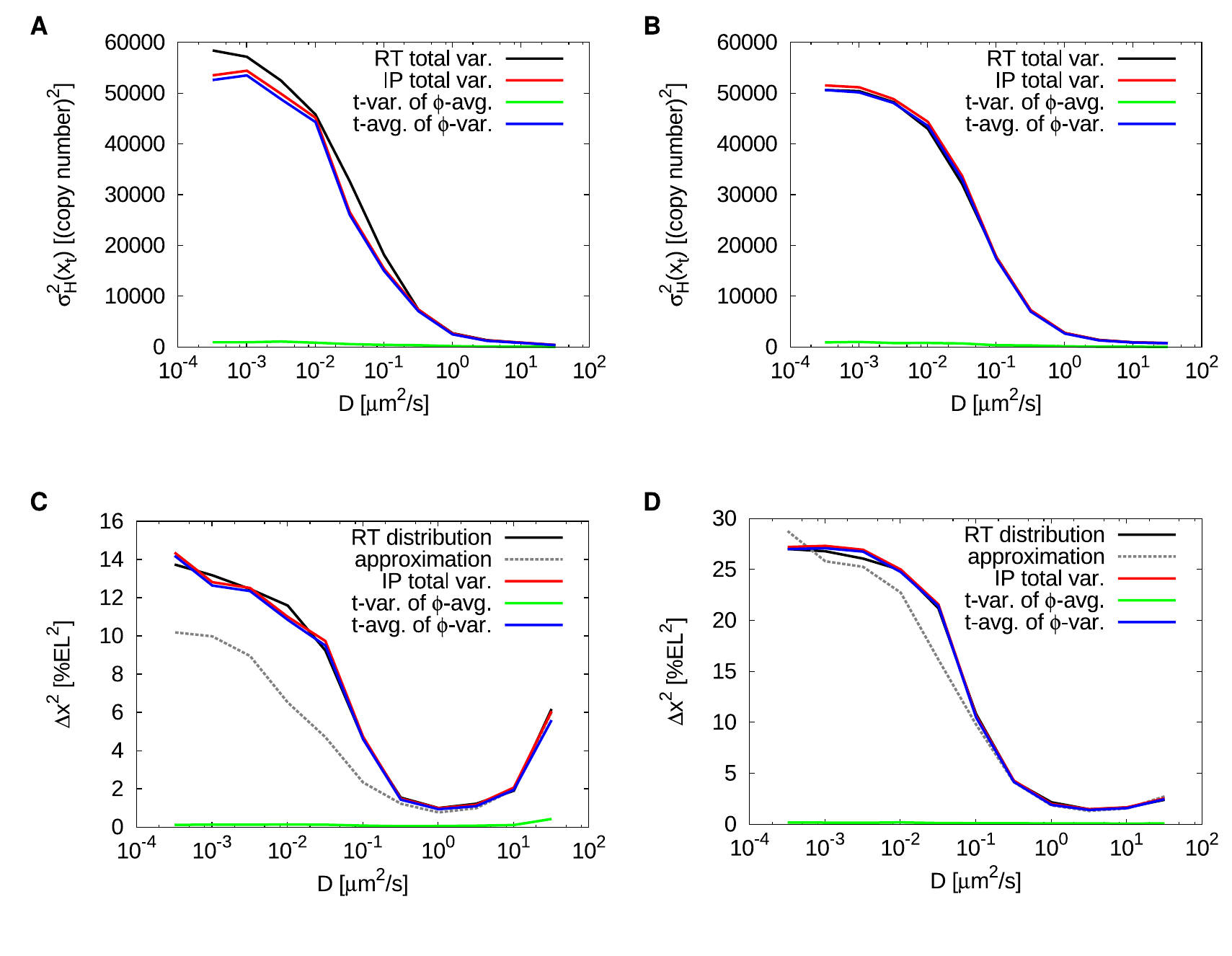}
\end{center}
\caption{ {\bf Decomposition of variances at the boundary.}
  \subfig{A} Decomposition of the total \Hb copy number variance at the average boundary position
  for the system with mutual repression as a function of the gap protein diffusion constant $D$.
  Plotted are: $\sigma^2_H(x_t)$ the total (time- and circumference-) variance measured during runtime (RT, black),
  the same quantitity determined from a set of 6400 instantaneous profiles (IP, red, 64 AP rows at 100
  different time points),
  $\sigma^2_{\Avg{H}_\phi}$ the variance in time of the circumference average of $H(x_t,\phi)$ (green) and
  $\overline{\sigma^2_{H}}$ the time average of the variance of $H(x_t,\phi)$ over the circumference (blue).
  \subfig{B} The same as (A) for the system without mutual repression.
  \subfig{C} The same variance decomposition as in (A) for the \Hb boundary position $x_t$ instead of the copy number.
  The black line shows the $\Delta x$ values measured as the standard deviation of the boundary position
  histogram accumulated during runtime (RT), the grey dashed line the corresponding values determined from
  the approximation $\sigma_H(x_t)/|\Avg{H(x_t)}'|$.  
  \subfig{D} The same as in (C) for the system without mutual repression.  
  In both cases, the main contribution to the total boundary variance $\sigma^2_{x_t}$ comes from $\overline{\sigma^2_{x_t}}$, implying that the
  blurring of the boundary is rather due to roughness than due to concerted boundary movements.
  }
\label{FigProfileAnalysisNoise}
\end{figure}

\begin{figure}[!ht]
\stepcounter{myc}
\begin{center}
\includegraphics[width=6in]{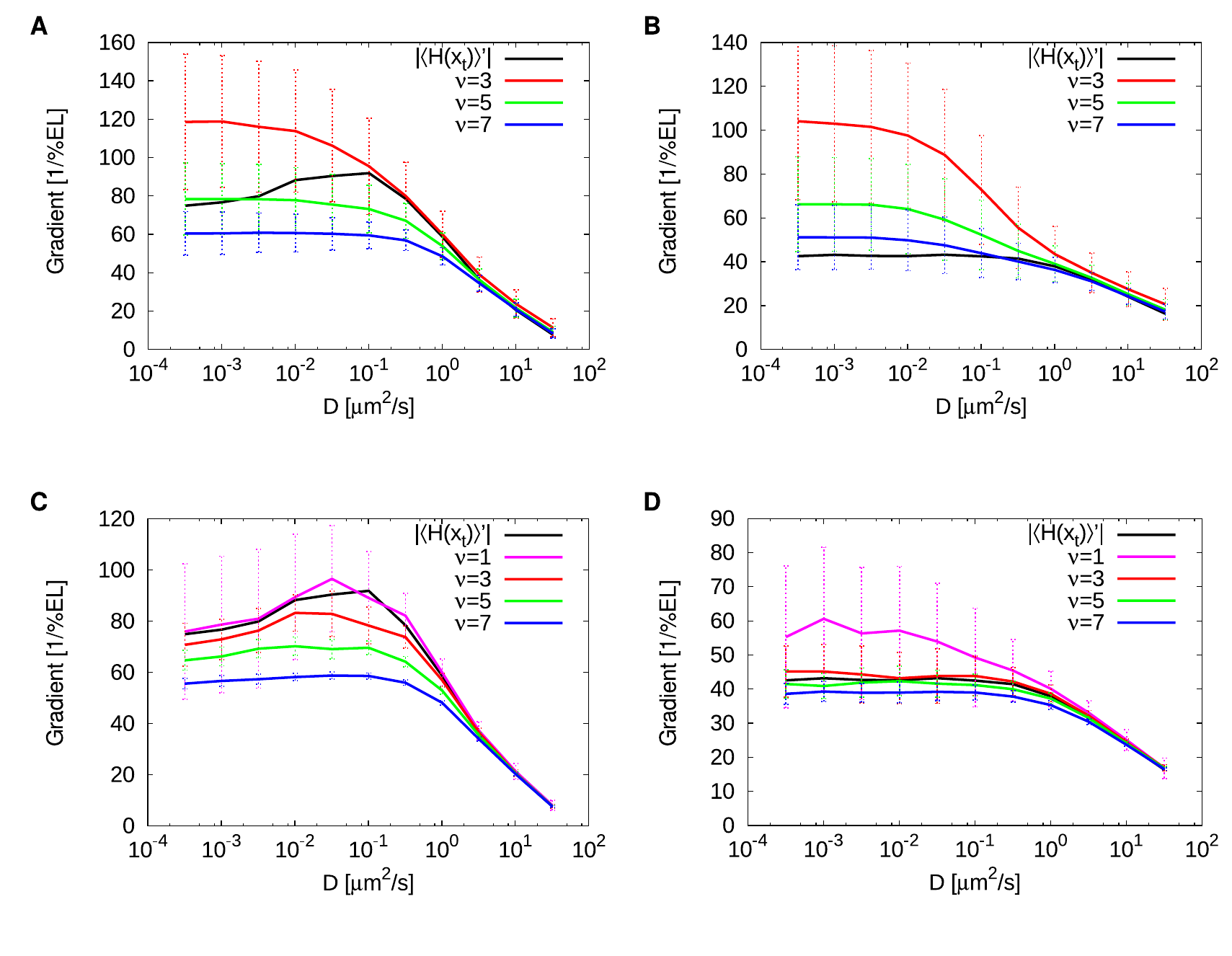}
\end{center}
\caption{ {\bf Microscopic properties of the boundary steepness.}
  Panels \subfig{A} and \subfig{B} compare for different gap protein diffusion constants $D$
  the average \Hb profile steepness at the boundary
  measured in a set of 6400 instantaneous profiles (64 AP rows at 100 different time points)
  to the steepness of the (time- and circumference-) average of the \Hb profile ($|\Avg{H(x_t)}'|$, black)
  for different numbers $\nu$ of neighboring data points used in calculating running averages over
  the instantaneous profiles for the system with (A) and without (B) mutual repression.
  Although for increasing $\nu$ the instantaneous profiles become less steep as a consequence of the smoothening,
  the values for $\nu=3$ indicate that the profiles at a given row and time instance are slightly
  steeper than the average profile.
  In panels \subfig{C} and \subfig{D} we show results of the same analysis performed
  on the 100 circumference-averages of the instantaneous profiles, again for the system with (C)
  and without (D) mutual repression. Here $\nu=1$ is the data obtained without calculating
  running averages (magenta).
  In both systems the steepness of the $\phi$-averaged profiles agrees reasonably well
  with the steepness of the average profile $|\Avg{H(x_t)}'|$.
  }
\label{FigProfileAnalysisSlope}
\end{figure}

\begin{figure}[!ht]
\stepcounter{myc}
\begin{center}
\includegraphics[width=5.2in]{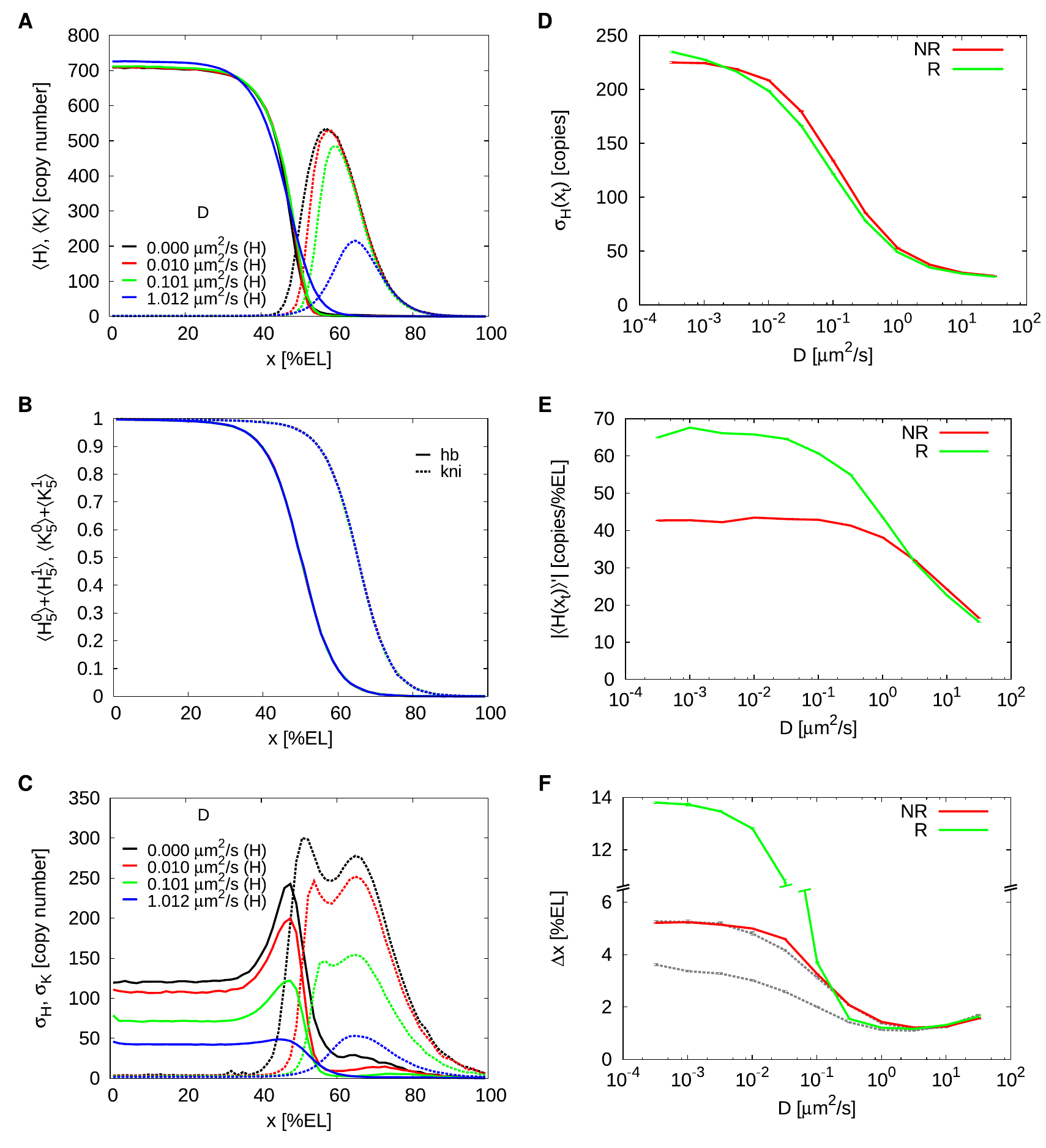}
\end{center}
\caption{{\bf The effect of mutual repression in a system where both
    \hb and \kni are activated by the \Bcd gradient.} See following page for description.}
\addtocounter{figure}{-1}
\label{FigOneGradient}
\end{figure}

\newpage
\FloatBarrier
\captionof{figure}{ {\bf The effect of mutual repression in a system where both
    \hb and \kni are activated by the \Bcd gradient.} \subfig{A} Time-
  and circumference-averaged \Hb ($\Avg{H}$, solid lines) and \Kni
  ($\Avg{K}$, dashed lines) total copy-number profiles along the
  AP-axis for various \Hb and \Kni diffusion constants $D$.
  \subfig{B} AP profiles of the average standard deviation of the total gap
  gene copy number for \Hb ($\sigma_H$, solid lines) and \Kni
  ($\sigma_K$, dashed lines). Note that the noise in $K$ in the \Kni
  domain is larger than that in $H$ in the \Hb domain.  \subfig{C} AP
  profiles of the probabilities $\Avg{H_5^0}+\Avg{H_5^1}$ and
  $\Avg{K_5^0}+\Avg{K_5^1}$ that the \hb (solid lines) and \kni
  (dashed lines) promoters have 5 Bcd molecules bound to them,
  respectively; in the absence of mutual repression between \hb and
  \kni, these profiles would directly determine the expression of \hb
  and \kni.  \subfig{D} The noise in the \Hb copy number at the \hb
  expression boundary as a function of the \Hb and \Kni diffusion
  constant $D$. \subfig{E} The steepness of the \hb expression
  boundary as a function of the diffusion constant of the gap
  proteins. \subfig{F} The width $\Delta x$ of the \hb expression
  boundary as a function of the \Hb and \Kni diffusion constant. The
  grey line corresponds to the approximation $\Delta
  x\approx\sigma_H(x_t)/|\Avg{H(x_t)}^\prime|$, which we consider to
  be more reliable than $\Delta x$ as measured from the distribution
  of threshold crossings, $p(x)$; the latter suffers from sporadic but
  strong suppression events of \hb by \kni in the anterior, which leads to a
  long tail of $p(x)$, increasing $\Delta x$. It is seen that while
  mutual repression has hardly any effect on the noise in the copy
  number at the boundary, it does markedly enhance the steepness of
  the boundary, and thereby its precision.  The ratio of the
  \Hb--\kni-promoter dissociation rate over the \Kni--\hb-promoter
  dissociation rate is $k_{off}^{\rm R,H}/k_{off}^{\rm R,K}=1/3$.}
\FloatBarrier
\newpage

\begin{figure}[!ht]
\stepcounter{myc}
\begin{center}
\includegraphics[width=6in]{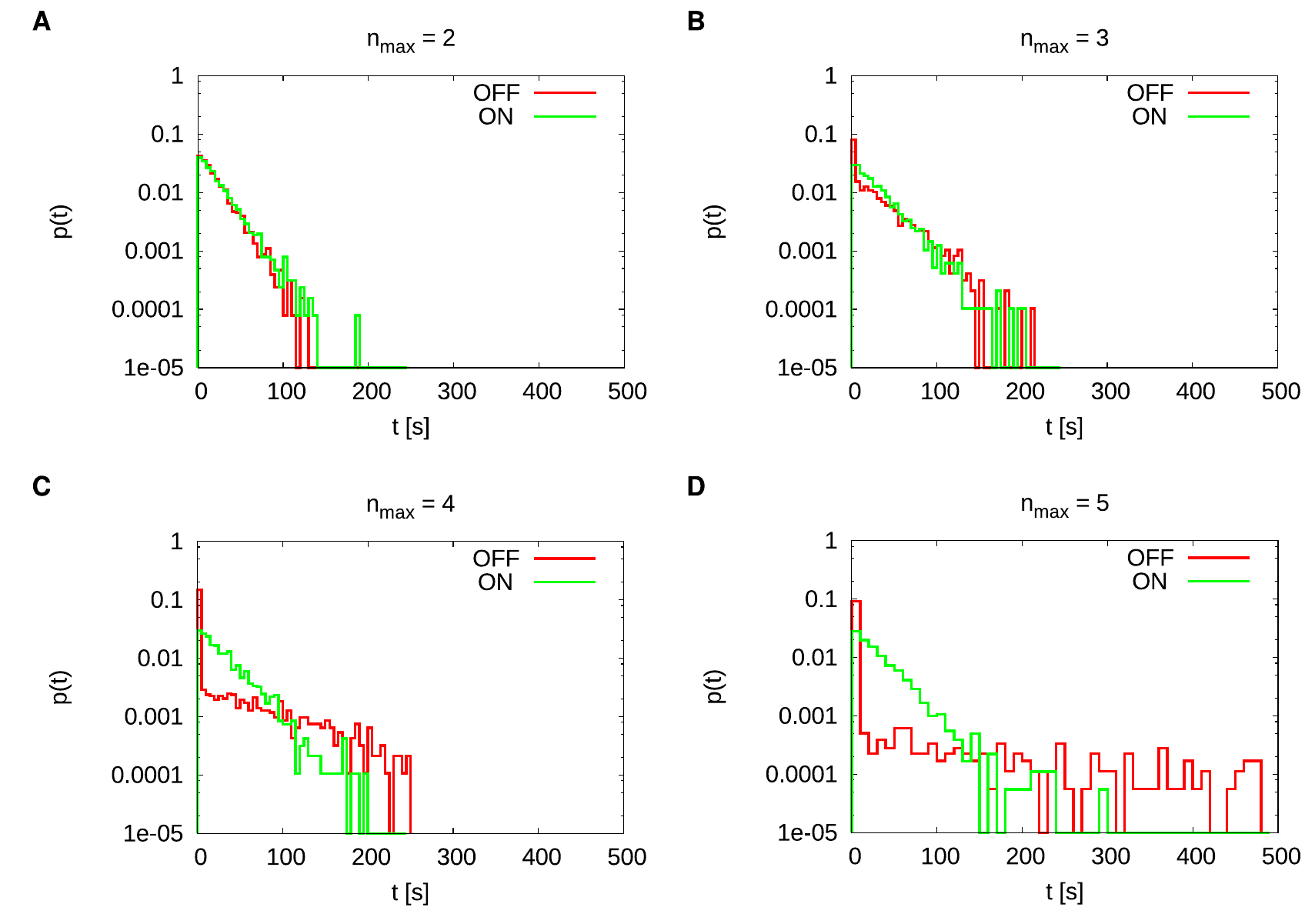}
\end{center}
\caption{ {\bf On- and off-times distributions of the \hb promoter
  for different Hill coefficients $\HC$ in a nucleus at midembryo.}
  The panels show normalized histograms of the times spent in the
  producing ($n=\HC$) promoter state (``ON'', green) and of the times
  spent in the non-producing ($n<\HC$) states (``OFF'', red) for
  \subfig{A} $\HC=2$,
  \subfig{B} $\HC=3$,
  \subfig{C} $\HC=4$ and
  \subfig{D} $\HC=5$ (standard case).
  It can be seen that with increasing Hill coefficient $\HC$
  the off-times distribution changes from an exponential to
  a non-exponential distribution with high weight on very short
  off-times (implying fast returns to the producing state) and
  a with a long tail of long off-times.
  Since the off-rate from the producing state is kept the same
  for all $\HC$ the on-times distributions remain unaltered.
  The on- and off-times have been determined from long
  time trajectories ($t_{total}=10^5~s$) of the occupancy of
  the producing state with a sampling resolution of $0.5~s$.
  }
\label{FigHillCoeffDistrib}
\end{figure}

\begin{figure}[!ht]
\stepcounter{myc}
\begin{center}
\includegraphics[width=6in]{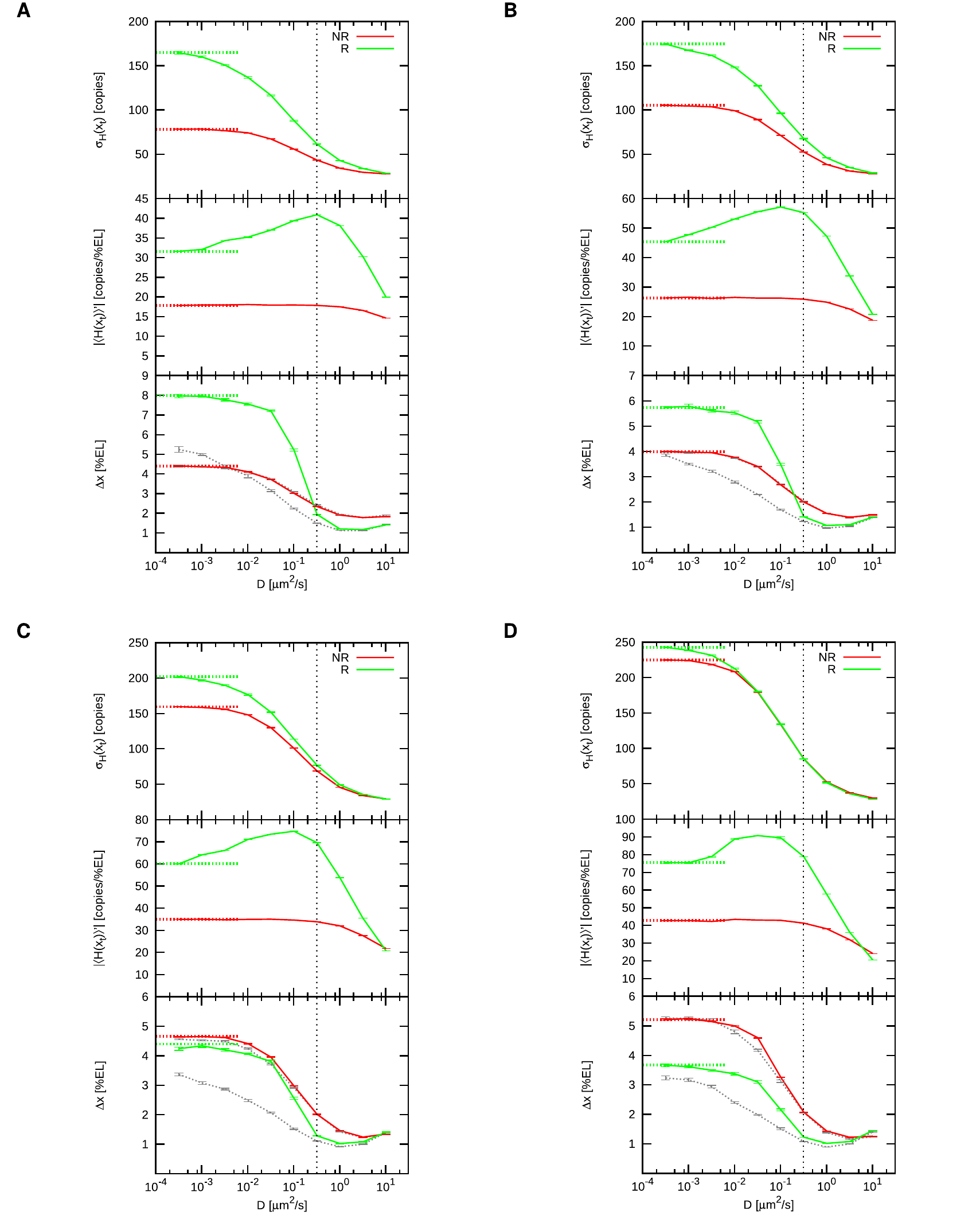}
\end{center}
\caption{
{\bf Boundary characteristics for reduced Hill coefficients $\HC$.}
See following page for description.
}
\addtocounter{figure}{-1}
\label{FigHillCoeffBoundary}
\end{figure}

\newpage
\FloatBarrier
\captionof{figure}{ {\bf Boundary characteristics for reduced Hill coefficients $\HC$.}  
  The standard deviation of the total \Hb copy number at the boundary ($\sigma_H(x_t)$, upper panels),
  the gradient of the average \Hb total copy number gradient at the boundary ($|\Avg{H(x_t)}^\prime|$, middle panels)
  and the boundary width ($\Delta x$, lower panels) as a function of the
  gap protein diffusion constant $D$ for the systems with (green) and without (red) mutual repression
  and Hill coefficients
  \subfig{A} $\HC=2$,
  \subfig{B} $\HC=3$,
  \subfig{C} $\HC=4$ and
  \subfig{D} $\HC=5$ (standard case).
  Grey dashed lines are values determined from the approximation
  $\Delta x = \sigma_H(x_t)/|\Avg{H(x_t)}^\prime|$, solid lines are
  values calculated from the distributions of $x_t$.
  Broad dashed lines are the values for $D=0$.
  Black dotted lines mark the $D$-value
  where the boundaries are both steep and precise due to mutual repression.
  }
\FloatBarrier
\newpage

\begin{figure}[!ht]
\stepcounter{myc}
\begin{center}
\includegraphics[width=6in]{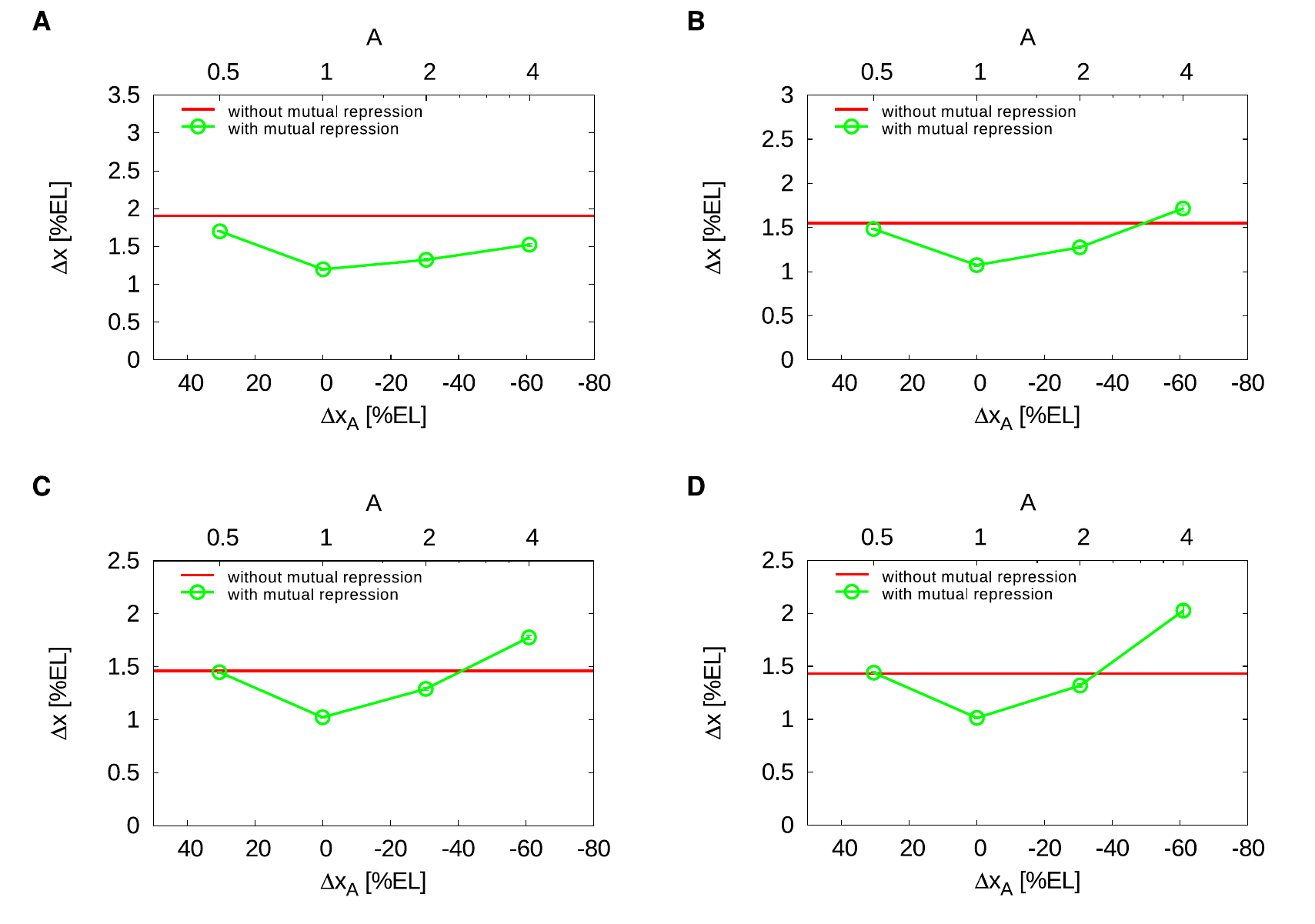}
\end{center}
\caption{ {\bf The effect of changing the activator amplitude $A$ on the
  boundary precision for reduced Hill coefficients $\HC$.}
  Shown are the the boundary width $\Delta x$ with (green) and without
  (red) mutual repression as a function of $\Delta x_A$, the separation between the
  \Hb and \Kni boundaries expected in the system without mutual repression,
  and the corresponding activator amplitude $A$ for Hill coefficients
  \subfig{A} $\HC=2$,
  \subfig{B} $\HC=3$,
  \subfig{C} $\HC=4$ and
  \subfig{D} $\HC=5$ (standard case).
  In all cases $D=1.0~\umsps$.
  }
\label{FigHillCoeffVarA}
\end{figure}

\begin{figure}[!ht]
\stepcounter{myc}
\begin{center}
\includegraphics[width=3.27in]{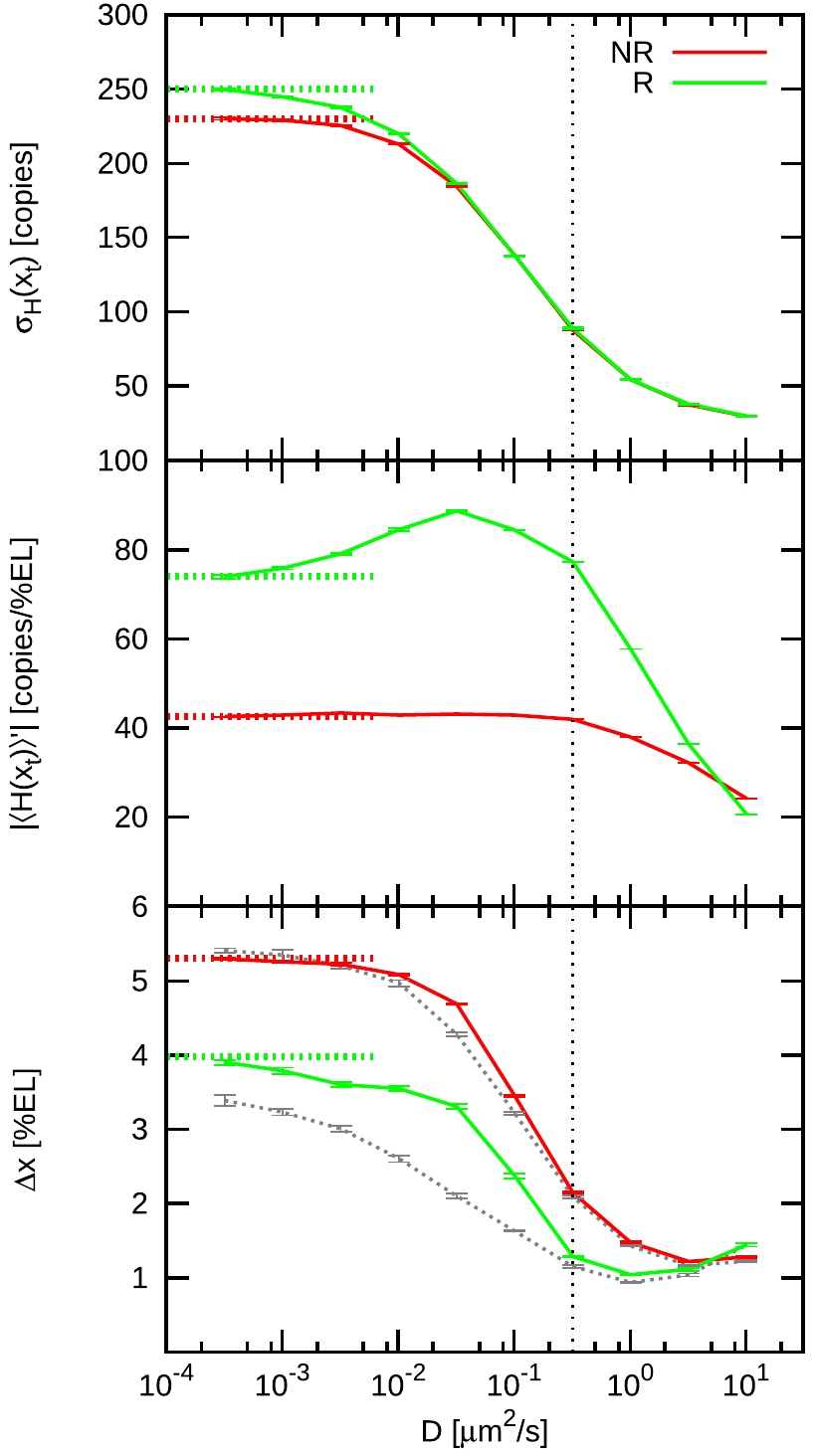}
\end{center}
\caption{ {\bf The effect of bursty gap protein production on the
  \Hb boundary precision.}
  The plot shows $\sigma_H(x_t)$ the standard deviation of the total \Hb copy number at the boundary, the
  steepness $|\Avg{H(x_t)}'|$ of the average total \Hb copy number profile at
  the boundary and the boundary width $\Delta x$ with (green) and without
  (red) mutual repression as a function of the gap protein diffusion constant
  $D$ for a system in which the gap proteins are produced in bursts of 10 at
  a time with decreased production rate $\beta=\beta_0/10$.
  The grey dashed lines are the values obtained from the approximation
  $\Delta x = \sigma_H(x_t)/|\Avg{H(x_t)}'|$.
  Thick dashed lines are values for $D=0$.
  Error bars were obtained from block averages over 10 independent samples.
  The black dotted line marks the $D$-value where the boundary is
  both steep and precise due to mutual repression.
  }
\label{FigBurstSize}
\end{figure}

\begin{figure}[!ht]
\stepcounter{myc}
\begin{center}
\includegraphics[width=3.27in]{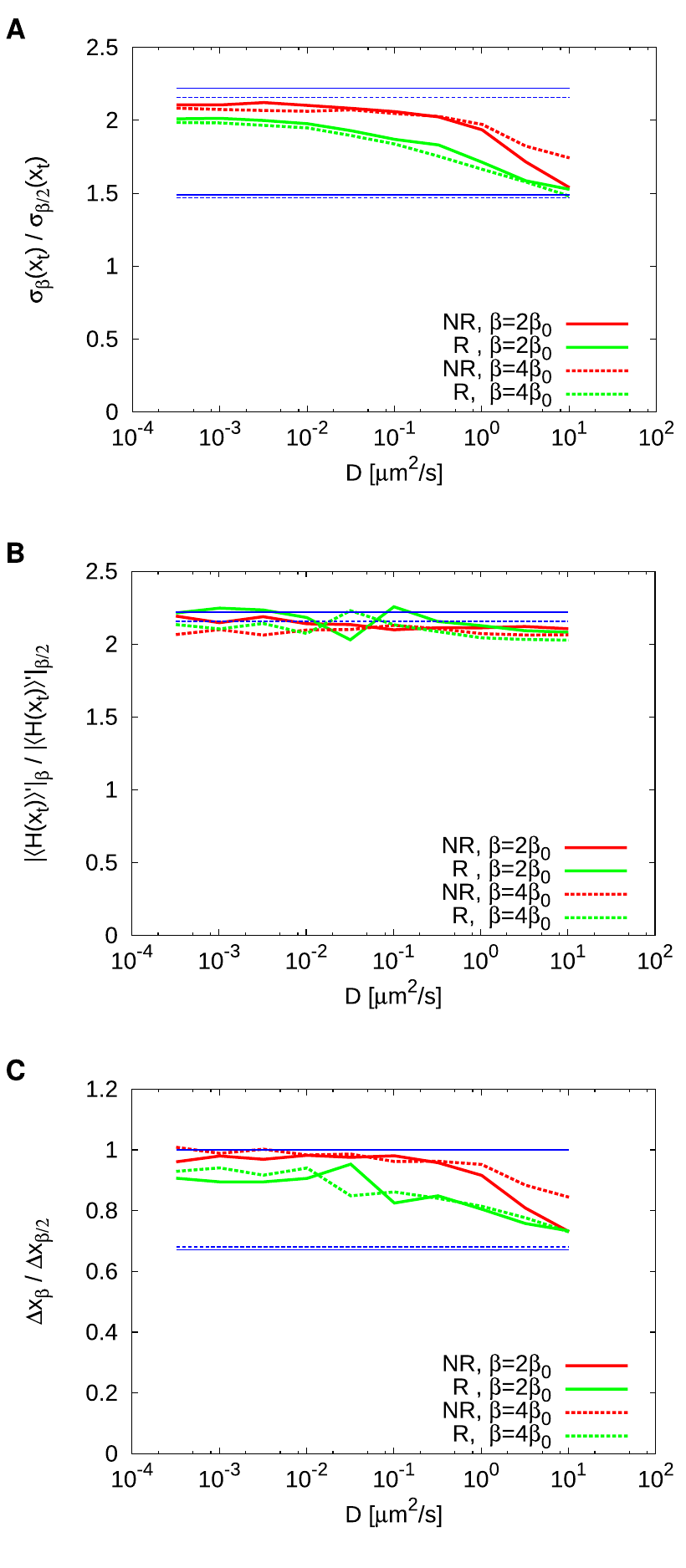}
\end{center}
\caption{ {\bf The effect of increased copy number on the
  \Hb boundary precision.}
  See following page for description.
}
\addtocounter{figure}{-1}
\label{FigProdRate}
\end{figure}

\newpage
\FloatBarrier
\captionof{figure}{ {\bf The effect of increased copy number on the
  \Hb boundary precision.}
  Shown are the value ratios of important boundary properties
  for production rates $\beta>\beta_0$ as compared to $\beta/2$ for
  \subfig{A} the total \Hb copy number noise $\sigma_H(x_t)$ at the boundary,
  \subfig{B} the steepness of the average \Hb profile at $x_t$, and
  \subfig{C} the resulting width $\Delta x$
  with (green) and without (red) mutual repression.
  Solid lines are for $\beta=2\beta_0$, dashed lines for $\beta=4\beta_0$.
  Blue lines depict the ratios as predicted from the expected scaling
  behavior for the limits of $D\rightarrow0$ (upper line pairs) and
  $D\longrightarrow\infty$ (lower line pairs).
  The steepness is expected to scale precisely with the increased
  copy number in both limits.
  Note that the expected factor of copy number increase upon doubling $\beta$
  is not precisely two because of the nontrivial dependence of the monomer-dimer
  equilibrium on the production rate.
  }
\FloatBarrier

\newpage
\section*{SUPPORTING VIDEOS}
\pdfbookmark[0]{SUPPORTING VIDEOS}{BookmarkSupportingVideos}
\newcounter{video}
\renewcommand{\thefigure}{S\arabic{video}}
\renewcommand{\figurename}{Video}
\begin{figure}[!ht]
\stepcounter{video}
\caption{ {\bf Establishment of gap gene expression patterns for a low diffusion constant of the gap proteins.}
  Movie of the total concentration of \Hb as a function of time for $D=0.01~\umsps$ and morphogen dosage factor $A=8$,
  starting from zero concentration of both \Hb and \Kni. Note that initially small ``crystallites'' are formed in the
  overlap region where both \hb and \kni are activated by their respective morphogens, \Bcd and \Cad.
  These crystallites then coarsen and join the \Hb domain formed near the anterior pole.
  The green line marks the positions where the total \Hb concentration crosses the boundary threshold value.
  }
\label{VidS1}
\end{figure}
\begin{figure}[!ht]
\stepcounter{video}
\caption{ {\bf Establishment of gap gene expression patterns for a low diffusion constant of the gap proteins.}
  Movie of exactly the same system trajectory as in Video S1, only now showing the difference between the total \Hb and
  total \Kni copy number.
  }
\label{VidS2}
\end{figure}
\begin{figure}[!ht]
\stepcounter{video}
\caption{ {\bf Establishment of gap gene expression patterns for a high diffusion constant of the gap proteins.}
  Movie of the total concentration of \Hb as a function of time for $D=0.32~\umsps$ and morphogen dosage factor $A=8$,
  starting from zero concentration of both \Hb and \Kni. 
  Note that the \Hb domain emerges at the anterior pole and progresses into the overlap region.
  The green line marks the positions where the total \Hb concentration crosses the boundary threshold value.
  }
\label{VidS3}
\end{figure}
\begin{figure}[!ht]
\stepcounter{video}
\caption{ {\bf Establishment of gap gene expression patterns for a high diffusion constant of the gap proteins.}
  Movie of exactly the same system trajectory as in Video S3, only now showing the difference between the total \Hb and
  total \Kni copy number.
  }
\label{VidS4}
\end{figure}

\pagebreak

\end{document}